\documentclass[traditabstract]{aa} 

\usepackage{hhline}
\usepackage{booktabs}
\usepackage{bm}

\usepackage[table, x11names]{xcolor}
\usepackage{array, booktabs, boldline} %
\usepackage{cellspace}

\usepackage{soul} % Used to strikethrough the text (to highlight corrections e.g. 
\usepackage{graphicx}
\usepackage{verbatimbox}
\usepackage{longtable,lscape}
\usepackage{rotating}
\usepackage{amssymb}
\usepackage{txfonts}

\usepackage{fixltx2e}

\usepackage{adjustbox}
\usepackage{amsmath}
\usepackage{relsize}
\usepackage{cleveref}
\usepackage{marginnote}
\usepackage{threeparttable}
\usepackage{comment}

\newcommand{\myemail}{antonello.calabro@cea.fr}

\definecolor{jf}{rgb}{0,0.75,0.5}

\begin{document}
%\setlength{\parskip}{0.1pt}


\title{Merger induced clump formation in distant\\infrared luminous starburst galaxies}

\author{Antonello Calabr{\`o}\inst{1} 
\and Emanuele Daddi\inst{1}
\and J{\'e}r{\'e}my Fensch\inst{2,3}
\and Fr{\'e}d{\'e}ric Bournaud\inst{1}
\and \\Anna Cibinel\inst{4}
\and Annagrazia Puglisi\inst{1}
\and Shuowen Jin\inst{5,6}
\and Ivan Delvecchio\inst{1}
\and Chiara D'Eugenio\inst{1}
}

% Antonello Calabro, Emanuele Daddi, Jeremy Fensch, Frederic Bournaud, Anna Cibinel, Annagrazia Puglisi, Shuowen Jin, Ivan Delvecchio, Chiara D'Eugenio

\institute{AIM, CEA, CNRS, Universit{\'e} Paris-Saclay, Universit{\'e} Paris Diderot, Sorbonne Paris Cit{\'e}, F-91191 Gif-sur-Yvette, France (\myemail)
\and European Southern Observatory, Karl-Schwarzschild-Stra\ss e 2, D-85748 Garching, Germany
\and Univ. Lyon, ENS de Lyon, Univ. Lyon 1, CNRS, Centre de Recherche Astrophysique de Lyon, UMR5574, F-69007 Lyon, France
\and Astronomy Centre, Department of Physics and Astronomy, University of Sussex, Brighton, BN1 9QH, UK
\and Instituto de Astrof{\'i}sica de Canarias (IAC), E-38205 La Laguna, Tenerife, Spain 
\and Universidad de La Laguna, Dpto. Astrof{\'i}sica, E-38206 La Laguna, Tenerife, Spain
}
\date{Received XXX}
%\and 18, INAF-Osservatorio Astronomico di Roma, via di Frascati 33, 00078, Monte Porzio Catone, Italy

\abstract 
{
While the formation of stellar clumps in distant galaxies is usually attributed to gravitational violent disk instabilities, we show here that major mergers also represent a competitive mechanism to form bright clumps. Using $\sim0.1''$ resolution ACS F814W images in the entire COSMOS field, we measure the fraction of clumpy emission in  $109$ main sequence (MS) and $79$ Herschel-detected starbursts (off-MS) galaxies at $0.5 < z < 0.9$, representative of  normal  versus merger induced star-forming activity, respectively. We additionally identify merger samples from visual inspection and from Gini-M20 morphological parameters. Regardless of the merger criteria adopted, the clumpiness distribution of merging systems is different from that of normal isolated disks at $> 99.5\%$ confidence level, with the former reaching higher clumpiness values, up to $20\%$ of the total galaxy emission. We confirm the merger induced clumpiness enhancement with novel hydrodynamical simulations of colliding galaxies with gas fractions typical of $z\sim0.7$. Multi-wavelength images of 3 starbursts in the CANDELS field support the young nature of clumps, which are likely merger products rather than older pre-existing structures. Finally, for a subset of $19$ starbursts with existing near-IR rest frame spectroscopy, we find that the clumpiness is mildly anti-correlated with the merger phase, decreasing towards final coalescence. Our result can explain recent ALMA detections of clumps in hyperluminous high-z starbursts, while normal objects are smooth. This work raises a question on the role of mergers on the origin of clumps in high redshift galaxies in general.
} 


\titlerunning{\footnotesize Clumpiness}
\authorrunning{A.Calabr\`o et al.}
 \maketitle
\section{Introduction}\label{introduction}

%Galaxy morphology provides an important observational evidence of galaxy evolution in cosmic history. 
While local galaxies have well defined morphological types described by the so called Hubble sequence, higher redshift systems are more irregular and clumpy, which makes it increasingly more difficult to associate one of the Hubble classes to them. At redshift $\sim0.6$ for example, the fraction of irregular systems increases by $40\%$ compared to the local Universe \citep{delgadoserrano10}.
The clumpy substructures detected in the average star-forming galaxy population at intermediate and high redshift represent aggregations of relatively young stars (depending on the observed spectral window) arising from a smoother and fainter disk luminosity profile.
They have stellar masses ranging $10^7$-$10^9$ M$_\odot$ and sizes of approximately $100$-$1000$ pc \citep{elmegreen05}, and are typically identified through highly resolved observations in the rest-frame UV \citep[e.g.][]{chapman03,puech10}, optical \citep[e.g.][]{murata14} and near-IR \citep{forster06,forster09}. 
%, thus they are usually identified in rest-frame UV \citep[e.g.][]{chapman03,puech10}, optical \citep[e.g.][]{murata14} and near-IR \citep{forster06,forster09} HST observations %or, very recently, also in high-resolution sub-mm ALMA images \citep{hodge18}. 

Clumpy galaxies at redshift $>0.5$ with stellar masses of $10^{10}$-$10^{11}$ M$_\odot$ are thought to be the progenitors of present-day spirals \citep{elmegreen05,bournaud07,ceverino10,elmegreen13}. %(Bournaud et al. 2007, Ceverino et al. 2010, Elmegreen et al. 2013). 
Moreover, typical clumps at $z>1$ might live between $100$ and $650$ Myr, depending on their stellar mass \citep{zanella15,zanella19}. As a consequence, they could migrate toward the center and eventually contribute to the formation of a central bar \citep{immeli04b,sheth12,kraljic12} or to the stellar bulge growth \citep{noguchi99,elmegreen08A,bournaud14,bournaud16}, %(Immeli et al. 2004; Sheth et al. 2012; Kraljic et al. 2012),
% The latter references come from Murata et al. 2014.
% which can shut down star-formation (morphological quenching, \citet{martig09}) and 
% jeremy.f: morphological quenching is due to bulge formation ! (actually, sheroidal stellar component due to major mergers)
contributing to stabilize the disks \citep[e.g.,][]{ceverino10} and to give the final imprint to the morphological shapes encoded in the Hubble sequence. According to \citet{kraljic12}, today Milky-way like spirals acquired their disk morphology at $z\sim0.8$-$1$, and they completely stabilized at redshift $0.5$ or lower \citep{cacciato12}. Alternatively, clumps in distant galaxies may lead to the formation of super-star-clusters and globular clusters \citep{shapiro10}. %(Shapiro et al. 2010). % or dwarf galaxies (according to Barnes \& Hernquist 1992 and Elmegreen et al. 1995). 

The formation and origin of clumps at all redshifts is not completely assessed, as it can be ascribed to different triggering mechanisms. Usually, in clumpy galaxies at $z>1$, they are thought to be triggered by violent disk instabilities in highly gas-rich, dense and turbulent disks, and they are continuously fed by cold gas streams from the circumgalactic medium (CGM) and the cosmic web \citep{dekel09}. These can sustain the high gas fractions of the order of $\sim0.4$-$0.5$ that are typically found in high redshift galaxies \citep{daddi10,tacconi10,rodrigues12}. %(ref.\footnote{(ref. Evolution of the Gas Mass Fraction of Progenitors to Today's Massive Galaxies: ALMA Observations in the CANDELS GOODS-S Field)}). Perche' Erb06 ???

Below $z\sim1$, cosmological simulations predict a strong cut-off of cold-flow accretion into galaxies \citep{keres05,dekel06}, which might indirectly result in the decrease of the average SFR density in the Universe \citep{madau14} and of the gas content in galaxies down to f$_\text{gas}\sim0.2$-$0.4$. This strong suppression of gas fraction may then affect the physical properties and abundance of clumps, and could require alternative mechanisms of formation and evolution.  
For example, smooth accretion of gas from tidally disrupted companions or stripped satellites in cluster environments have been suggested to feed some low-mass, local analogs of high-$z$ clumpy galaxies \citep{garland15}. 
Instead, a more relevant additional channel for producing clumps at any host stellar mass and epoch is represented by major mergers, as proposed in \citet{somerville01,lotz04,dimatteo08}. 
% Need to explain, because Jeremy finds no clumpiness enhancement in mergers at redshift 2 (Fensch +2017).
However, at intermediate redshifts, there is no general consensus yet about the role and importance of mergers for clumps formation. Additionally, while mergers produce very dense and compact starbursting cores, it is yet unclear what fraction of star-formation occurs in off-nuclear regions, possibly in the form of clumps.

% Murata 14 compares fraction of mergers from literature to fraction of clumpy galaxies, but do not compute directly if a galaxy is a merger or not. 
%If we assume a typical merger timescale of ∼0.5 Gyr, the expected fraction of clumpy galaxies becomes much lower than the observed fraction at z   0.5 (the dashed line in Figure 13). It seems to be difficult to explain the fraction of clumpy galaxies at 0.2 < z < 1.0 only by the wet major merger.

From the observational point of view, \citet{puech10} analyzed UV rest-frame clumps in $11$ clumpy galaxies at $z\sim0.6$. He claimed that interactions may be the dominant driver for clumps formation at that epoch, because of the complex kinematic structure observed for half of them. %, suggestive of a major-merger ongoing process. % while he also found that the majority of UV clumps tend to vanish when looking at longer wavelengths (i.e., in the optical).
In contrast, \citet{guo15} and \citet{murata14} argue that major mergers have a negligible role for explaining the fraction of clumpy galaxies at all masses at $z\lesssim1.5$, suggesting instead a prevalent role of violent disk instabilities or minor mergers.  %are the main triggering mechanism at high stellar masses (M$_\ast > 10^{10.6}$ M$_\odot$), while minor mergers may contribute the most for intermediate mass systems with $10^{9.8}<$ M$_\ast < 10^{10.6}$ M$_\odot$. 
In the same direction, \citet{bournaud12} found that clumpy low-mass star-forming galaxies at $z\sim0.7$ are similar to gas-rich turbulent disks observed at higher redshifts, suggesting that gravitational instabilities are the most important cause of gas fragmentation and clumps formation. 

However, even with different sample selections, clumps identification methods and observed photometric bands, these studies are focused on the main origin of clumpy galaxies among the whole population. Since they do not perform a merger identification, they neither analyze the implications that mergers have on clumps formation, nor they quantitatively measure the clumpiness parameter (i.e., fraction of light in clumps) in comparison to normal disks. \citet{puech10} show a possible connection between mergers and UV clumps, but many of their substructures are not detected anymore in the optical, %which arises questions about their real nature and 
thus cannot be directly compared with our sample. Moreover, their subset is not representative of the whole merger population at intermediate redshifts, given the strict requirements of their selection criteria. %being selected also with very restrictive requirements. %and above a stellar mass of $10^{10}$ M$_\odot$, being selected also with very restrictive requirements. 

The connection between mergers and clumps is not clear also in hydrodynamical simulations: while some studies suggest that merger events can trigger turbulent modes in the ISM that lead to rapid gas fragmentation and clump formation \citep{teyssier10,bournaud11,renaud14} %, others ?)
, other morphological studies on simulated galaxies do not see any enhancement at all of clumpy emission during the merger \citep[e.g.,][]{nevin19}. % jeremy.fensch: you can add Oklopcic+17, Buck+17 (cosmo sims: FIRE and NIHAO)
The reason of these discrepancies resides in the different resolutions adopted, in feedback and turbulence modeling and in the specific initial conditions considered, such as the merger geometry, the stellar mass ratio and the initial gas fraction f$_\text{gas}$ of colliding galaxies \citep[e.g.,][]{dimatteo08,governato10,fensch17}. %, others ?). 
Most of the simulations focus on local or high redshift galaxies, thus they have lower or higher gas fractions than typical values at intermediate redshifts. % (f$_{gas}\sim0.3$). 
The few simulations with similar f$_{gas}$ values \citep{cox06,dimatteo08} have not allowed so far the gas to cool down below $10^4$~K, which is necessary to study the evolution of the gas structure during the interactions \citep{teyssier10, bournaud11a}. In other cases, the resolutions are too low for the sub-kpc spatial scales we want to investigate for detecting clumpy structures \citep{sparre16,rodriguezgomez16}. 
Given all these uncertainties, more observations and constraints are needed to clarify whether mergers can trigger clumps formation at intermediate redshift, and put constraints on feedback models in simulations of galaxy collisions. 

In order to test this connection with observations, multiple merger identification criteria are applicable.
In the local Universe, for example, a deep connection has been established between mergers and ultra-luminous infrared galaxies (ULIRGs) \citep{sanders96,clements96}, which are systems with a total infrared luminosity $>10^{12}$ L$_\odot$, reflecting a high and obscured star-formation activity. 
At higher redshifts, dusty starburst (SB) galaxies, defined by their SFRs much higher compared to normal star-forming systems on the main sequence (MS) \citep{noeske07,daddi07,elbaz07}, could be the analogs of local ULIRGs.
However, their nature in the distant Universe is still strongly debated: some studies claim they may be gas rich galaxies undergoing anomalous gas accretion events \citep[e.g.,][]{scoville16}, while other works show that the most extreme cases are mergers, displaying disturbed morphologies \citep{elbaz03}. According to the latter scenario, starbursts and main sequence galaxies are associated, respectively, to a highly efficient star-formation mode and gas consumption induced by major merger events, and to a quasi-steady star-formation activity with much longer gas depletion timescales of $\sim1$-$2$ Gyr \citep{sargent14,silverman15}.
Building on this previous knowledge, \citet{calabro18} found that starbursts at $0.5<z<0.9$ are mostly merger triggered, as revealed either by their morphology or by the presence of extremely obscured starbursting cores. In addition, \citet{cibinel19} showed that the merger fraction above the main sequence is of approximately $80 \%$.
Even though the longstanding debate has not yet concluded, these latter studies give a strong motivation for searching ongoing merging systems among the starburst galaxy population. This approach would be complementary to the visual inspection and to non parametric merger estimators \citep{conselice03,lotz04}, which could be highly incomplete and very difficult to perform for distant objects. In addition, it could be also more reliable compared to the kinematics: \citet{law09} show indeed that clear merger candidates at $z>1$ can have regular rotation patterns typical of disk galaxies.

The COSMOS field allows to select the largest (so far) statistical sample of starbursts with plenty of ancillary data, including an almost complete imaging coverage in at least one HST band (F814W), photometric data ranging from UV to far-IR and sub-mm \citep{laigle16,jin18}, and IR-based SFRs \citep{jin18}.  
In this paper, we exploit high resolution ($0.095''$) HST/ACS F814W public images over the whole $\simeq2$ deg$^2$ COSMOS field to build a large sample of starburst galaxies and investigate with unprecedented statistics the role of mergers on the formation of clumps at intermediate redshifts. To this aim, we compare the starbursts to a control sample of randomly selected isolated main sequence galaxies at the same cosmic epoch.
At these redshifts, i-band images probe the optical rest-frame spectral range that, compared to UV emission (dominated by young massive O-B stars), is more sensitive to the light of intermediate age (A to G type) stars, and thus are more sensitive to the stellar mass than to the ongoing SFR. In addition, these observations are less affected by dust attenuation compared to UV rest-frame images of clumpy galaxies studied systematically at high redshift ($z>>1$).
%Given the longer rest-frame wavelengths covered, we can study the clumpy morphology in a way which is less affected by dust attenuation. This is particularly important in infrared luminous mergers, typically highly obscured, where their UV emission may not reflect the true stellar distribution. 

We also present a set of numerical simulations of merger interactions between typical z$\sim$0.7 galaxies, which could help to interpret our observational results. %in order to provide an independent check of the merger effect on clump formation. 
To this aim, these simulations consider the proper gas fractions for this cosmic epoch (f$_\text{gas}\simeq0.3$), while simultaneously allowing the gas component to cool below $10^{4}$ K and to be resolved on small spatial scales of $6$ pc, necessary to study its evolution during the merger.

%This study may pose the basis for more detailed studies of clumps properties (e.g., stellar mass, age), which may require additional bands (or emission line maps) at similar spatial resolution, and that will be addressed in future papers.

The paper is organized as follows. In Section~2 we describe our selection of starburst galaxies and of a control sample of main sequence systems in COSMOS field. We also present their HST F814W images, from which we derive basic morphological properties and an estimation of the clumpiness parameter. After showing our observational results in Section~3, we describe in Section~4 the hydrodynamical simulations used to verify the connection between mergers and clumps at intermediate redshift. Then we discuss possible physical interpretations of our findings, and we show the summary and conclusions in Section~5. 
We adopt the \citet{chabrier03} initial mass function, AB magnitudes, and standard cosmology ($H_{0}=70$ $\rm km s^{-1}Mpc^{-1}$, $\Omega_{\rm m} = 0.3$, $\Omega_\Lambda = 0.7$). % We also assume by convention a positive equivalent width (EW) for emission lines and a negative EW for lines in absorption.

\section{Methodology}\label{methodology}

In this section we describe our sample selection and the derivation of the morphological properties from single-band HST images, also quantifying, through the clumpiness parameter, the contribution of off-nuclear clumps to the galaxy emission.  

\subsection{Starbursts and main sequence galaxies selection}\label{selection}

\begin{figure*}[t!]
    \centering
    \includegraphics[angle=0,width=1\linewidth,trim={0.6cm 11cm 0.7cm 7.7cm},clip]{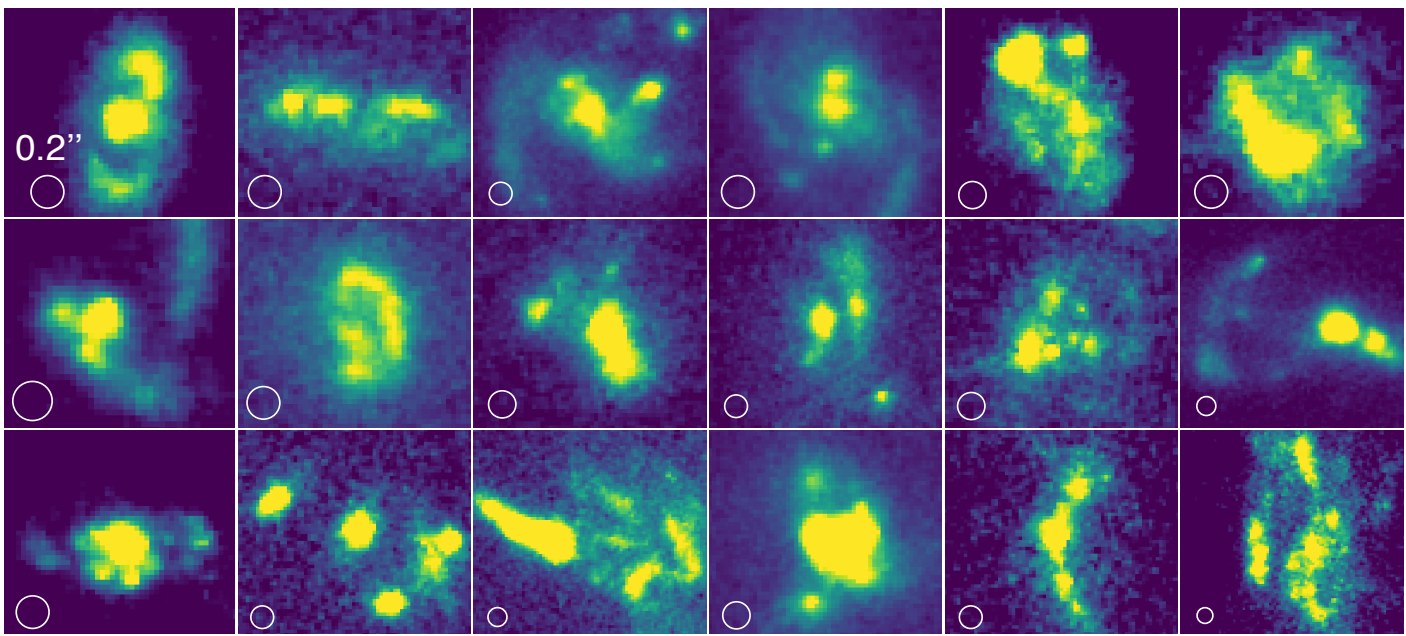}
    \caption{\small Representative sample of clumpy starbursts at $0.5<z<0.9$ in COSMOS field, observed by HST-ACS in the F814W band. The white circles in the bottom left corner have a diameter of $0.2''$. 
    }\label{clumpy_galaxies}
\end{figure*}

\begin{figure*}[t!]
    %\vspace{+0.4cm}
    \centering
    \includegraphics[angle=0,width=1\linewidth,trim={0.1cm 0.1cm 0.1cm 0.1cm},clip]{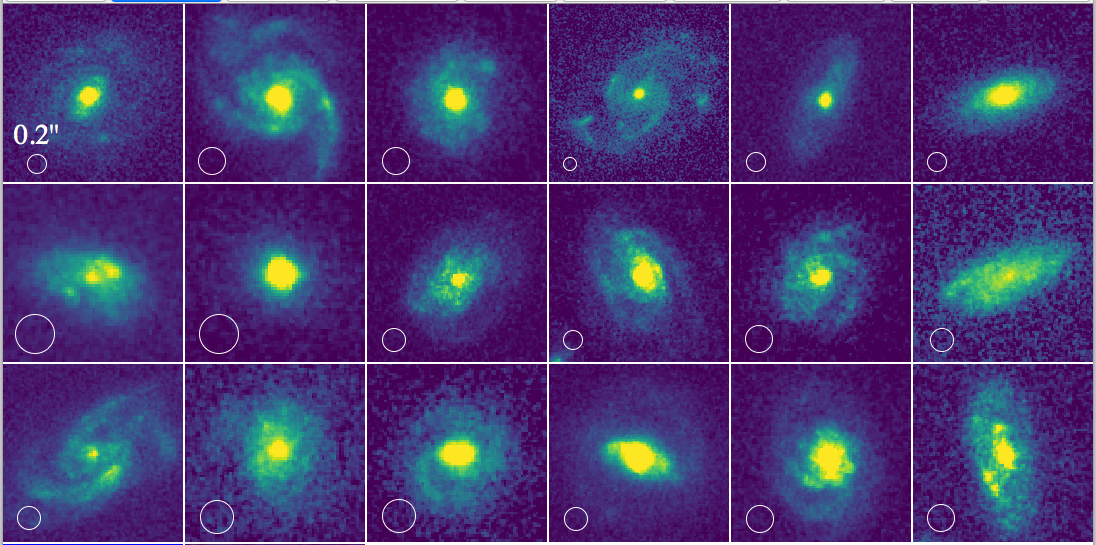}
    \caption{\small Representative sample of typical (randomly selected) main sequence disks at $0.5<z<0.9$ in COSMOS field. The image details are the same of Fig. \ref{clumpy_galaxies}. 
    }\label{clumpy_galaxies2} 
\end{figure*}

As mentioned in the Introduction, even though the nature of distant starbursts is still debated, in a previous work \citep{calabro18} we argued that luminous-infrared starbursts at $0.5<z<0.9$ are mostly major mergers. For this reason, focusing on the same redshift range, we decided to start by analysing all starbursts in the full COSMOS field to test the effect of mergers on clumps formation and evolution, while using main sequence isolated galaxies as a control sample. %. At the same time, main sequence isolated systems provide a control sample for this analysis. 

All the galaxies in this work were selected from the IR+(sub)mm catalog of \citet{jin18}, which includes $\sim 15$k star-forming galaxies in COSMOS, with deblended photometry ranging from Spitzer $24\mu m$ to VLA $1.4$ GHz bands.
The stellar masses M$_\ast$ of the sources are computed by \citet{laigle16} at the photometric redshifts by fitting $16$ bands from near-ultraviolet to mid-infrared. As explained in \citet{calabro18}, using the spectroscopic redshift (when available) does not affect significantly the stellar mass derivations, as they are in general agreement within the uncertainties of $0.1$ dex reported by the same authors. 

We adopted the infrared SFRs (= SFR$_\text{IR}$) derived by \citet{jin18} by fitting multi-wavelength broad-band photometric data from IRAC to radio VLA bands. Four components were used to determine the best-fit SED: a \citet{bruzual03} stellar model (at different ages and metallicities, constant SFH, Chabrier IMF and \citet{calzetti00} attenuation law), a mid-infrared template for AGNs \citep{mullaney11}, and \citet{draine07} dust emission models. When available, the spectroscopic redshift was fixed in the fit. For main sequence galaxies that are undetected in the $100\mu m$ - $850 \mu m$ bands (including Herschel and SCUBA2 photometry) at S/N$_\text{IR} < 5$, the SFRs were computed from their $24\mu m$ fluxes as explained in \citet{jin18}. %, after an upward rescaling by a factor of $1.5$ for systematics, consistently with \citet{jin18}. %From the SED fitting explained above, IR-based photometric redshifts were also derived, using the \citet{laigle16} photometric estimates as a first guess, while it was fixed to the spectroscopic value when available. 
Finally, we calculated the total SFRs as SFR$_\text{TOT}$=SFR$_\text{IR}$+SFR$_\text{UV,unobscured}$, where the UV unobscured SFRs were inferred from their \citet{laigle16} u-band total magnitudes, as shown in \citet{calabro18}.    

Following the procedure described in \citet{calabro18,calabro19}, we compared the M$_\star$ and SFR of our galaxies to the star-forming main-sequence of \citet{schreiber15} (which was shown to agree well with our data) in order to derive the distance from the main sequence dist$_\text{MS}$ $=$ log$_{10}$ (SFR/SFR$_\text{MS}$). %In practice, for each source at a different redshift, we estimated this quantity considering a MS evolving with redshift \citep{sargent15} 
Finally, starbursts are taken above the typical threshold of dist$_\text{MS}>0.6$ (i.e., above a factor of four), following \citet{rodighiero11}.

We required also the redshift, either spectroscopic or photometric \citep[from][]{jin18}, to be in the range between $0.5$ and $0.9$. As mentioned in the Introduction, we want to focus on the intermediate redshift regime to build-up on the growing understanding generated by our previous works \citep{calabro18,calabro19}.  %and also the feedback and ISM modeling in simulations is particularly challenging. 
%This constrain is additionally motivated to build-up on the growing understanding generated by our previous works \citep{calabro18,calabro19}. 
Indeed, it would allow to follow-up some of the clumps with near-infrared integral-field spectrographs to detect H$\alpha$ and Pa$\beta$ within Y and K bands. In addition, it is in this cosmic epoch that we may have the final imprint on the morphology of present-day galaxies.

As an additional constraint for our sample, we imposed the stellar mass M$_\ast$ to be greater than $10^{10}$ M$_\odot$, since above this limit and up to $z=0.9$, all SBs would be Herschel-detected at S/N$_{IR}$ $> 5$ (see Figure 13 in \citet{jin18}), and we have a mass-complete sample of normal star-forming galaxies down to a factor of three below the main sequence.

These three criteria yielded a subset of $118$ Herschel-detected (S/N$_\text{IR}$) starbursts, from which we discarded two residual quasar-like objects and additional $20$ galaxies without HST-ACS coverage, leaving a sample of $96$ starbursts in total. 

Afterwards, we selected a control sample of $145$ MS galaxies with HST F814W images, in order to have a larger statistics and avoid that additional sample cuts in our analysis would produce a lower number of MS systems than SBs. Our normal star-forming galaxies were randomly selected within $\Delta$dist$_\text{MS}$ of $\pm0.47$ dex (a factor of three) from the MS, in the same redshift and stellar mass range defined for SBs, and requiring that they are star-forming according to the NUV-R/R-J diagram to avoid quiescent galaxies \citep{laigle16}.

Even though this latter subset is representative of secularly evolving star-forming disks, we remember that it may contain also a fraction of ongoing mergers. Indeed, at our redshifts, the merger fraction is expected to be higher than the relative number of starbursts \citep{schreiber15}, both because of SFR fluctuations during the merger process itself \citep{dimatteo08} and because it might be more difficult for mergers to trigger starbursts as in the local Universe \citep{fensch17}. We will discuss about several additional methods to identify mergers in Section \ref{ginim20section}.

\subsection{HST images and morphology}\label{hstimages}

In order to study the morphology and the presence of clumpy structures in our galaxies, we need high resolution images that can probe spatial scales significantly below $1$ kpc.
For the sample analyzed in this paper, we adopted F814W ACS images \citep{koekemoer07}, which we retrieved from the COSMOS web service (\url{http://irsa.ipac.caltech.edu/data/COSMOS/}{http://irsa.ipac.caltech.edu/data/COSMOS/}). They represent so far the deepest and highest resolution publicly available data in this field, with $\sim 2000$ s time integration, a magnitude limit of $25.61$ mag for extended sources (assuming a circular aperture radius of $0.3''$) and a median FWHM resolution of $\sim 0.095''$ (with a pixel scale of $0.03''$/pixel). 
At our redshifts, this means that we are able to distinguish substructures with separations of at least $\simeq600$ pc at $z=0.5$ to $\simeq800$ pc at $z=0.9$, and thus detect clumps on this size scale (FWHM), which is appropriate given those typically found in high-redshift clumpy galaxies. %and even resolve the biggest clumps thus enough to detect t clumps that are typically observed both in lower and higher redshift clumpy galaxies (see Section \ref{Introduction}). 

In the first step of our analysis, we created a segmentation map of the HST F814W ($15''\times15''$) cutouts using the python package \textsc{photutils}\footnote{\url{https://doi.org/10.5281/zenodo.2533376}{https://doi.org/10.5281/zenodo.2533376}}. In brief, the code identifies the sources as groups of connected pixels having a flux higher than a constant threshold. The latter value is calculated at a given S/N per pixel above the background, which is estimated from the entire cutout using a sigma-clipped statistics. We found that a S/N threshold of $1.3$ works well in all the cases, including the low surface-brightness external regions and wings, while separating different galaxies in the same cutout region. The few cases where two close-in-sky but spatially unrelated galaxies (i.e., located at completely different redshifts) are selected as the same source in the segmentation map, we applied the \textit{deblending} function inside \textsc{photutils}, keeping only the central object. The location of the final source was saved into a mask M$_0$: we assigned a value of 1 to all the pixels inside the galaxy contours identified by the segmentation map, and 0 otherwise.

Afterwards, we run the galaxy morphology tool \textsc{statmorph} \citep{rodriguezgomez19} to derive the elongation of the galaxy contained in the previously selected region, which we expect can crucially affect the detectability of clumpy structures. \textit{E} is defined as $\frac{A}{B}$, where A and B are, as in SExtractor, the maximum and minimum $rms$ dispersion of the object profile along all directions. Equivalently, they can be considered as the semi-major and semi-minor axis lengths of the ellipse that best describes the galaxy shape. 

\subsection{Morphological merger classification}\label{ginim20section}

In addition to the elongation, better characterizing the morphological properties is essential to identify which galaxies in the main sequence could be possibly mergers. To this aim, we pursued two approaches: one relying on non-parametric quantitative estimators, and the second based on a visual analysis. 

In the first approach, the Gini and M20 coefficients, defined by \citet{lotz04}, are usually adopted for selecting possible mergers and interacting systems. 
Gini (G) measures the degree of inequality of the flux distribution among the pixels in the image, and is higher for galaxies with bright clumps and nuclei. For our HST cutouts, we computed G with the following formula: 
\begin{equation}\label{eq1l}
G=\frac{1}{\bar Xn(n-1)}\sum_i^n(2i-n-1)X_i ,
\end{equation}
where n is the number of pixels of the galaxy (defined by the mask M$_0$), X$_i$ are the counts in each pixel $i$ sorted in increasing order and $\bar X$ is the mean pixel value \citep{glasser62}. 

On the other hand, M20 is defined as the normalized second order moment of the brightest $20\%$ pixels of the galaxy:
\begin{equation}
\begin{aligned}
M_{20}=log_{10}\left(\frac{\sum_i M_i}{M_{tot}}\right)\text{, with} \sum_i f_i < 0.2 f_{tot} \\
\text{with } M_{tot}=\sum_i^n M_i = \sum_i^n f_i [(x_i-x_c)^2+(y_i-y_c)^2]
\end{aligned}\label{eq2l}
\end{equation}

In the above formula, $x_i$ and $y_i$ are the pixel coordinates, and $x_c$ and $y_c$ represent the galaxy's center, such that M$_\text{tot}$ is minimized. $f_i$ are the counts in each pixel, while $f_\text{tot}$ symbolizes the total counts in the galaxy pixels identified by segmentation map derived before. 
In short, this quantity measures the relative concentration of light around the position that minimizes M$_{20}$ itself. It is higher in the presence of bright bars, spiral arms, tidal tails, off-center clumps, and it is very sensitive, for example, to multiple nuclei. Both parameters were calculated with python code routines by applying the equations \ref{eq1l} and \ref{eq2l}. 
\begin{figure}[t!]
    \centering
    \includegraphics[angle=0,width=\linewidth,trim={1.6cm 0.5cm 3.5cm 2cm},clip]{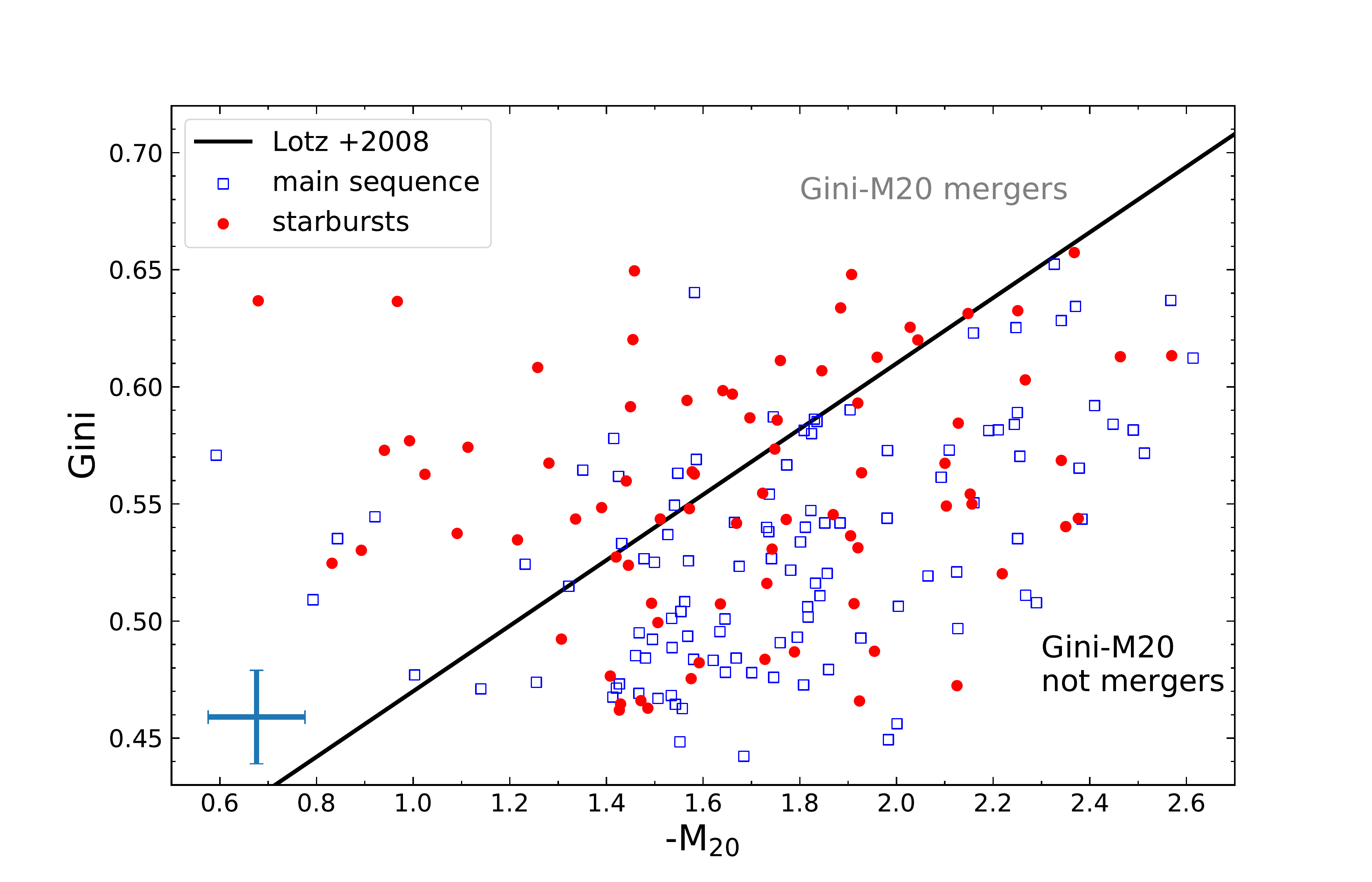}
    \includegraphics[angle=0,width=\linewidth,trim={1.6cm 0.5cm 3.5cm 2cm},clip]{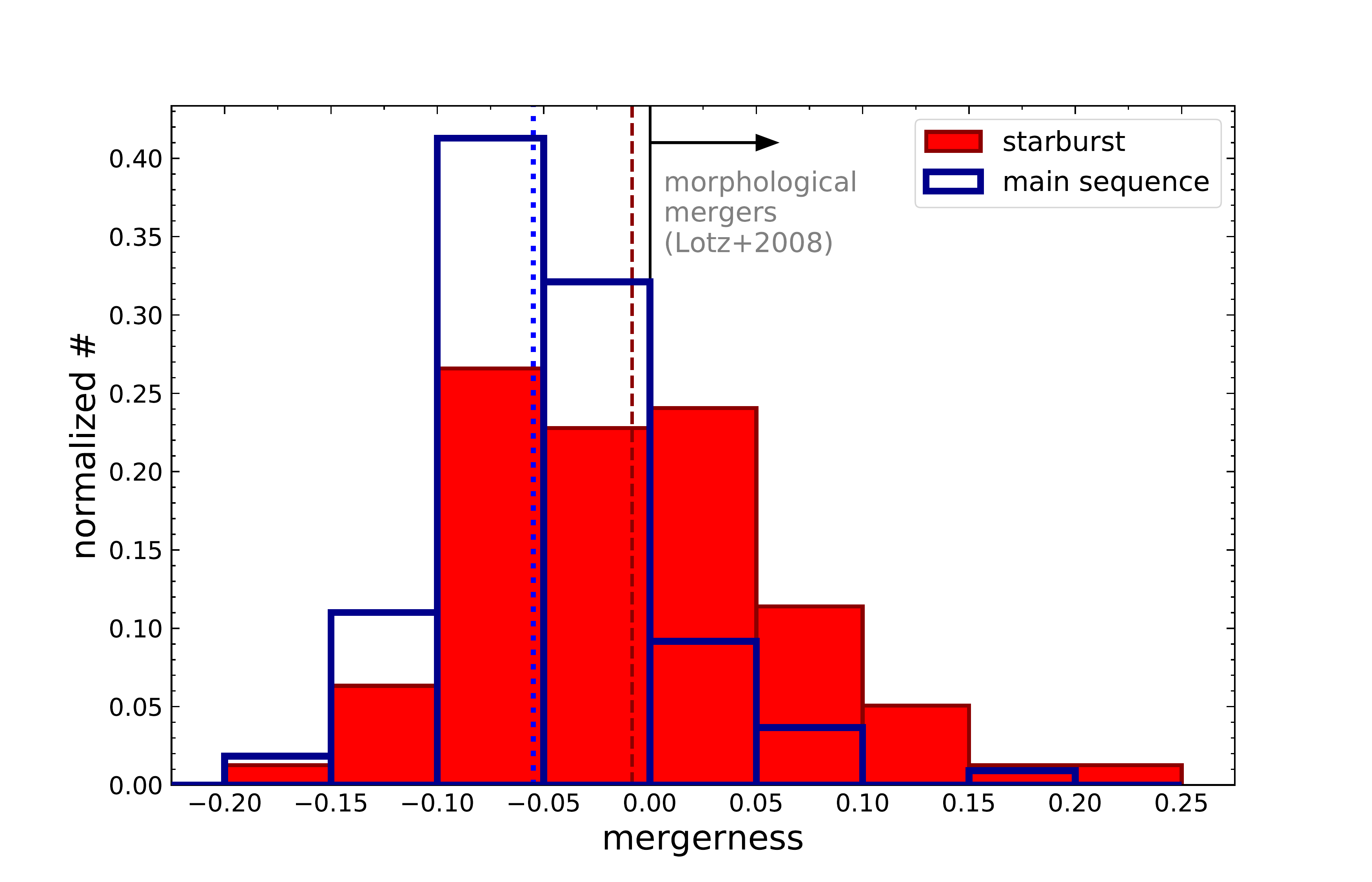}
    \caption{\small \textit{Top:} Gini-M$_{20}$ diagram for the final sample of starburst and main sequence galaxies analyzed in this work (in red filled circles and blue empty squares, respectively). Galaxies above the black continuous line are morphological mergers, according to \citet{lotz08}. The typical uncertainties of Gini and M$_{20}$ are shown by the representative error bars in the lower left corner. \textit{Bottom:} Histogram distribution (for SBs and MS) of the $mergerness$ parameter, defined as the difference between the measured Gini coefficient and that required to classify the galaxy as a morphological merger. It shows that the majority of galaxies classified as mergers according to this criterion are also starbursts.}\label{gini_m20}
\end{figure}

The typical uncertainties of Gini and M$_{20}$ estimations are of $0.02$ and $0.1$, respectively. They were estimated by \citet{lotz08} from ACS F814W galaxy images at our same redshifts in the EGS field, at a depth comparable with our analysis ($\sim 2000$ s of integration). In addition, the ULIRGs in their sample, to which they apply the classification, have I$_\text{F814W}$ $<23$ mag, thus largely applicable to our case.

Since Gini and M$_{20}$ are very sensitive to typical merger features, increasing when these signatures become stronger, we can use both to identify a subset of galaxies with merger morphologies. Following the classification criteria of \citet{lotz08}, we defined a `mergerness' parameter $m$ as:
\begin{equation}\label{mergerness}
mergerness = Gini + 0.14 \times M_{20} - 0.33 ,
\end{equation}
where the coefficients were calibrated by \citet{lotz08} and do not vary with redshift (up to $z \simeq 1.2$). This quantity represents the difference between the estimated Gini and that required to classify the system as a merger. Therefore, according to this criterion, galaxies with $m>0$ will be identified as morphological mergers throughout the paper. Among the MS population, $15$ objects satisfy this condition. The exact $mergerness$ values of all the galaxies are listed in Table~A.1 in the Appendix. 

We remark that for our sample we cannot apply the morphological analysis performed by \citet{cibinel19} on resolved stellar mass maps, since it requires multi-wavelength images. However, our single band optical rest-frame images are sensitive to the stellar mass of the system more than UV rest-frame observations, and this dataset represents so far the best compromise if we need high spatial resolution information.   

We show in Fig. \ref{gini_m20}-$top$ the distribution of the Gini and M$_{20}$ coefficients for our galaxies, and highlight with the black continuous line the merger threshold of \citet{lotz08}. In the bottom panel of Fig. \ref{gini_m20} we display instead the histogram of mergerness parameters, separately for main sequence and starburst galaxies. These two populations have overall a different distribution of $m$ and different medians, where starbursts tend to have a higher mergerness compared to normal star-forming galaxies. % in the same mass and redshift range.  

We notice that the threshold criterion of \citet{lotz08} defines a mergerness parameter space where SBs start to dominate in number over the MS population. In our case, $76\%$ of the morphological merger systems ($m>0$) turn out to be starbursts.  %choosing a random sample of morphological mergers ($m>0$) in the stellar mass and redshift range defined in Section \ref{selection}, we found that $76\%$ of them are starbursts. 
However, this subset is not complete, representing a minor fraction ($43\%$) of all the starburst population in our sample. %, which contains a significant number of objects with Gini-M$_{20}$ properties similar to normal star-forming disks. 
This result should not be surprising. %, as mergers do not always appear morphologically disturbed. 
Indeed, the observability timescale of a merger in the upper part of the Gini-M$_{20}$ diagram may not coincide with the starburst phase duration and depends on many factors, including the mass-ratio and the gas fraction of colliding galaxies, the viewing angle, the impact geometry, the dynamics of the disks (e.g., rotation direction) and the extinction. 
For example, the Gini parameter is more sensitive to face-on systems \citep{lotz08}. In addition, the surface brightness at redshift $z>0.5$ decreases by more than one order of magnitude compared to the local Universe, making more difficult for interacting signatures to emerge from the noise. All these mechanisms are thus likely responsible for the fraction of low-$mergerness$ SBs identified in our sample. 

\subsection{Visual classification}\label{visual_classification}

In order to mitigate previous effects and select a more complete subset of merging systems, we also performed a visual classification. Based on a one-by-one inspection of MS and SB galaxies, we flagged as `visual mergers' all the systems with a disturbed morphology because of the presence of clear tidal tails, shells, bridges or collisional rings. In addition, we included in this class all the pairs within a projected separation of $<20$ kpc and photometric redshift difference $<0.08$.  
Compared to the previous criterion, this likely identifies mergers on a longer timescale. On the one hand, the pairs select a sample of very-early stage mergers that are going to coalesce in $\sim 0.5$ Gyr or more. On the other hand, `visual mergers' also comprise already coalesced colliding systems, if residual merger signatures are sufficiently bright to be detected by eye. These features are typically too faint to contribute significantly to enhance the Gini and M20 parameters above the \citet{lotz08} limit, so these mergers are likely missed from the automatic procedure.

% Cosa hai trovato
This classification yielded a sample of $78$ visually selected mergers: $24$ of them are in the main sequence ($\sim20\%$ of the whole MS population), while the remaining $54$  objects lie in the starburst regime. Therefore, we obtain in this case that the majority of starbursts ($\sim 68\%$) are also visual mergers.
% Sono di piu' rispetto ai morphological mergers - a causa di quanto detto prima. cioe' del longer timescale probed. 
We also notice that the sample of mergers identified visually is more numerous compared to morphological mergers, which is likely an effect of the different timescales probed by the two diagnostics, as explained above. 

\subsection{Clumpiness measurement}\label{clumpiness}

In order to quantify the contribution of young stellar clumps to the total galaxy emission, we adopted the clumpiness parameter $c$, which measures the fraction of light residing in high spatial frequencies structures. % oppure "Previous studies have defined a clumpiness"
For its estimation, we follow the approach described in \citet{conselice03} and \citet{lotz04}. 

Firstly, we smoothed the original ACS F814W images (I$_0$) using a gaussian filter with a radius of $5$ pixels. This corresponds to an angular scale of $0.15''$ and a physical radius of $\simeq1$ kpc at $z\simeq0.7$ ($0.9$ to $1.16$ kpc in our full redshift range), which is the approximate size of the clumps we want to detect (see Section \ref{introduction}). Then we subtracted the smoothed image I$_\text{smoothed}$ from the original image, imposing $0$ for all the pixels with a negative value in the residual image I$_\text{res}$ $=$ I$_0$ - I$_\text{smoothed}$, as done by \citet{conselice03}. 

In a second step, following the procedure adopted in \citet{salmi12}, we selected all the pixels in I$_\text{res}$ which are at least $5\sigma$ above the background of the galaxy, in order to reduce the noise contamination. The background level was estimated with a $\sigma$-clipping statistics procedure applied on all the regions of I$_0$ which have not been assigned to any sources in the segmentation map. 
The threshold limit was chosen empirically, and we found it was the minimum and best value allowing to recover clumpy structures that would have been identified also on a visual inspection. 
For all the pixels above the $5\sigma$ threshold, we assigned them a value of $1$ ($0$ otherwise), in this way defining a mask for the clumps (M$_\text{clumps}$). We did not put any constraints on the number of connected pixels to be part of the clumps. However, even when requiring a small amount (e.g., 2-10) of connected pixels, the results are not significantly affected. 

\begin{figure}[t!]
    \centering
    \includegraphics[angle=0,width=\linewidth,trim={0.1cm 0.cm 0.1cm 0cm},clip]{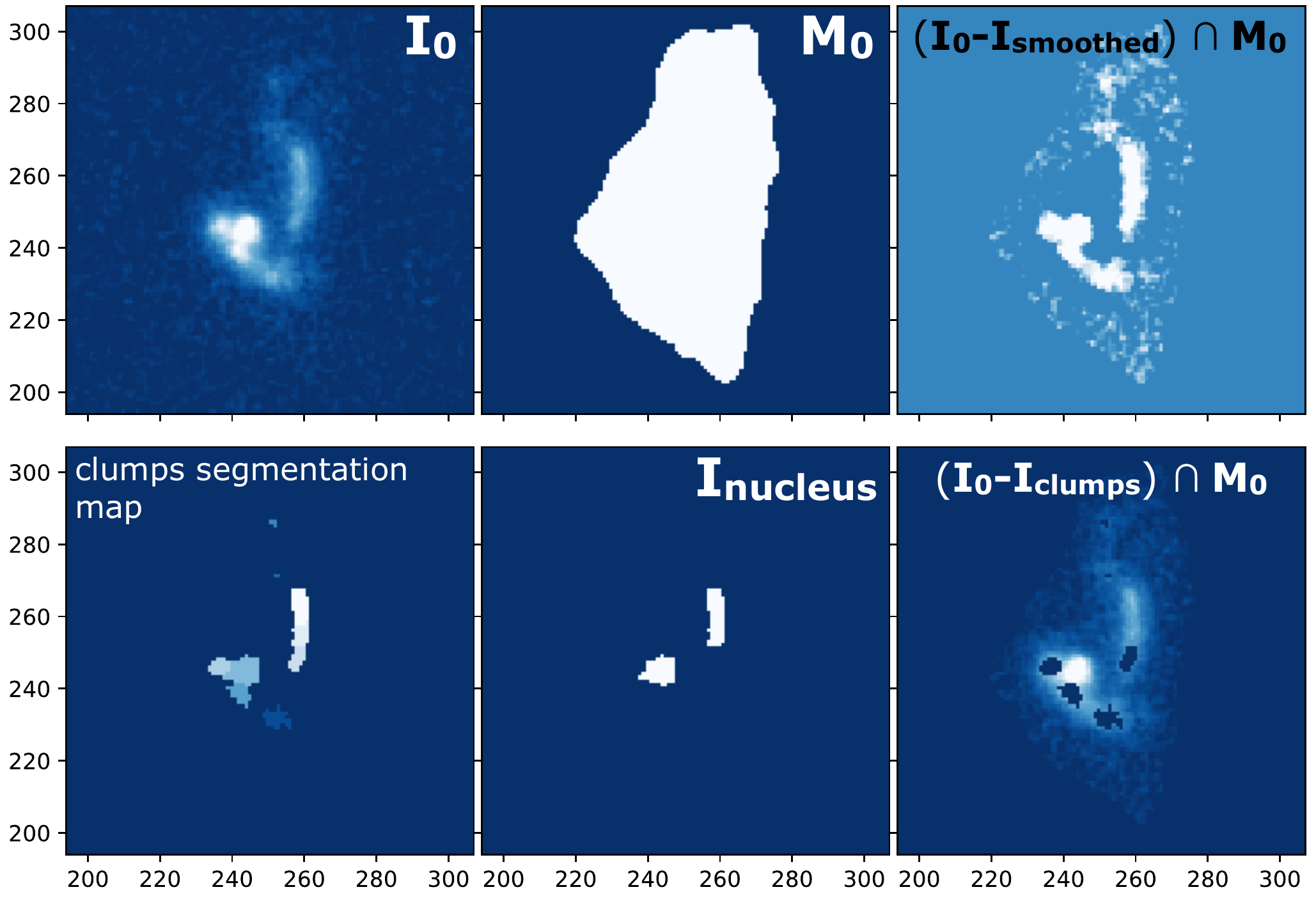}
    \caption{\small Flowchart of the clumpiness estimation procedure for one galaxy in our sample (ID 412250): (1) original HST F814W cutout image; (2) segmentation image identifying the galaxy contours; (3) original minus smoothed image, enhancing the visibility of high spatial frequency components; (4) clump detection after applying a $5\sigma$ threshold; (5) nuclei visual identification; (6) residual image after clumps light subtraction, which appear as black regions superimposed to the original galaxy image. This figure highlights the power of our approach. We are able to detect clumps very close to the nucleus that would instead have been removed by masking systematically a circular region around the center. It shows also the important role of the deblending function to separate multiple clumps based on the presence of multiple peaks in a single segmentation region, as explained in the text. This case additionally illustrates our conservative approach: we select two nuclei in the clumps segmentation map even though we are not sure about the second on the right (which in alternative could be part of a tidal tail).}\label{flowchart}
\end{figure}

In the third step, we removed the galaxy nuclei from the clumpiness calculation, which, by definition, contains only off-nuclear clumpy structures. For example, the nuclei of spiral galaxies are usually made of old stellar population bulges that we do not want to consider in the above parameter. %\textcolor{blue}{there is another reason to remove them (written in some Lotz papers), luminosity profile diverges there and this can create problems}. 
For this scope, we derived the segmentation map of the clump mask, deblending the regions containing more than 2 local luminosity peaks, by using the same \textit{python} codes applied to the original image in the first step. 
Afterwards, we created the nucleus mask M$_\text{nucleus}$, setting M$_\text{nucleus}=1$ for the clumps identified as nuclei by a visual inspection of the original i-band HST images, and $0$ otherwise. %The nucleus is centrally located and usually (but not necessarily) coincides with the most luminous concentration of stars. This regions will thus be excluded from the clumpiness calculation. 
In most of the cases, the centrally located nuclei correspond to the brightest clumps identified through our routine. However, this is not a necessary condition, since the nuclei (especially the merger cores) can be very obscured and even undetected in optical. Additionally, simulations have shown that the luminosity of newly formed clumps in merger systems can easily overcome that of the nuclei in the two colliding disks. 
Despite these uncertainties, we conservatively selected and removed at least two nuclei in all the starbursts and in clear MS mergers identified in Section \ref{hstimages}. This represents a limiting case, since we expect that a large fraction of starbursts may actually be fully coalesced systems, even though it is hard to securely isolate them with our data. 

Other works systematically mask an inner circle region with a fixed angular aperture for all the galaxies when computing the clumpiness \citep[e.g.,][]{lotz04}. However, this method may remove clumps that are very close to the nucleus. In addition, it is difficult for any automatic procedure to identify the nucleus, especially in case of disturbed morphologies like in mergers, where a careful visual inspection can be more reliable. %Furthermore, similarly to other automatic ways to indentify a clump as the galaxy nucleus (e.g., based on its position within the galaxy or the total luminosity), , which rather requires a careful visual inspection for each source to remove the nuclei. 
For these reasons, we believe that our method is more precise and can be easily kept under control, assuming that the size of our sample is not excessively large. We show in Fig. \ref{flowchart} a representative example (galaxy ID 412250) of the full identification procedure of our clumps. 

Finally, the clumpiness parameter $c$ was derived in a standard way by dividing the total flux residing in previously detected clumps and the total flux of the galaxy, after masking the nucleus. This calculation can be written explicitly as:
\begin{equation}\label{equation11}
c=\mathlarger{\mathlarger{\sum}}\limits_{M_0 \bigoplus M_\text{nucleus}} \frac{M_\text{clumps}(i,j) \times  I_0(i,j)}{I_0(i,j)}
\end{equation}
where I$_0$ is the original image and M$_{0}$, M$_\text{nucleus}$ and M$_\text{clumps}$ are masks, already introduced before. The sum is done over all the galaxy pixels defined by the segmentation map (M$_{0}$), excluding the nucleus.
We consider this quantity appropriate for our work, since it compares the flux of the clumps (outside the nucleus by definition) to the total flux in the same off-nuclear regions. 

Another viable option is to compare the emission residing in high spatial frequencies to the total object emission (including the nucleus) as: $c'=$ $\sum$ [ $|M_\text{nucleus}(i,j)-1|$  $\times$ $M_\text{clumps}(i,j) \times  I_0(i,j)$] / $\sum$ $I_0(i,j)$, where the sum is over M$_0$. For clarity, since this would not change the results of our analysis, we adopt uniquely the definition in Eq.\ref{equation11} throughout the paper, and include both quantities $c$ and $c'$ in Table~A.1 in the Appendix.

\subsection{Magnitude and elongation cuts: building the final sample}\label{cuts}

\begin{figure}[t!]
    \centering
    \includegraphics[angle=0,width=\linewidth,trim={1.3cm 3.cm 0.9cm 7.3cm},clip]{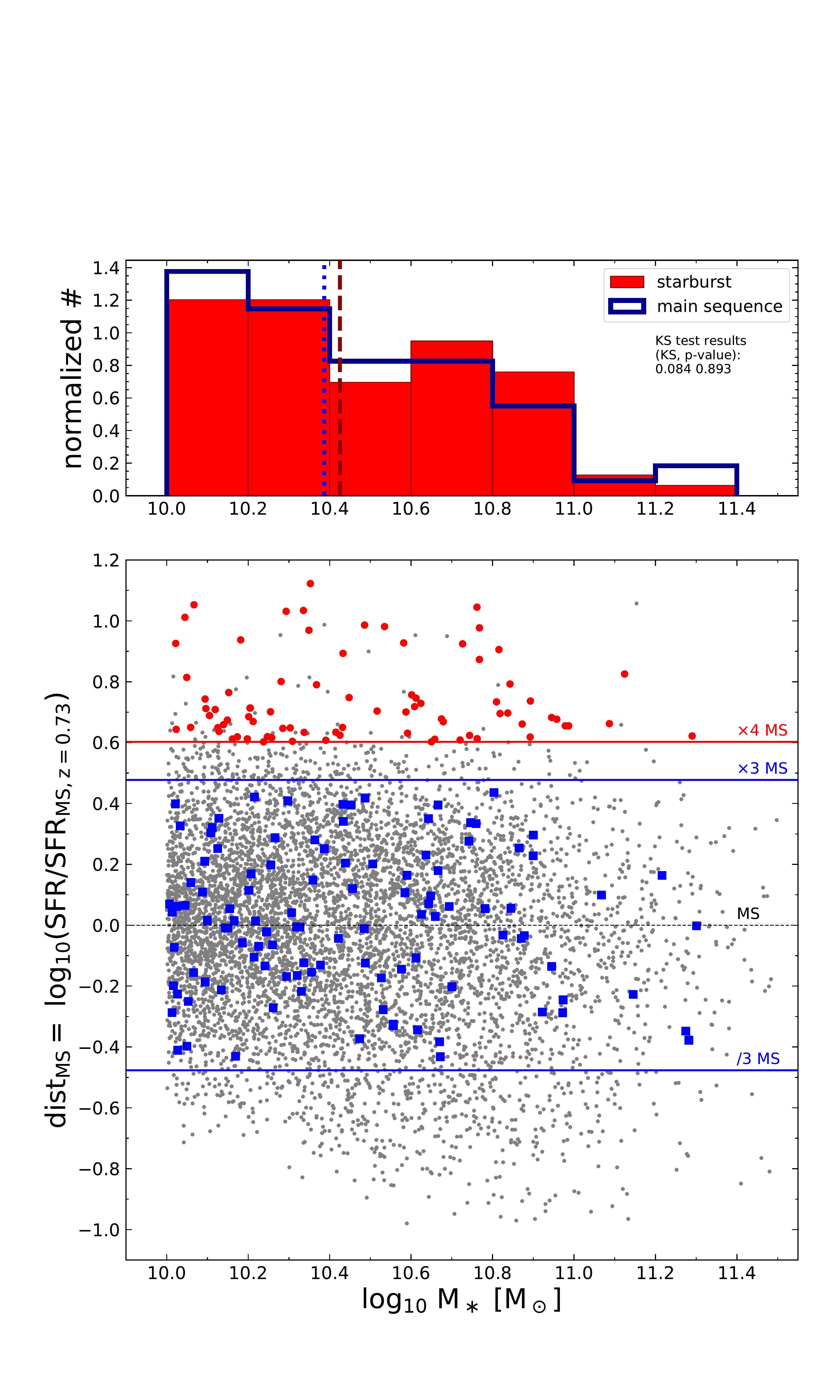}
    \caption{\small \textit{Bottom:} Diagram showing for our parent sample of star-forming galaxies in COSMOS ($0.5<z<0.9$) their distance from the main sequence as a function of stellar mass. We highlight with a black horizontal line the 0 level. The blue and red lines indicate the limits taken for our main sequence and starburst selection, respectively. The final selected sample is highlighted with blue squares and red circles, correspondingly. The SFR used in the $y$-axis is normalized to the median redshift of our sample ($0.73$), following the SFR-redshift evolution of \citet{sargent14}. \textit{Top:} Histogram distribution (normalized to unit area) of the stellar masses of our selected sample of starbursts (red) and main sequence galaxies (blue), showing that the two subsets have similar distributions. The two vertical lines indicate, according to their corresponding colors, the median values of M$_\ast$ of SB and MS galaxies.}\label{selection_mass}
\end{figure}

At $z\sim0.5$, the brightness of all the objects is approximately one order of magnitude lower (at fixed luminosity) compared to the local Universe, so it becomes increasingly difficult at higher redshifts to identify internal structures of galaxies, such as clumps. Moreover, when the galaxies become too faint, the visibility and detection of clumps is automatically affected, so that our method returns systematically lower $c$ values close to $0$.
For this reason, we used the i-band magnitudes of \citet{laigle16} and applied a threshold as $i_\text{mag} < 22.5$, beyond which the average clumpiness of galaxies (computed in bins of $0.5$ in $i_\text{mag}$) drops by $>50\%$ compared to the median value and becomes closer to zero (see Fig. \ref{sample_cuts} in Appendix). In addition, for all the galaxies with $i_\text{mag}>22.5$, it was harder to distinguish their internal morphology on a visual inspection. 

Another important aspect that hampers clumps identification is the inclination of the galaxy.
For an edge-on system, the standard detection method can erroneously consider the whole disk as a single elongated (and eventually multiply deblended) clump, producing an artificial enhancement of the clumpiness up to a value close to 1. In some cases, the edge-on disk is very faint, probably attenuated by the increased dust column density along the line of sight, so that no substructures are detected and the clumpiness is 0. However, the majority of these objects would be removed by the first $i_\text{mag}$ threshold. 

In order to avoid all these cases, we performed a visual inspection and removed all the edge-on galaxies that suffered from these problems. Since inclination effects can be different for each galaxy and both reduce or enhance their clumpiness, it was not possible to apply the same threshold procedure used for $i_\text{mag}$. However, we found that almost all ($96\%$) of the objects discarded by eye have an elongation $>3.5$, so we can consider this value as a representative threshold for our selection. For the full original sample, a comparison between the clumpiness and both the i-band magnitude and the elongation is shown in Appendix \ref{additional_plots}. 

We remark that the i$_\text{mag}$ and elongation cuts remove similar fractions of starburst and main sequence systems ($\sim10\%$ in each case), and also the same percentages of morphologically classified mergers and not mergers (according to Gini-M$_{20}$ diagram), thus no systematic biases are introduced against either of the two populations. We also notice that almost all of the objects removed by this procedure have very low clumpiness, below $0.05$. 
After applying these cuts, we also verified that starburst and main sequence galaxies have similar histogram distributions in i-band magnitude and elongation in the allowed ranges, and very close medians of the two quantities. In particular, for SBs and MS galaxies, the medians i$_\text{mag}$ are $21.25$ and $21.35$, respectively, while the median elongations are $1.55$ and $1.59$. In any case, we verified that applying lower, more conservative thresholds in $i_\text{mag}$ and elongation would not alter the conclusions of this paper.

After cleaning the sample from the contamination of faint or very elongated objects, we derived a final subset of $79$ starbursts and $109$ main sequence galaxies, that we will analyze in the following sections. 
The final starbursts and main sequence sample selection can be visualized in Fig. \ref{selection_mass}. The stellar mass histogram, visible on the top of the figure, shows that the two selected populations have a similar distributions of M$_\ast$, with medians of log$_{10}$(M$_\ast$)$=10.42$ and $10.39$, respectively, thus our cuts do not introduce systematic biases in M$_\ast$ against one of the two populations. 

\section{Results}\label{results}

In this Section we present the results of our clumpiness measurements, and compare the properties of the starburst and main sequence populations, that we have taken as representatives of two star-formation modes: a higher efficiency stellar production induced by merger events in the first case, and a normal star-formation activity associated with secularly evolving disks in the latter.  % two star-formation modes: 

\begin{figure}[t!]
    \centering
    \includegraphics[angle=0,width=\linewidth,trim={2.2cm 0.1cm 3.5cm 2cm},clip]{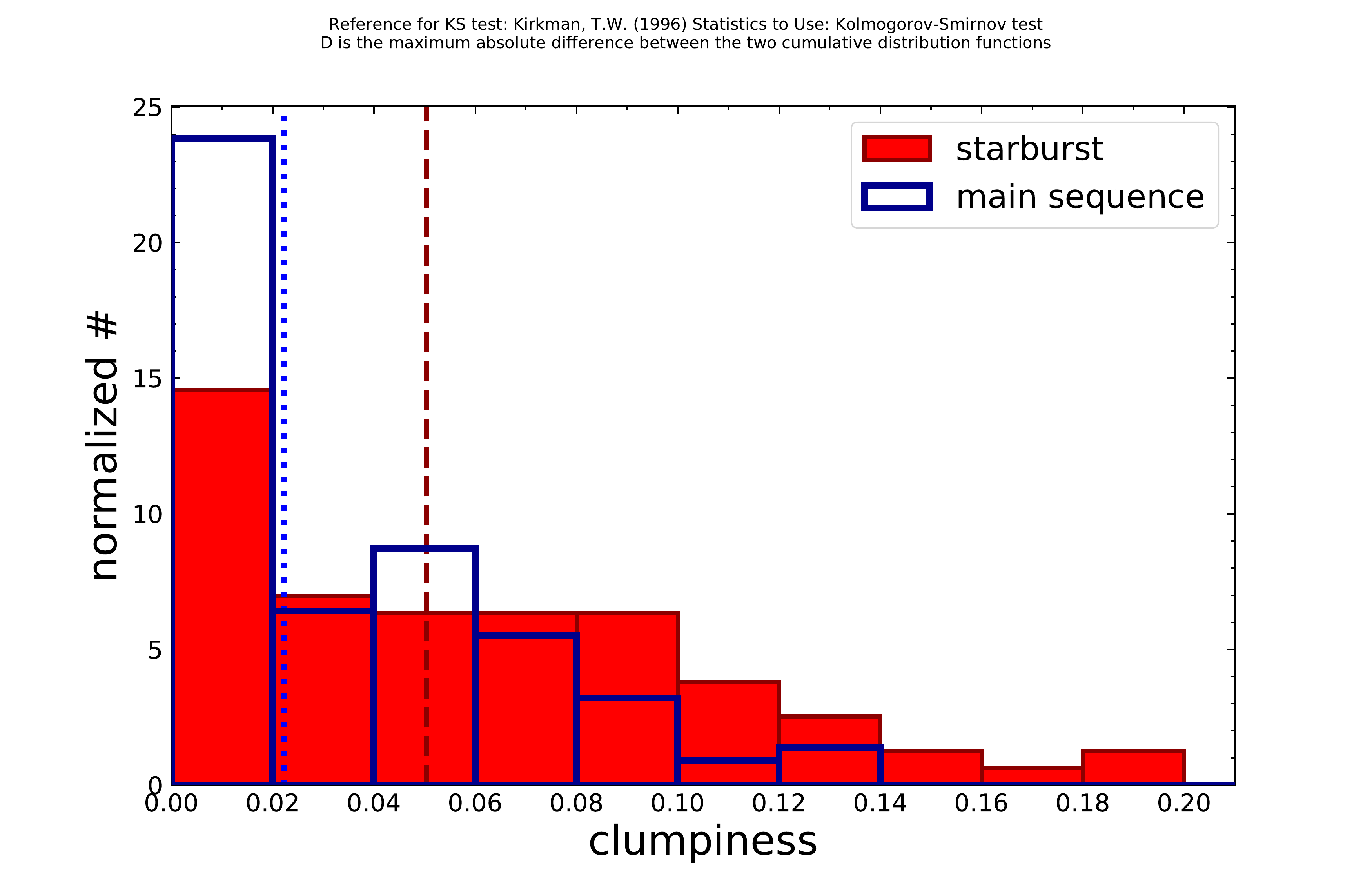}
    \caption{\small Distribution of the clumpiness parameter (normalized to unit area) for starburst galaxies (red filled) and main sequence systems (blue edges), showing that starbursts dominate in the high clumpiness tail. The results of the Kolmogorov-Smirnov test (KS statistics and p-value) between the two sample distributions are $0.26$ and $0.0029$, respectively. The median clumpiness for SBs and MS ($0.05$ and $0.022$) are drawn with dashed and dotted lines, respectively, with their corresponding colors.}\label{histogram_clumpiness1}
\end{figure}

In Fig.\ref{histogram_clumpiness1} we show the histogram distribution of the clumpiness for starburst galaxies (in red) and main sequence galaxies (in blue). 
The clumpiness parameter spans a range between $0$ and $0.20$, meaning that clumps can contribute at maximum to one fifth of the total off-nuclear galaxy emission at this redshift. 
We can see that the distributions of both subsets are peaked at low clumpiness ($c<0.02$) and, after this excess, they follow an approximately constant and then declining trend. However, the two histograms differ for many aspects. Main sequence systems are dominant in the first clumpiness bin, while, after a region where the relative abundances are consistent ($0.02<c<0.08$), starbursts are systematically more numerous above $c\simeq0.08$. This translates into more than a factor of two higher median clumpiness for SBs compared to the MS population ($0.05$ and $0.022$, respectively). 

In order to test whether the two distributions ($\chi_1$ and $\chi_2$) are significantly different, we run a Kolmogorov-Smirnov (KS) test, which yields both the maximum difference $D_{\chi_1,\chi_2}$ between the two cumulative distribution functions and the probability (p-value) to obtain the same $D_{\chi_1,\chi_2}$ under the assumption that the two underlying one-dimensional probability distributions are equal. We found a $D_{\chi_1,\chi_2}=0.26$ and p-value of $0.029\%$, thus the identity hypothesis can be rejected at $>99.97\%$ confidence level, suggesting that the two subsets are intrinsically different. 
We also characterized the two tails in the high clumpiness regime (excluding the objects falling in the first bin) for SB and MS galaxies separately. The first follows a smooth decreasing trend, reaching a maximum clumpiness of $0.194$, while $c=0.14$ comprises $90\%$ of the starbursts in the tail. On the other hand, for main sequence systems, the highest clumpiness observed is $0.126$, and in this case $90\%$ of the population in the tail has $c<0.1$.  
\begin{figure}[t!]
    \centering
    \includegraphics[angle=0,width=\linewidth,trim={1.0cm 0.1cm 2.85cm 3.cm},clip]{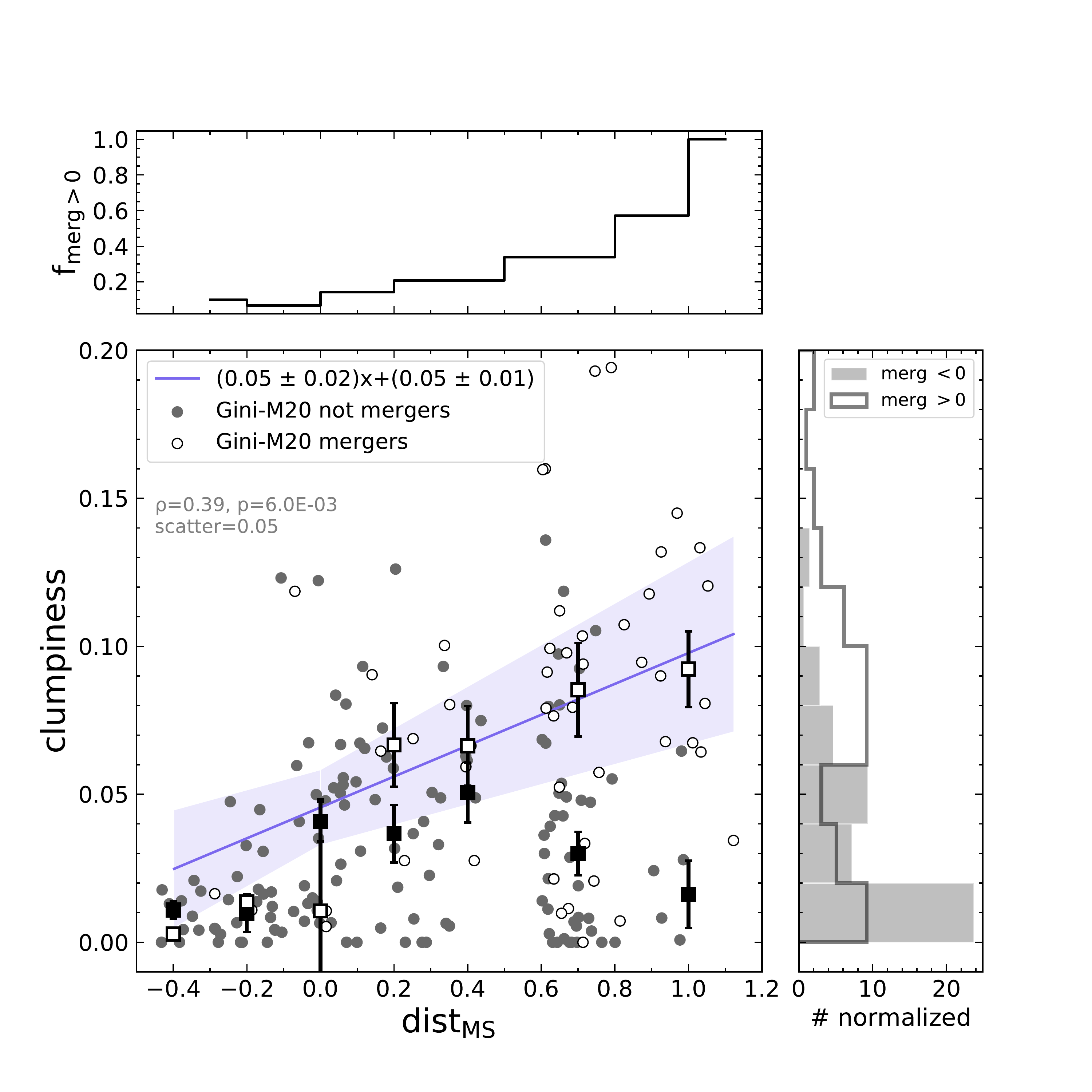}
    \caption{\small Scatter plot between the clumpiness and the distance from the main sequence ($c$ vs. dist$_\text{MS}$) for our sample. Each galaxy in the plot is color-coded according to its mergerness value. Morphological mergers identified in the Gini-M$_{20}$ diagram ($mergerness$ $>0$) are shown with white filled circles, and their fraction f$_\text{merg}$ (compared to the whole population) constantly increases toward higher dist$_\text{MS}$ (upper histogram). Median clumpiness values and $1\sigma$ errors are calculated for mergers and not merger systems in seven bins of dist$_\text{MS}$ of $0.2$ dex approximately. The violet line and shaded area indicate, respectively, the best-fit linear correlation for Gini-M$_{20}$ mergers (whose equation is reported in the legend) and the corresponding $1\sigma$ error region. The annotation below the legend indicates the Pearson correlation coefficient $\rho$ and p-value for the same galaxies, along with the $1\sigma$ scatter of morphological mergers around the best-fit line.
    The histogram on the left compares instead the clumpiness distribution for morphologically selected mergers ($mergerness>0$, in white) and not mergers (gray region, derived from the black circles in the scatter plot). In this case, a larger discrepancy is observed between the two distributions compared to Fig. \ref{histogram_clumpiness1}, confirmed by the higher significance of the KS test (KS$=0.432$, p-value$=2\times 10^{-8}$).}\label{distMS_clumpiness}
\end{figure}

The detailed properties of the clumpiness distribution among SB and MS systems in our sample can be better visualized in a scatter plot (Fig. \ref{distMS_clumpiness}) by comparing their clumpiness to the distance from the main sequence distance (dist$_\text{MS}$). By definition, starbursts occupy the right part of the diagram at dist$_\text{MS}>0.6$, while main sequence objects span the range $-0.47<$ dist$_\text{MS}$ $<0.47$. For the whole sample, we also computed the median clumpiness (shown with gray squares) in seven bins of dist$_\text{MS}$ with bin size of $\simeq0.2$ dex, and the error on the median (shown with black symmetric error bars). % ($-0.47$:$-0.3$,$-0.3$:$-0.1$,$-0.1$:$+0.1$,$0.1$:$0.3$,$0.3$:$0.47$,$0.6$:$0.8$,$0.8$:$1.2$)
We also flagged all galaxies with mergerness $>0$ as empty circles, and the remaining not merging systems with black filled circles. 

Overall, we can see that there is a large diversity among the main sequence population. Galaxies with dist$_\text{MS} < -0.1$ are mostly isolated main sequence galaxies, with a low median clumpiness of $\simeq 0.01$, reaching a maximum value of $0.05$. Above that threshold, a different behavior of the clumpiness could be noticed between Gini-M$_{20}$ mergers and not merger systems. In particular, the clumpiness of the first subset rises on average by a factor of six within the main sequence, and up to a factor of nine at dist$_\text{MS}>0.6$. A statistical analysis reveals the existence of a significant correlation between the two quantities, with a Pearson correlation coefficient of $0.39$ and p-value $=0.006$, although it may not be very strong ($\sim 2.5 \sigma$) considering the S/N of the best-fit angular coefficient.
We can also see that the highest clumpiness values ($c>0.14$) in Fig. \ref{histogram_clumpiness1} and \ref{distMS_clumpiness} are found only for mergers that are simultaneously starbursts. 
Interestingly, this upward trend mimics the increase of the morphological mergers fraction. Indeed, as showed in the upper histogram in Fig. \ref{distMS_clumpiness}, the relative number of Gini-M$_{20}$ mergers rises along the main sequence, and then goes from $35\%$ to $\sim100\%$ at the rightmost extreme of the starburst regime.

On the other hand, galaxies that are not identified as morphological mergers follow a different trend and do not display any correlation. In particular, their median clumpiness increases by a factor of $\sim 4$ compared to the less star-forming systems, and stays always between $0.02$ and $0.05$ in the `upper main sequence' part. Finally, it decreases at dist$_\text{MS}>0.6$, even though we have less statistics in the last bin.
%`upper main sequence' galaxies increases monotonically by a factor of three at least, which then remains approximately constant (within the uncertainties) throughout the starburst regime. 
Similar trends were also found for our sample when comparing the clumpiness to the specific SFR.

If we select galaxies according to their mergerness parameter and compare their clumpiness distributions (right histogram in Fig. \ref{distMS_clumpiness}), we find a more clear separation compared to previous histograms (KS$=0.432$, p-value$=2\times 10^{-8}$). Interestingly, now the clumpiness distribution of morphological mergers is approximately constant over the whole range up to $c\simeq0.12$, since the counts in all the bins are consistent with poissonian statistic fluctuations with $95\%$ confidence level. 
If we consider together MS visual mergers and starburst galaxies, and compare them with isolated, not visually interacting main sequence systems, we obtained a similar result of Fig. \ref{distMS_clumpiness} (KS$=0.307$, p-value $\simeq0.1\%$), with the latter prevailing at $c<0.06$ and starbursts being dominant at higher clumpiness. 

\begin{figure}[t!]
    \centering
    \includegraphics[angle=0,width=\linewidth,trim={2.2cm 0.1cm 3.5cm 2cm},clip]{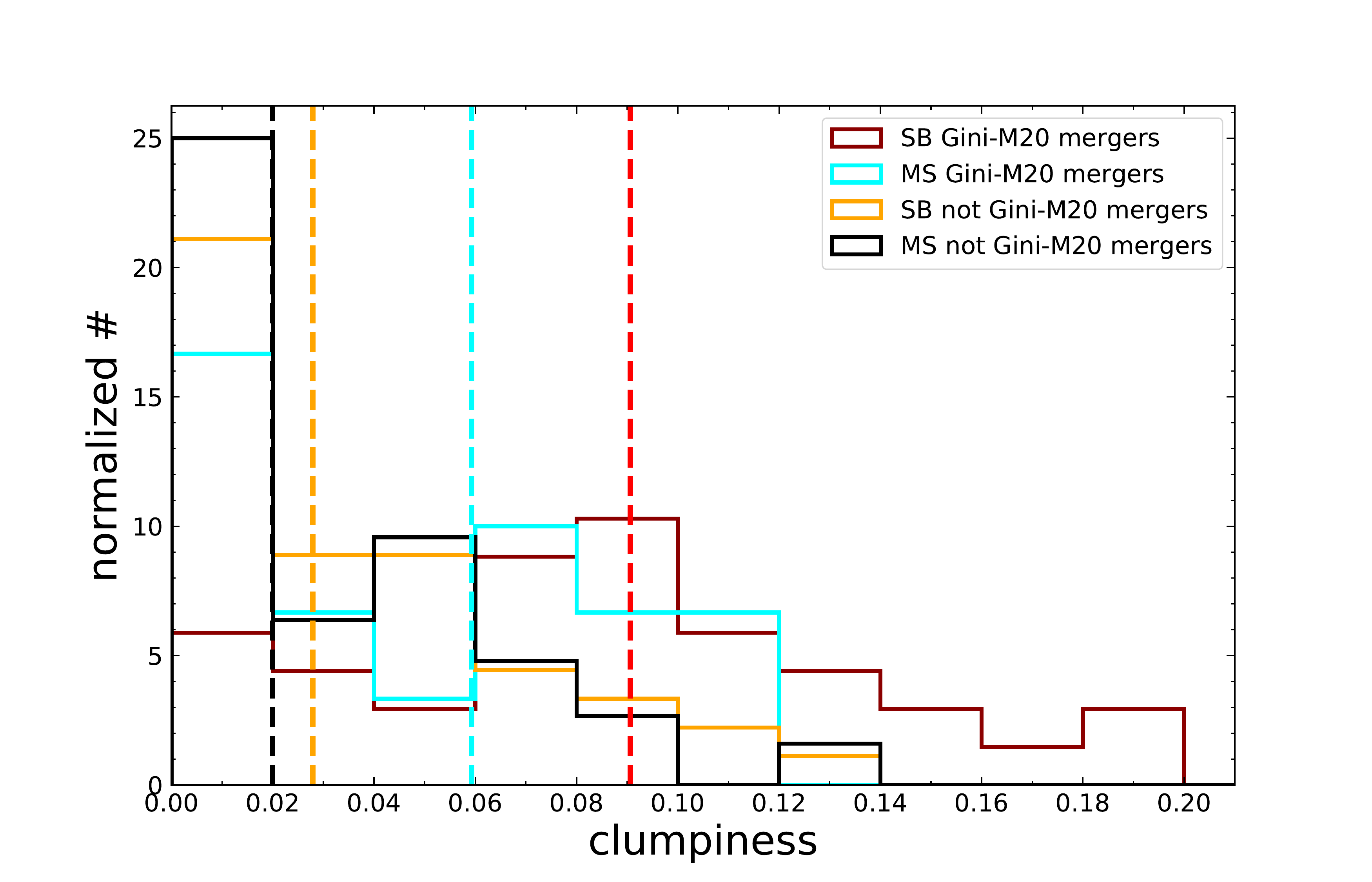}
    %\vspace{+0.15cm}
    \includegraphics[angle=0,width=\linewidth,trim={2.2cm 0.1cm 3.5cm 2cm},clip]{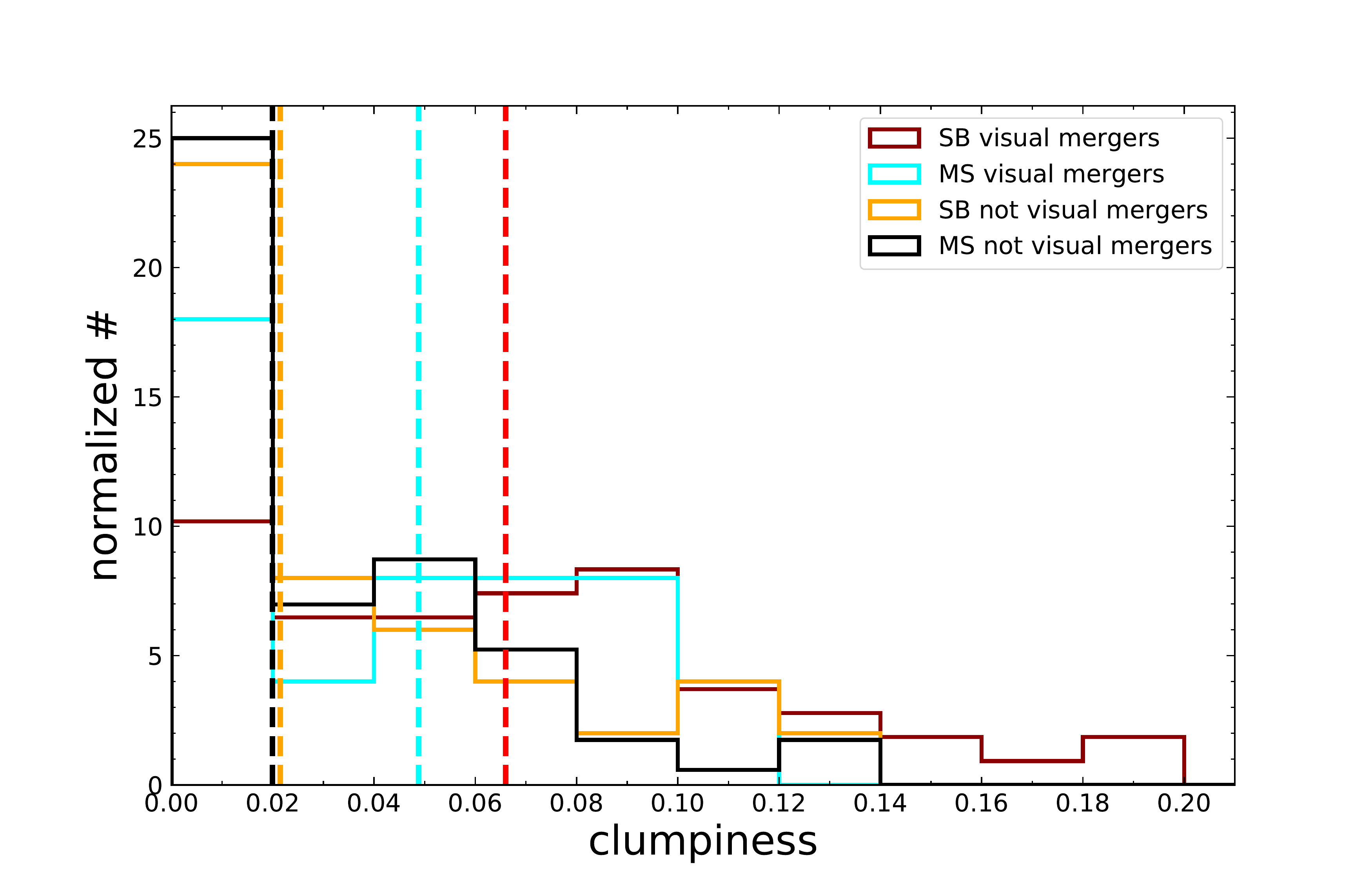}
    \caption{\small \textit{Top:} Normalized distribution (continuous line) and medians (vertical dashed lines) of the clumpiness for four types of galaxies: main sequence galaxies with mergerness $<0$ and $>0$ (in black and cyan, respectively), and starbursts (in orange those with mergerness $<0$ and in red the morphological merger subset). 
    We notice that the merger-selected starbursts dominate the high clumpiness tail, while the remaining SBs are basically indistinguishable from the population of main sequence galaxies with $mergerness$ $<0$. A substantial contribution to the high clumpiness population also comes from mergers inside the MS, as shown by the cyan excess at $0.06<c<0.12$. A Kolmogorov-Smirnov (KS) test between merger and not-merger MS systems (cyan and black subsets) yields $0.302$ (p-value $=0.094$).
    \textit{Bottom:} Same histograms as above, but considering visual mergers instead of the automatic morphological classification. 
    }\label{histogram_clumpiness4}
\end{figure}

The impact of the morphological merger classification on the clumpiness distributions suggests that we can decompose the entire population in four classes according to their mergerness and dist$_\text{MS}$, to look for any trend in the SB and MS subsets.
This exercise is made in Fig. \ref{histogram_clumpiness4}-\textit{top}, showing that, in this case, there is a more striking difference between morphological merger starbursts and not morphological merger systems, regardless of their distance from the MS, with the first having a median clumpiness more than a factor of three higher ($c_\text{median}=$ $0.09$). 
The clumpiness distribution of MS morphological mergers is also slightly different (at $>90\%$ confidence level) from the rest of the MS population (KS$=0.34$, p-value$=0.075$), with a clumpiness distribution skewed toward a larger median ($c_\text{median}=0.06$), and a big contribution in the higher clumpiness tail (cyan line at $0.06<c<0.12$) compared to other not-merging systems, suggesting that mergers might increase the clumpiness even with a moderate enhancement of the SFR. 
%We remember also that the morphological merger classification may not be complete, and we will discuss possible physical explanations of these findings (/trends) in Section \ref{mergerphase}.
A similar result is found when using the visual classification to identify merger systems from their morphology (Fig. \ref{histogram_clumpiness4}-\textit{bottom}). A KS test performed on visual mergers and not mergers in the main sequence yields a KS$=0.27$ and p-value of $0.1$ ($\simeq 90\%$ confidence level).

On the other hand, visual merger starbursts are significantly different from isolated undisturbed objects, showing on average higher clumpiness parameters (KS$=0.38$, p-value$=0.01$). In this case, the median clumpiness values of merging galaxies ($0.066$ and $0.049$ for SB and MS, respectively) are slightly lower than in the upper panel of Fig. \ref{histogram_clumpiness4}, which is somehow expected, given that the two diagnostics probe in general different merger timescales. In particular, as mentioned in Section \ref{visual_classification}, visual mergers contain also pairs and already coalesced systems with faint (although unambiguous) residual interacting features, in which the clumps may have still not formed or already disappeared. These objects all contribute to increase the fraction of mergers in the first clumpiness bin ($c<0.05$).

\begin{figure}[t!]
    \centering
    \includegraphics[angle=0,width=\linewidth,trim={1.1cm 0.cm 2.8cm 2cm},clip]{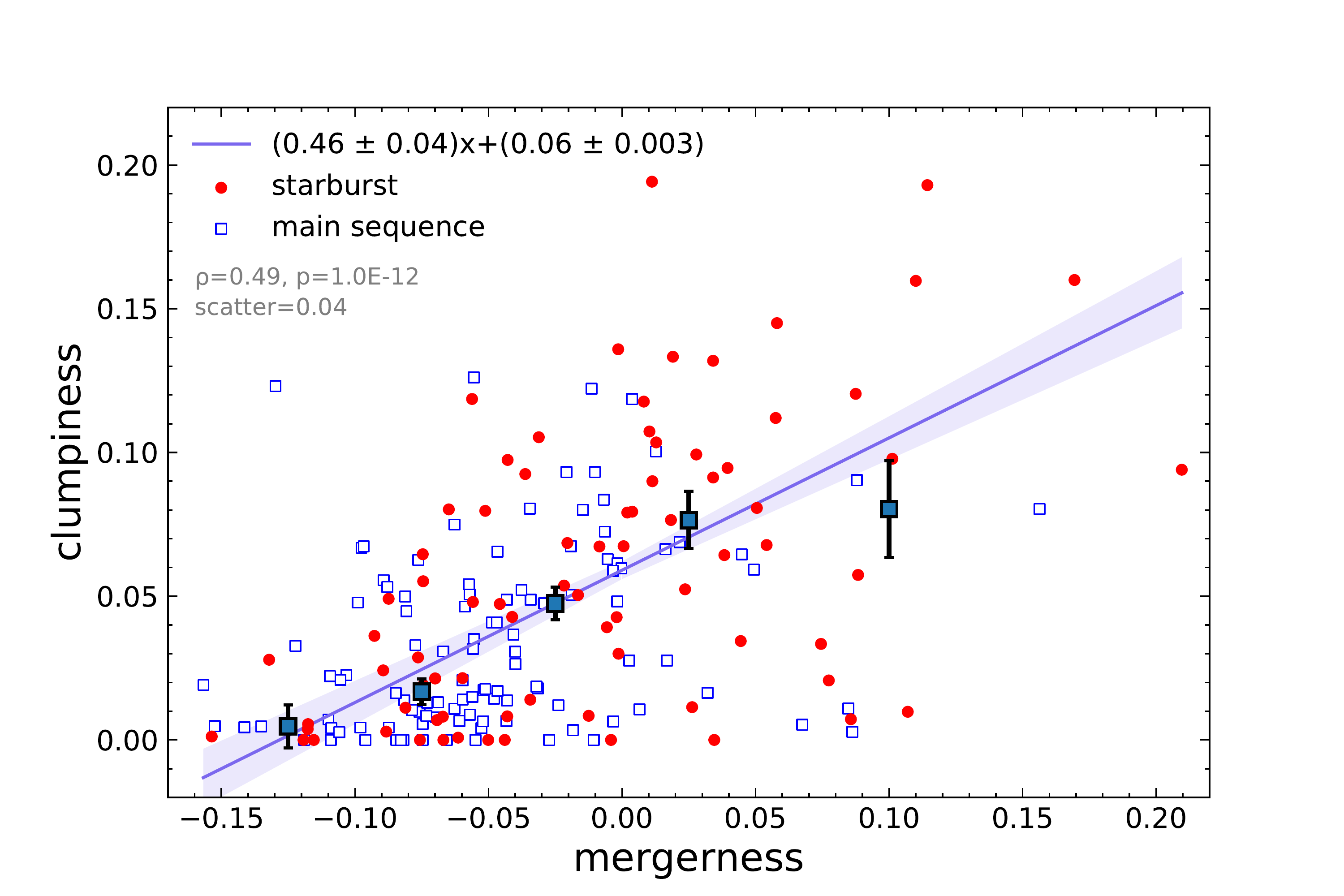}
    \caption{\small Comparison between the clumpiness and the mergerness parameters for our sample of SB and MS galaxies (with red filled circles and blue empty squares, respectively). The clumpiness medians with corresponding errors are computed for the whole sample and shown in $5$ different bins of mergerness with square symbols and vertical black bars. The best-fit linear correlation and $1 \sigma$ error are indicated by the violet line and violet shaded area. In the upper-left part of the plot we include the equation of the best-fit line, the Pearson correlation coefficient $\rho$, the p-value and the $1 \sigma$ scatter of the relation.}\label{mergerness_vs_clumpiness1}
\end{figure}

Given the appreciable clumpiness enhancement in all types of mergers (including MS systems, Fig. \ref{histogram_clumpiness4}), we also checked for a possible direct relation between clumpiness and mergerness. In Fig. \ref{mergerness_vs_clumpiness1} we show this comparison for our sample, finding a strong correlation between the two quantities (Pearson correlation coefficient $=0.49$, p-value $=10^{-12}$), even though with a relatively large dispersion ($0.04$). %This diagram represents an alternative way to visualize our results of Fig. \ref{distMS_clumpiness} and \ref{histogram_clumpiness4}, corroborating again that a physical connection between mergers and starbursts should exist.} 

We may wonder whether the two axis presented in Fig. \ref{mergerness_vs_clumpiness1} are independent or, in other words, whether the merger classification (i.e., $mergerness>0$) is affected by the presence of clumps in the galaxy. However, we have reasons to think that this is not likely the case. First, we did not find a direct relation between the clumpiness and either the Gini parameter or M$_{20}$ separately (Fig. \ref{clumpiness_correlations} in the Appendix). %, which are the first parameters measured from the images. 
Even though a slightly increasing trend of the clumpiness is observed on average with higher Gini, many galaxies (both MS and SBs) have a low clumpiness value despite their high M$_{20}$ or Gini coefficients. 
The presence of clumpy isolated galaxies with relatively high clumpiness but mergerness below 0 indicates itself that clumps cannot be responsible for erroneously selecting the galaxy as a morphological merger. 
Later in the paper in Section \ref{simulations} (Fig.~14), we will corroborate this also through numerical simulations, further showing that the mergerness is independent from the clumpiness. 
We remind that our non-parametric morphological procedure has been widely tested and validated at all redshifts of our interest \citep[e.g.,][]{lotz04,lotz08}, and we have already removed faint sources and highly inclined disks for which these tools may not work. 
Furthermore, we are using multiple merger classifications. In particular, the visual approach classifies galaxies regardless (in principle) of the presence of clumps, and it reinforces the results found with other methods.

Our results do not depend significantly on the mergerness threshold used in Section \ref{ginim20section} to identify morphological mergers. To be conservative, if we consider a lower threshold between $-0.1$ and $0$, this would strengthen the result of Fig. \ref{histogram_clumpiness4}, since the median clumpiness of main sequence mergers (cyan vertical line) would move rightwards. For example, if we choose a threshold of $-0.05$, we would obtain for the above subset a median $c_\text{med}=0.048$, and the difference with respect to not merging MS sources would be more significant (KS$=0.364$, p-value $<0.001$). 

We also remark that these results are not affected by the choice of the bin size and by the sample cuts. Indeed, similar histogram distributions, trends and significances (Fig. \ref{histogram_clumpiness1} to \ref{histogram_clumpiness4}) are obtained when varying the first parameter by small amounts within a factor of 2 of the chosen bin size.
The same conclusion holds for different thresholds of the i-band magnitude for the final sample selection, as in case of more conservative choices, to keep only the brightest sources (e.g., i-mag cut $<22.5$). 

Furthermore, we have so far analyzed a limiting, very conservative situation, according to which we systematically searched and removed two nuclei in all ongoing merging starbursts. On the contrary, only one nucleus is removed for main sequence galaxies, unless we could clearly see two distinct merging components, in which case we also removed the two nuclei. This approach implies that we are likely underestimating the clumpiness of starburst systems, while simultaneously overestimating that of normal star-forming galaxies, because in many cases one or more of the nuclei might be too obscured to be seen.

Finally, we found no correlation between the stellar mass and the clumpiness (Fig. \ref{redshift_scatter}-top in the Appendix), indicating that the stellar content is not the main driving parameter for the increasing patchy morphology among our sample. Similarly, there is no significant evolution of the clumpiness with redshift, as shown in Fig. \ref{redshift_scatter}-bottom in the Appendix.

%%%%%%%%%%%%%%%%%%%%%%%%%%%%%%%%%%%%%%%%%%%%%%%%%%%%
%%%%%%%%%%%%%%%%%%%%%%%%%%%%%%%%%%%%%%%%%%%%%%%%%%%%
%%%%%%%%%%%%%%%%%%%%%%%%%%%%%%%%%%%%%%%%%%%%%%%%%%%%

\section{Discussion}\label{discussion}

Our analysis has shown that merging driven starbursts are significantly more clumpy than normal main sequence galaxies at $z\simeq0.7$. The question thus arises on which is the physical reason of this difference. We discuss first the possible role of dust in driving these results, and conclude this is minor. We then use numerical simulations of merger systems at $z\sim0.7$ in the following subsection, showing that mergers can effectively induce clumps formation and increase the measured clumpiness. Later in this section, we support the young nature of the clumps in three galaxies for which multi-wavelength HST images are available. We conclude by discussing a possible merger evolutionary trend of the clumpiness and suggesting a possible extension of our results to interpret clumpy galaxies observed at significantly higher redshifts.

%%%%%%%%%%%%%%%%%%%%%%%%%%%%%%%%%%%%%%%%%%%%%%%%%%%%%%%%%%%%%%%
%%%%%%%%%%%%%%%%%%%%%%    SIMUS   %%%%%%%%%%%%%%%%%%%%%%%%%%%%%
%%%%%%%%%%%%%%%%%%%%%%%%%%%%%%%%%%%%%%%%%%%%%%%%%%%%%%%%%%%%%%%

\subsection{Effects of dust attenuation on the observed clumps}\label{dust_effects}

Several studies of the outcoming UV radiation from high redshift clumpy galaxies have shown that UV clumps may not trace the stellar mass but rather reflect the patchy distribution of dust attenuation \citep{moody14,cochrane19}, with FUV emission tracing holes in the dust. One may thus ask what is the effect of dust on the clumps that we see in our galaxies.

We remark that our clumps are detected in the optical rest-frame, at an average wavelength of $4700$ \AA\ at $z\sim0.7$, which is subject to attenuation factors that are significantly lower (by a factor of three or more, assuming for example a Calzetti law with A$_\text{V} < 1.5$ in off-core regions) with respect to the UV regime, on which the above studies are focused. Hence their results cannot be applied straightforwardly to our case. % Even though the extinction from dust cannot be completely eliminated in the optical, it is currently the best compromise if we want to achieve the highest spatial resolutions for large samples of objects before the JWST era.}

Furthermore, our clumps are off-nuclear structures by definition, that is, they are situated several kpc away from the galactic nuclei, which we systematically removed. Hence, they are significantly less attenuated (or nearly unattenuated) compared to the central cores, where most of the IR luminosity is produced.
This has been shown in spatially resolved studies of HII regions in massive star-forming galaxies ($9.8< log_{10}$ M$_\ast$ $<11$) at $z\sim1.4$ (Nelson et al. 2015): outside of the central $1$ kpc, the attenuation A$_\text{V}$ decreases at least by a factor of $5$ compared to the center, and is generally low ($\simeq 0.5$ mag for optical emission lines). %We remind that $1$ kpc is the typical size of bulges in spiral galaxies, or that of obscured IR-luminous cores in ULIRGs, and is slightly above our resolution element at $z\sim0.7$. Our clumps are thus further away by definition. 
The fact that a subset of our galaxies are highly infrared-luminous is not relevant, as the dusty SFR activity in these systems is mostly concentrated in the central, obscured, usually sub-kpc cores while the extinction is low in external regions (A$_\text{V}<0.5$) \footnote{If the real cores were completely obscured in the optical, we could conservatively remove at least two nuclei in starburst objects (and MS mergers) when computing the total flux in clumps, and thus the intrinsic clumpiness would be even higher, which will reinforce our conclusions.}, as shown by spatially resolved studies of local ULIRGs, (e.g., Scoville et al. 1997, Alonso-Herrero et al. 2006, Garcia-Marin et al. 2009, Piqueras-Lopez et al. 2016). %, while our clumps are further away from the cores by definition.} 
%By the way, Garcia-Marin et al. 2009 showed for a representative sample of ULIRGs at $z<0.1$ that the extinction in the external regions is substantially lower than in the nuclei, with $A_\text{V}$ of the order of $<0.2$ mag, even though these values might be higher at earlier cosmic epochs.} 

%\textbf{All in all, studies on the same kind of objects suggest that clumps observed at optical wavelengths are real. For instance, Miralles-Caballero et al. (2012) detected H$\alpha$-emitting star-forming clumps at kpc distances from the nuclei of local LIRGs or ULIRGs (typically along the tidal tails), demonstrating that they trace the same star-forming clumps identified at much longer wavelengths in Br$\gamma$ or Pa$\alpha$ maps, and in the H-band continuum (Alonso-Herrero et al. 2006). 
%}

\subsection{Confirming merger-induced clumps formation with simulations}\label{simulations}

\begin{figure*}[]
    \centering
    \includegraphics[angle=0,width=0.91\linewidth,trim={0.cm 0cm 0.cm 0cm},clip]{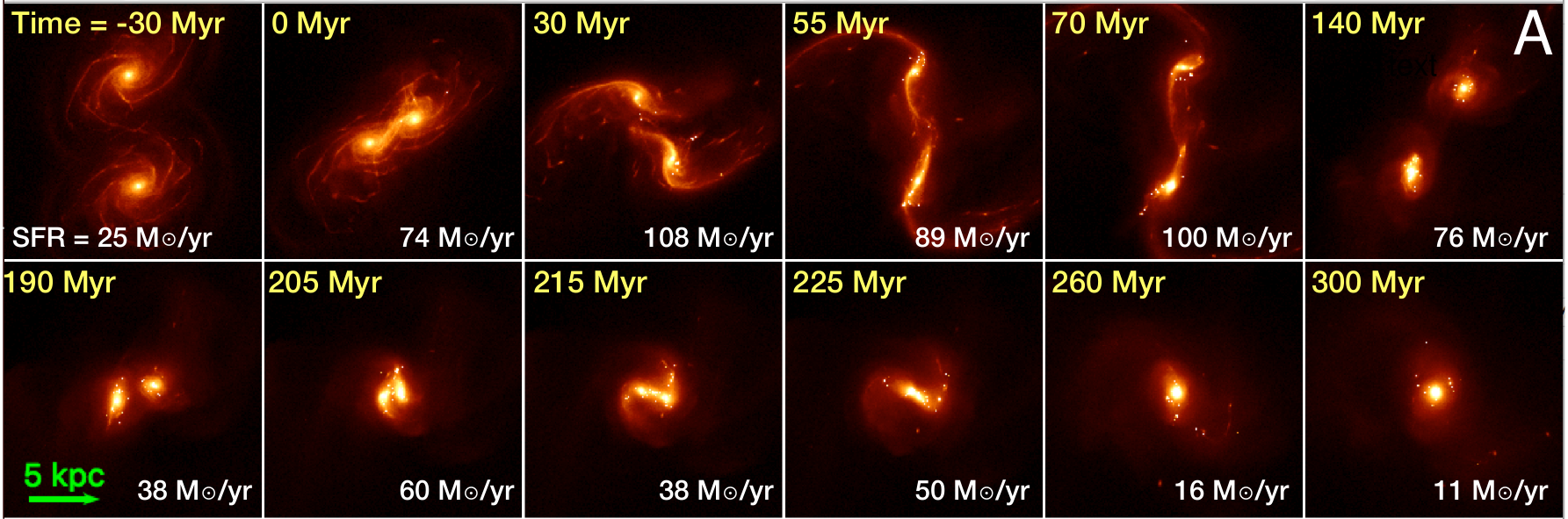}
    \includegraphics[angle=0,width=0.91\linewidth,trim={0.12cm 0.1cm 0.12cm 0.1cm},clip]{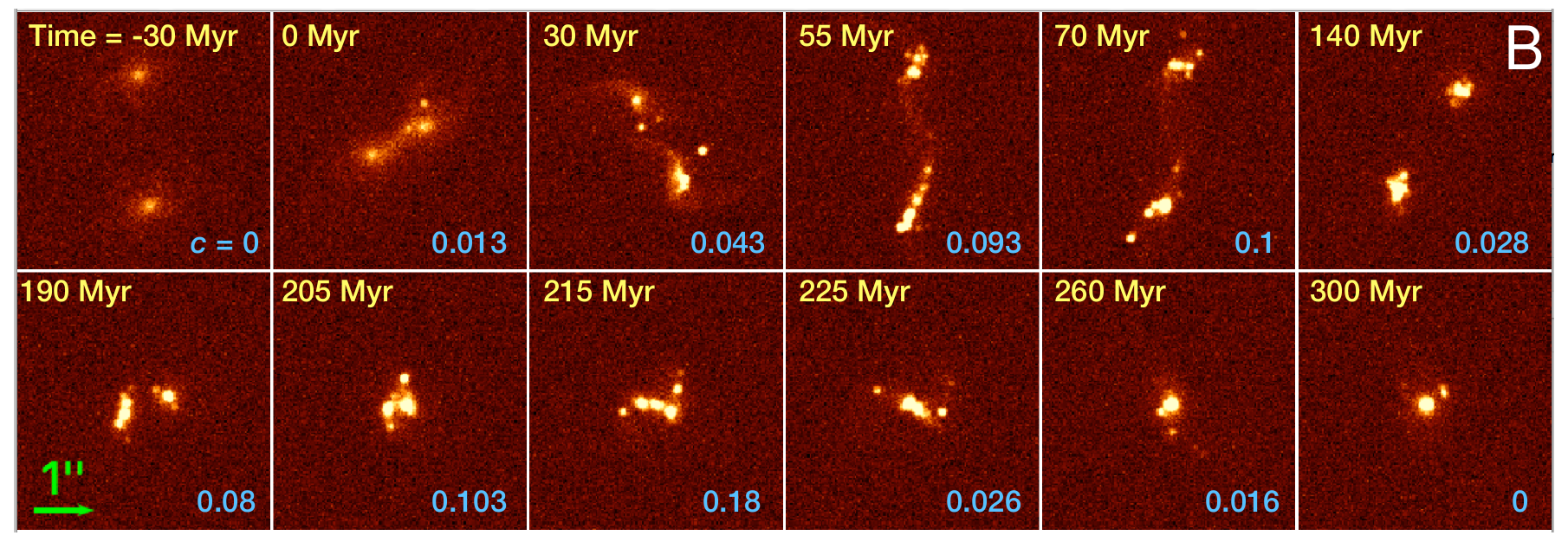}

    \smallskip
    \includegraphics[angle=0,width=0.91\linewidth,trim={0.cm 0.1cm 0.cm 0.1cm},clip]{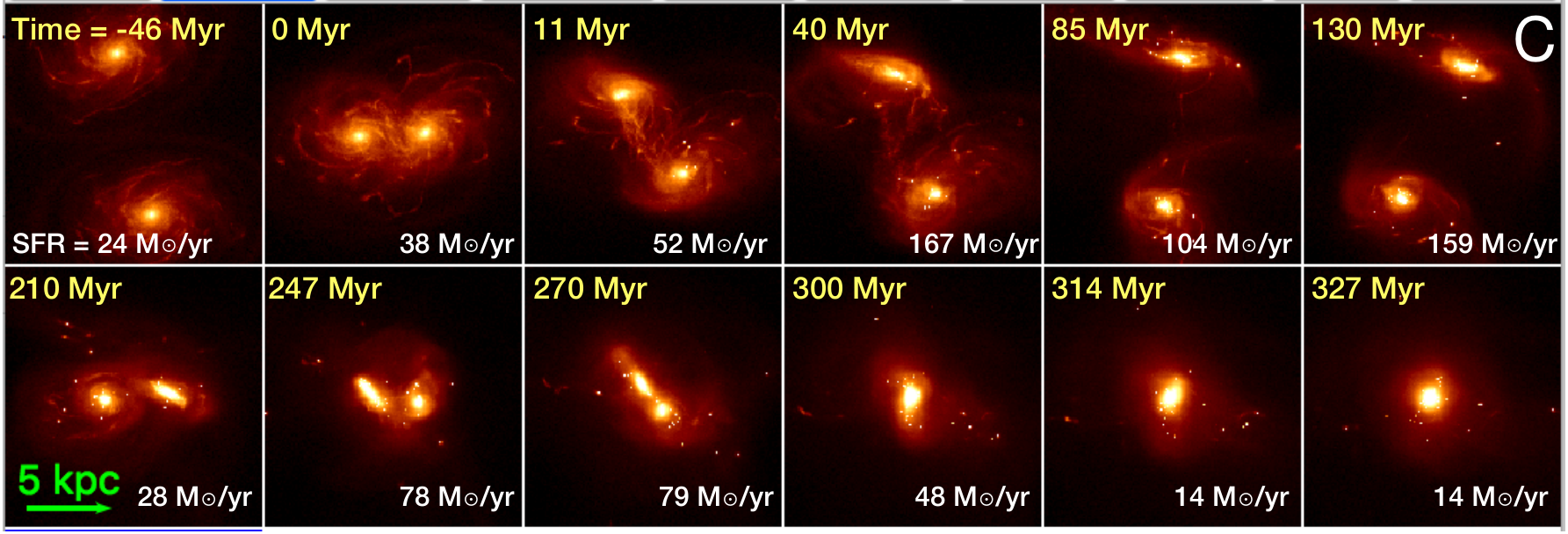}
    \includegraphics[angle=0,width=0.91\linewidth,trim={0.1cm 0.1cm 0.1cm 0.1cm},clip]{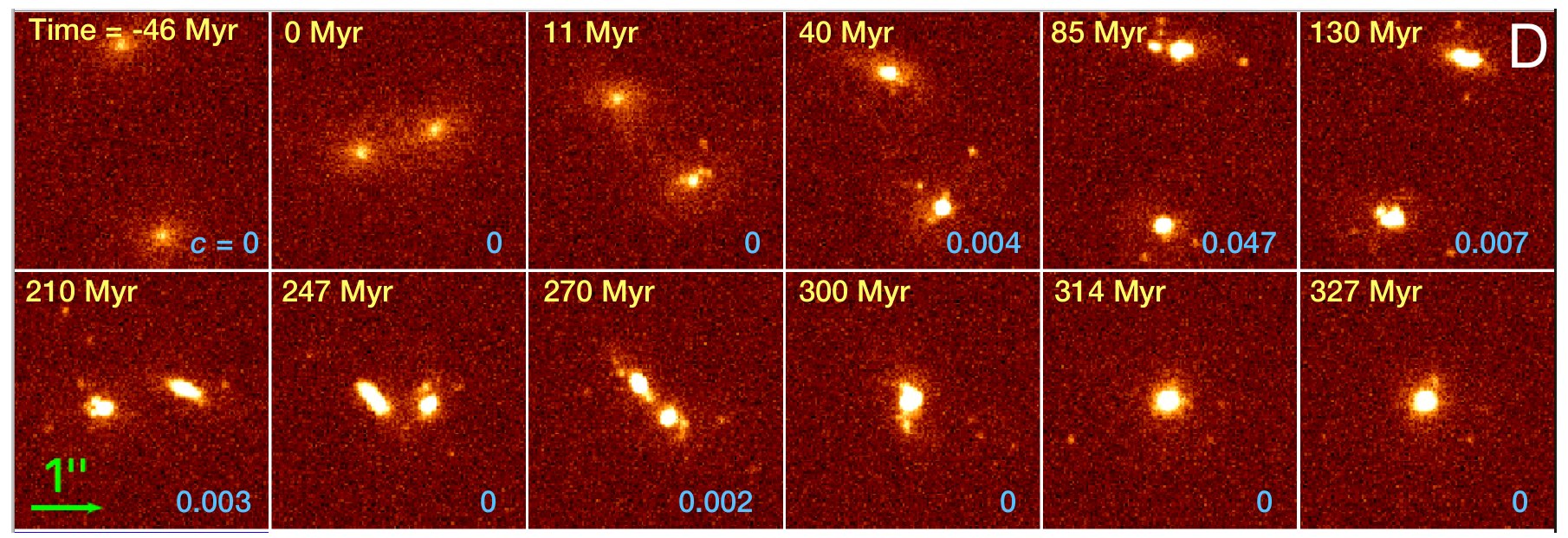}

    \caption{\small \textit{A:} Snapshots of the stellar mass density from the simulation of two colliding galaxies with prograde-prograde orbit coupling, at different times. The images have a physical scale of $200$ pc/pixel, comparable to our observations (assuming the median redshift of our galaxies $\simeq0.7$). \textit{B:} Mock HST F814W observations obtained from the above simulated cutouts after conversion to the $\sim0.095''$ PSF resolution of our images (through a gaussian filter) and addition of the noise. 
    The instantaneous SFR and the estimated clumpiness for each time step is indicated in the corner. $24$ M$_\odot$/yr is the SFR of two isolated disks, according to our fiducial run.
    \textit{C-D:} Same as panels A-B, but for a retrograde-retrograde collision.
    }\label{simulations_cutouts}
\end{figure*}

%\begin{figure*}[h!]
    %\centering
    %\includegraphics[angle=0,width=\linewidth,trim={0.5cm 9.55cm 0.5cm 4.45cm},clip]{Figure1_for_paperIII-compressed.pdf}
    %\includegraphics[angle=0,width=\linewidth,trim={0.5cm 8.15cm 0.5cm 3.1cm},clip]{Figure1_for_paperIII_seconda_reduce.pdf}
    %\includegraphics[angle=0,width=0.85\linewidth,trim={0.cm 0.1cm 0.cm 0.1cm},clip]{simulations1_RR.pdf}

    %\includegraphics[angle=0,width=0.85\linewidth,trim={0.1cm 0.1cm 0.1cm 0.1cm},clip]{simulations2_RR_croppedOK.pdf}
    %\caption{\small Same as Figure \ref{simulations_cutouts}, but for a retrograde-retrograde collision.}\label{simulations_cutouts2}
%\end{figure*}

In order to check whether mergers are the physical cause of the increased clumpiness, we performed hydrodynamical simulations of collisions, choosing initial conditions that are typical of galaxies in our redshift range. In particular, we set a gas fraction of $30\%$ (typical of z $\sim$ 0.7) \citep{combes13,freundlich19}. As mentioned in the Introduction, previous simulations of galaxy collisions with such high gas fraction did not reach the low gas temperatures needed for properly reproducing the gas distribution during merger events.

The setup is based on the simulations described in \citet{fensch17} using the adaptive mesh refinement code \textsc{RAMSES} \citep{teyssier02}. The galaxies have the same characteristics as the ones in \citet{fensch17}. The refinement strategy is based on the density and the highest resolution elements are 6~pc. Gas in cells that are denser than 10 cm$^{-3}$ and cooler than $2 \times 10^4$~K is converted into stars following a \citet{schmidt59} law, with an efficiency per free-fall time set to $10\%$. We include three types of stellar feedback, described in \citet{fensch17}. The energy output from SNII explosions is released by a kinetic kick and a thermal energy injection, each accounting for half of the total energy ouptut \citep{dubois08}. The HII regions are modelled by Str\"omgren spheres, whose sizes are done considering that the gas surrounding the source has a minimal density above 300 cm$^{-3}$. Gas inside the sphere is heated to $5 \times 10^{4}$~K and receives a radial velocity kick modeling the radiation pressure. % from the scattering of photons.  

We performed one isolated and two merger simulations. To account for numerical diffusion effect, in the isolated simulation the galaxy moves along the same orbit as one of the galaxies in the interaction orbits. The spin-orbit coupling plays a significant role in the interaction. For instance, only galaxies with spins aligned with that of the interaction, what is called a {\it prograde coupling}, can create tidal tails \citep[see review by ][]{duc13}. We run one prograde-prograde (or, equivalently, direct-direct) and one retrograde-retrograde encounter. The two orbits correspond to Orbit $\#1$ from \citet{fensch17}. %They have one pericenter at t=0, at which the galaxies are separated by XXX\footnote{being computed}~kpc. 
After an intermediate apocenter, they coalesce within $\sim230$ and $\sim300$ Myr, respectively.
Stellar density maps are shown in panels A and C of Fig.~\ref{simulations_cutouts}. On it we see the formation of stellar condensations during the interaction, similar to what is observed in collisions at low-redshift \citep{dimatteo08,renaud14,matsui19}.
Since we want to check the intrinsic clumpiness of the galaxies, we did not include dust in our simulations. %First, we are interested in the intrinsic clumpiness distribution of our galaxies, since we aim at testing a possible link between mergers and the intrinsic structure of stellar clumps.

\begin{figure*}[t!]
    \centering
    \includegraphics[angle=0,width=0.95\linewidth,trim={1.1cm 0.cm 3.cm 2cm},clip]{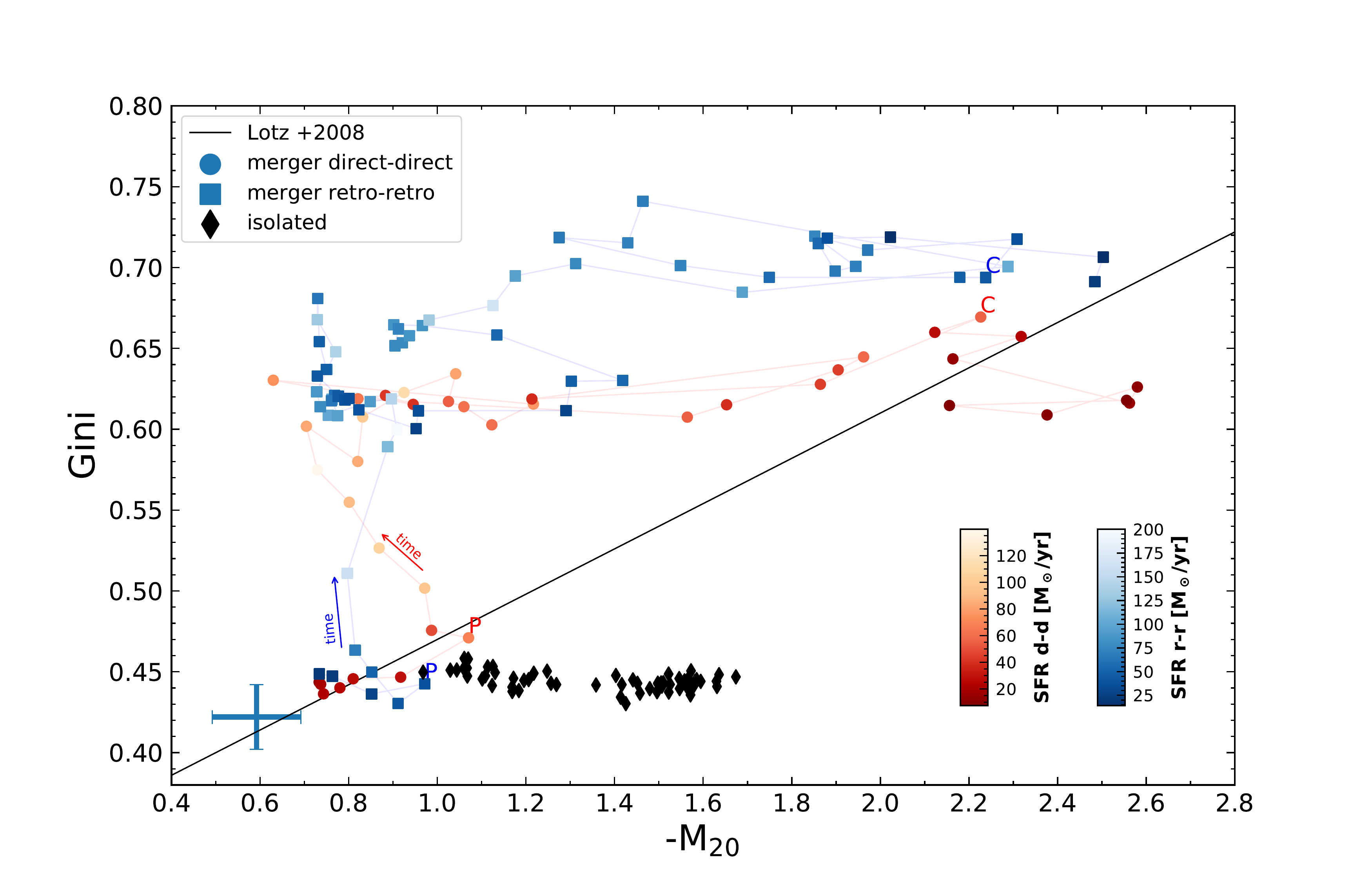}
    \caption{\small Time evolution of the Gini and M$_{20}$ parameters for simulated prograde-prograde and retrograde-retrograde collisions (respectively, orange and blue points or connecting lines). The P and C symbols indicate, respectively, the first pericenter passage and the coalescence, as defined in the text. We can notice that isolated galaxies spend all their lifetime in the not-merger region (according to the separation line by \citet{lotz08}), while ongoing merging systems are mostly located above the morphological merger criterion. For each simulation subset, the points are color-coded according to the specific SFR, while black arrows indicate the positive temporal direction.}\label{gini_m20_simulations}
\end{figure*}

\begin{figure}[t!]
    \centering
    %\includegraphics[angle=0,width=\linewidth,trim={0cm 0.cm 0.9cm 2cm},clip]{Av_clumpiness_new.pdf}
    %\includegraphics[angle=0,width=\linewidth,trim={0cm 0.cm 0.9cm 2cm},clip]{EWHa_clumpiness_new.pdf}
    %Clump_sfr_instant_paper.png
    \includegraphics[angle=0,width=\linewidth,trim={0.7cm 0cm 2.5cm 1.5cm},clip]{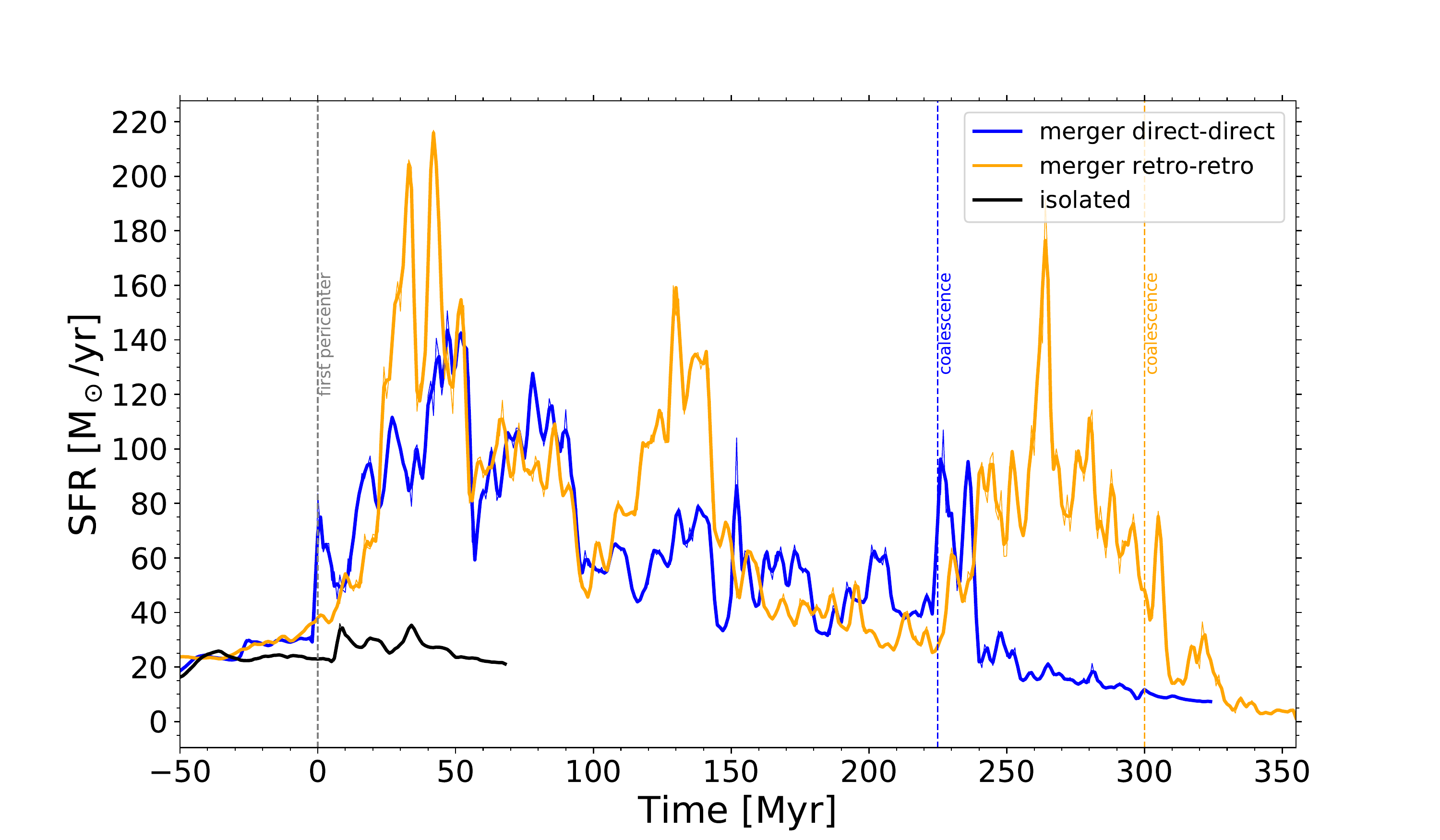}
    \vspace{+0.1cm}

    \includegraphics[angle=0,width=\linewidth,trim={0.7cm 0cm 2.5cm 1.5cm},clip]{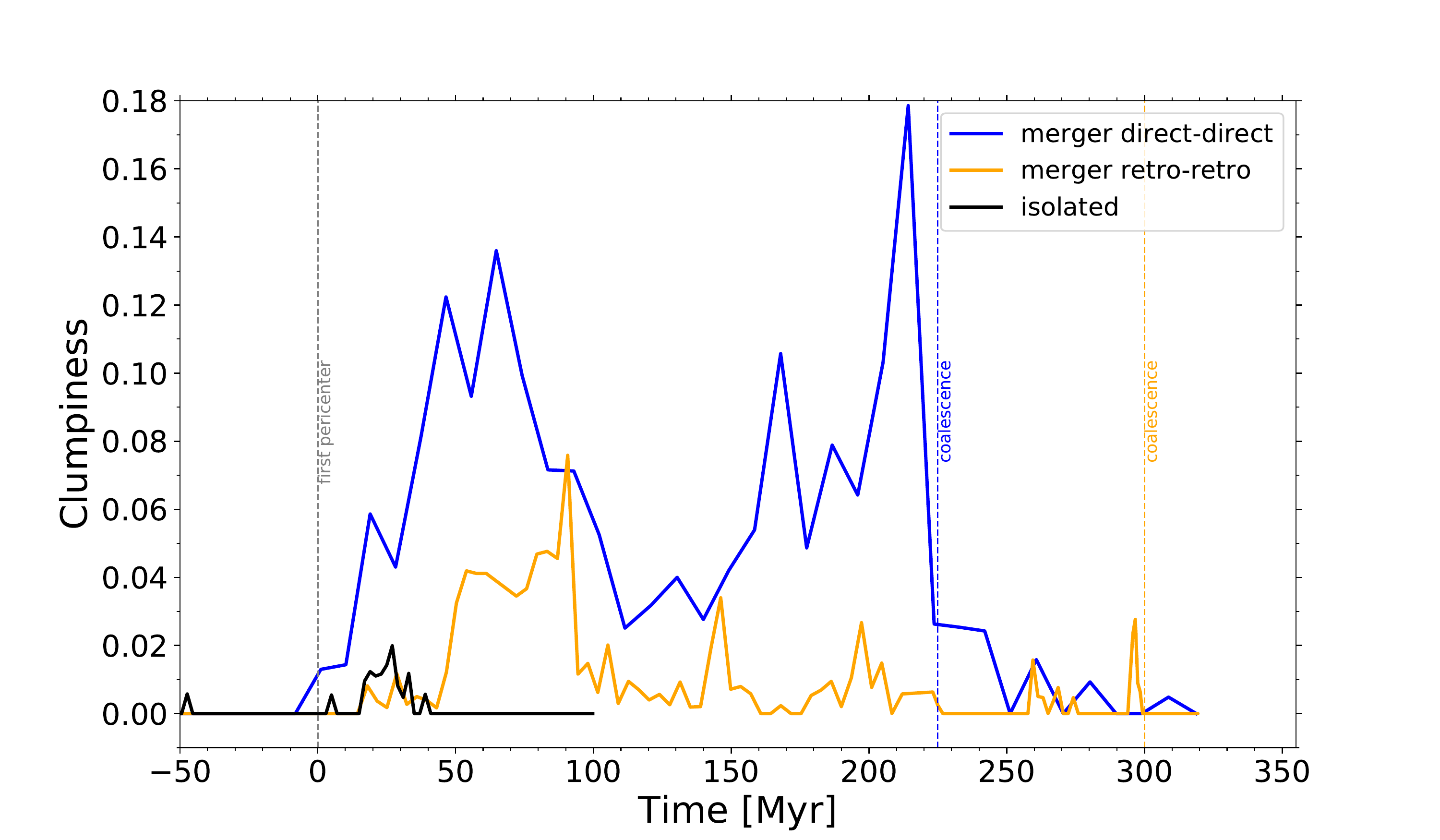}
    %\par\bigskip
    \caption{\small \textit{Top and bottom:} The SFR and the clumpiness (respectively) are shown as a function of time for the direct-direct (i.e., prograde-prograde) merger (in blue), retrograde-retrograde merger (in orange) and for the fiducial run of an isolated galaxy multiplied by two (in black). In the latter, only the evolution during $200$ Myr is shown as a representative case, while in the other two runs the time range considered ($\sim450$ to $\sim820$ Myr from the onset of the simulation) encompasses the effective merging period, from the first pericenter passage to the final coalescence (vertical dashed lines colored accordingly).}\label{simulations_results1}
    %In the first case, a correlation between the two quantities is found. In the r.r. merger simulation, the system reaches a higher SFR on average, but a much modest impact on the clumpiness is obtained compared to the previous case.
\end{figure}

\begin{figure}[t!]
    \centering
    %\includegraphics[angle=0,width=\linewidth,trim={0cm 0.cm 0.9cm 2cm},clip]{Av_clumpiness_new.pdf}
    %\includegraphics[angle=0,width=\linewidth,trim={0cm 0.cm 0.9cm 2cm},clip]{EWHa_clumpiness_new.pdf}
    %Clump_sfr_instant_paper.png
    %\par\bigskip
    \includegraphics[angle=0,width=\linewidth,trim={1.1cm 0cm 3cm 0.cm},clip]{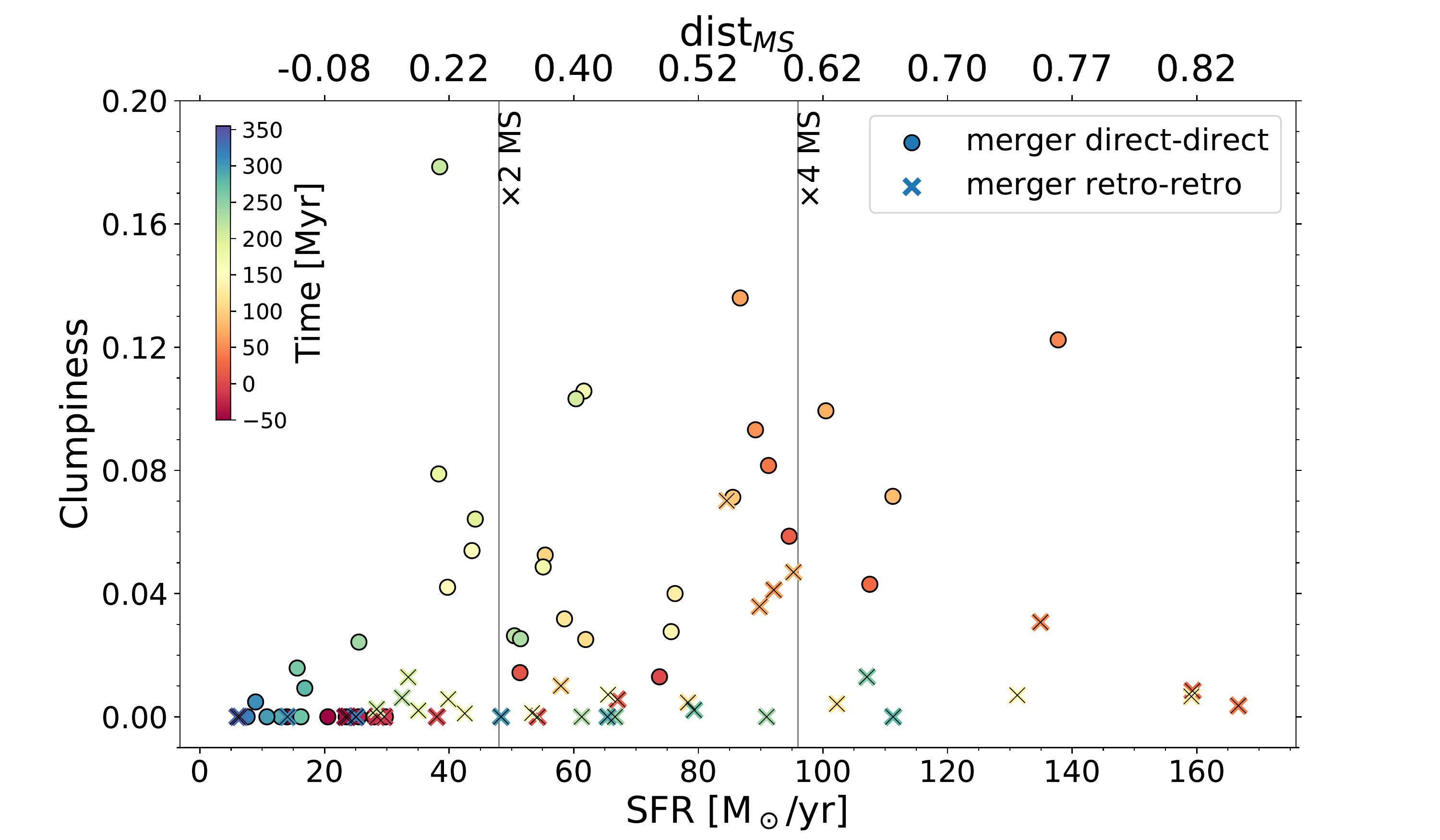}
    \vspace{+0.2cm}

    \includegraphics[angle=0,width=\linewidth,trim={1.1cm 0.cm 3.cm 2cm},clip]{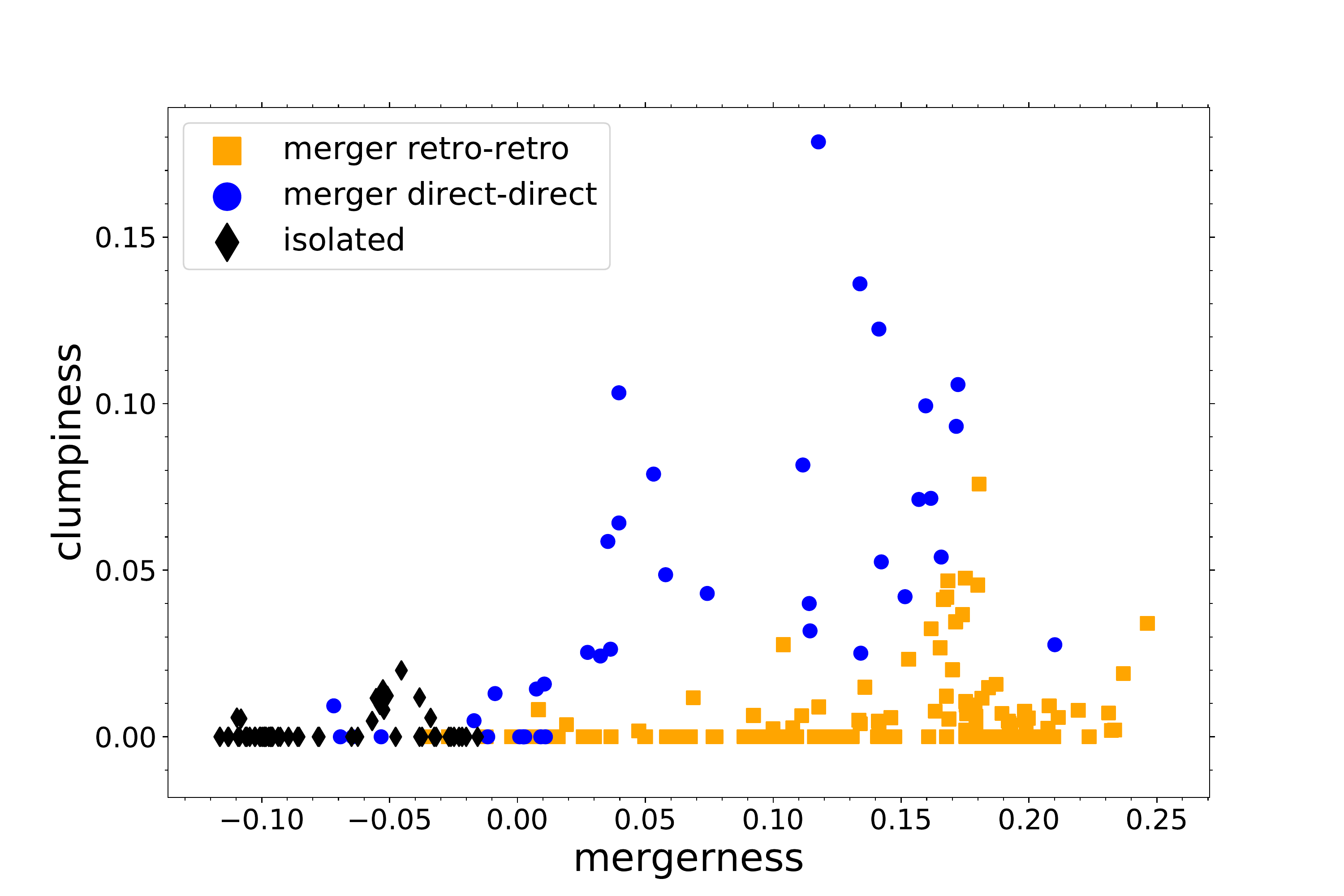}
    \caption{\small \textit{Top:} Comparison between the clumpiness and the SFR at different timesteps (colorbar on the left), for direct-direct and retrograde-retrograde mergers (with circles and crosses, respectively). This diagram should compared to the corresponding observational result in Fig. \ref{distMS_clumpiness}. \textit{Bottom:} The clumpiness is plotted as a function of the mergerness parameter, which is calculated from mock images at finite time intervals. This panel can be compared with the observational finding in Fig. \ref{mergerness_vs_clumpiness1}. In both cases, we can notice a correlation between the quantities in the x and y axis, supporting the observational results.}\label{simulations_results2}
    %In the first case, a correlation between the two quantities is found. In the r.r. merger simulation, the system reaches a higher SFR on average, but a much modest impact on the clumpiness is obtained compared to the previous case.
\end{figure}

We then mock HST observations in the F814W filter by assuming $z = 0.7$, the median redshift of our sample. We use the \citet{bruzual03} stellar evolution model, with solar metallicity and \citet{chabrier03} initial mass function. The stars from the initial conditions are given a random age between 500~Myr and 7~Gyr following a uniform law. The image is convolved to the HST resolution, and the noise corresponding to COSMOS field data acquisition is added to the images. The resulting images are shown in panels B and D of Fig.~\ref{simulations_cutouts}.

In the mock observations, the tidal tails created by the merger are not detected anymore. In contrast, bright clumps, corresponding to the blending of star clusters, are clearly visible at the new resolution in several steps of the evolution. Remarkably, some configurations are very similar to the morphology of our observed galaxies. In particular, the eighth cutout closely resembles the galaxy ID $705860$ shown in Fig.~\ref{hstimages} (the second of the third row), reinforcing the interpretation that the clumps (including the nuclei) are probably all part of the same `two-body' merger system, even though their physical association is not obvious by eye. Finally, we measured the clumpiness, Gini and M$_{20}$ parameters from the mock images for both collision geometries and for the isolated case, with the same procedure adopted for the observations. The $mergerness$ was also calculated using Equation \ref{mergerness}. 

The results on the morphological properties are shown in Fig. \ref{gini_m20_simulations}. In both merger simulations, the system starts from a low Gini parameter and high momentum of light M$_{20}$, because of the presence of two distant galaxy components. After the first pericenter passage (indicated with a P), Gini increases on average and the track moves toward the top, as the light becomes more widely distributed along the tidal tales and interacting features. Throughout the middle of the interaction, Gini and M$_{20}$ can oscillate depending on the relative position of the two galaxies and the number of close passages, although the system always remains in the merger region of the diagram, also regardless of the clumpiness value. 
The two main bodies will then approach for the last time and finally coalesce, which implies a rapid fading of interacting signatures, a net decrease of the M$_{20}$ parameter, and a return into the not-merging regime, although in a different region compared to the starting point. 
We remind that the coalescence (indicated with a C) corresponds to the timestep when only one nucleus is recognized in the segmentation map (instead of two). However, the real physical coalescence of the two unresolved cores may happen at slightly later times, even after the simulation interval analyzed here. 
For the isolated galaxy, we always stay in the not merger region, below the \citet{lotz08} separation line. In all cases, we notice that these simulation runs are performed with a particular (although representative) set of initial conditions, such as inclination and impact parameter, thus in reality we may populate the Gini-M$_{20}$ diagram in different ways.  

The time evolution of the SFR and the clumpiness for the three simulations is shown instead in Fig.~\ref{simulations_results1}. 
In the upper pannel of this Figure, we notice a rapid increase of the SFR after the first pericenter passage and at the coalescence, similarly to previous simulations of galaxy collisions. We can also see that the SFR increases by a factor four to five, which is an intermediate value between the high enhancement (10-100) that can occur with 10\% gas fraction, and the low enhancement obtained for the same orbits and a 60\% gas fraction (below 4) \citep{fensch17}. This effect will be discussed in more details in a companion paper (Fensch et al., {\it in prep.}). We additionally remark that a time averaging of the SFR (and of the clumpiness) in the past $50$ Myr could be applied if we want to match the average SFR timescale of our observations, even though the diagrams would not change qualitatively. 

The evolution of the clumpiness is displayed in the bottom pannel of Fig. \ref{simulations_results1}. While we see that for both collision orbits the clumpiness increases compared to the fiducial run, the relative values are quite different. The prograde-prograde orbit reaches high values of the clumpiness 50~Myrs after the first pericenter and right before the coalescence, and peaks at 0.17, which is in the range of observed clumpiness in the COSMOS field. On the contrary, the clumpiness does not increase much during the retrograde-retrograde orbit, reaching only 0.05, 75~Myrs after the first pericenter. 

On Fig.~\ref{simulations_cutouts}, we can see that a high clumpiness in the prograde-prograde case is obtained by the blending of star clusters, which tends to happen at the base of the tidal features. This blending results in the accumulation of star clusters in this region, as we can visualize in the fourth cutout. The formation of the tidal tail resulting from the prograde spin-orbit coupling, the accumulation of star clusters and the enhanced clumpiness do not happen in the retrograde-retrograde collision. 

Overall, numerical simulations are able to explain qualitatively several observational findings. 
First, our merger simulations can explain the increase of the clumpiness parameter (i.e., the fraction of light coming from clumpy structures) up to values that are similar to those observed. Simulations are also able to reproduce qualitatively the increasing trend of the clumpiness with specific SFR and mergerness, which we have seen observationally in Fig. \ref{distMS_clumpiness} (including the distance from the MS) and Fig. \ref{mergerness_vs_clumpiness1}, respectively.
In addition, the same simulations show that the beginning of the clumpiness enhancement occurs at around the same time of the first rise in SFR, and not before. 
These findings suggest that clump formation and starburst activity in mergers is driven by the same underlying mechanisms, that are, enhanced gravitational instability and collapse due to the merger perturbations.

Secondly, during a merger event, a significant increase of the clumpiness can be obtained even with a modest SFR enhancement, below the threshold for starburst classification (Fig. \ref{simulations_results2}-\textit{top}). This can explain the fraction of morphological mergers in the main sequence with a relatively large clumpiness parameter.

Furthermore, we found that the geometry of the interaction plays an important role. For example, in the case of a retrograde-retrograde merger, only a very small enhancement of the clumpiness is found during the starburst, or no increase at all, implying that a high clumpiness parameter is not obtained neither in all mergers, nor in all starbursting systems. We remark that very low clumpiness parameters ($<0.04$) in correspondence of elevated SFRs ($\times 4$ above the MS), are not inconsistent with our data, which comprise a significant number of starbursts with a similar low fraction of clumpy emission (cf. Fig. \ref{histogram_clumpiness1}). It is thus clear that the variety of spin-orbit coupling can explain part of the scatter of the clumpiness observed in our sample.
%To conclude, it is clear that the variety of spin-orbit coupling can explain part of the scatter of the clumpiness observed in our sample. 
A forthcoming paper will study the physical properties and evolution of simulated clumps with more detail, with a more diverse set of galactic disks (Fensch et al., in prep.).

Finally, we may wonder how simulation results would change if we include the dust attenuation. 
Several works have already done this, showing that the extinction has effects, in particular on the SFR estimates, but not on the existence of clumps themselves (Jonsson et al. 2010, Hayward et al. 2010). Clumps identified in stellar mass maps are still detected in the optical in simulation images of submillimeter-bright galaxies \citep{cochrane19}.
Furthermore, at higher redshifts ($z\sim1$), the extinction is not particularly high even in big UV clumps (regardless of being merger-induced or not), so the patchy radiation that we detect does not represent just holes in the dust distribution, but real clumps with high gas and SFR densities (Elmegreen \& Elmegreen 2005, Wuyts et al. 2012).
For these reasons, the dust attenuation would not change the results from our simulations.

\subsection{Clumpiness evolution during the merger}\label{mergerphase}
%Recall here the observability timescale of mergers (Gini-M20 mergers) - vedi Nevin 2019

In Section \ref{results} we have divided the sample in SBs and MS galaxies, finding that the two subsets have different clumpiness distributions, with the former prevailing in the high clumpiness end, while the latter are dominant at very low $c<0.02$. 
However, this classification is too simplistic, and the complexity of the MS and SB populations can be resolved in part by considering their apparent morphology. We have indeed seen in Fig. \ref{histogram_clumpiness4} that morphological merger starbursts, selected by their large Gini and M20 coefficients, are the major responsible for the high-end tail of the clumpiness distribution, with a median $c=0.09$ and a maximum fraction of light in the clumps of $20\%$. 
On the other hand, the subset of starbursts that are not merger selected from their morphology is essentially indistinguishable from that of isolated (not merger) main sequence systems (Fig. \ref{histogram_clumpiness4}). 

As highlighted in previous works, our morphological classification criterion, firstly defined by \citet{lotz04}, is able to identify mergers over a relatively short temporal window compared to the whole merger duration. The observability timescale of a merger in the Gini-M$_{20}$ diagram is approximately $0.2-0.3$ Gyr according to \citet{nevin19}, corresponding to the period when the interaction signatures are more evident in the form of bright tidal tails or very disturbed, elongated or asymmetric global structures. In contrast, at the coalescence, residual merging features rapidly fade below the surface brightness detection threshold, hampering its true nature recognition visually, especially at our redshifts. %In this phase though, the total SFR inferred from the infrared can still be large and above the starburst threshold. 
Their increasingly difficult identification at the coalescence is also seen in simulations.
This suggests that different time evolutionary merger phases can in part explain the large spread of mergerness and clumpiness in the SB distribution. 

In order to test this interpretation, we analyzed the subset of $19$ starbursts, with available HST images and in our redshift range, that were presented in \citet{calabro18} as representatives of off-MS systems at $0.5<z<0.9$ above a M$_\ast$ of $10^{10}$ M$_\odot$. We showed that this sample comprises a sequence of different evolutionary merger stages, which can be traced by the equivalent width (EW) of Balmer or Paschen lines, and by the total attenuation (A$_\text{V,tot}$) toward the center of the starburst core in a mixed dust and stars configuration \citep{calabro19}.

\begin{figure}[t!]
    \centering
    \includegraphics[angle=0,width=\linewidth,trim={0.7cm 1.5cm 2.5cm 3.8cm},clip]{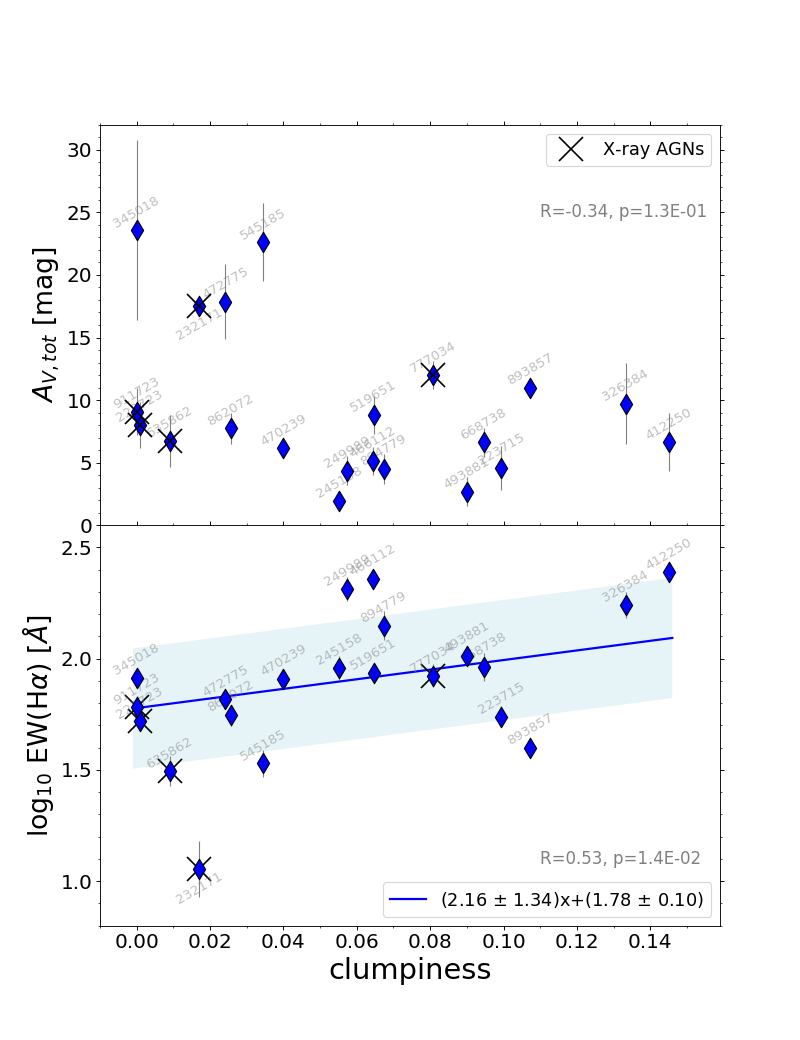}
    \caption{\small Comparison between the clumpiness and the mixed-model attenuation A$_\text{V,tot}$ toward the center (upper plot), and the equivalent width (EW) of H$\alpha$ (lower panel) for $21$ starbursts in our same redshift range $0.5<z<0.9$, and analyzed in \citet{calabro19}. The corresponding IDs are added to each galaxy, while black crosses indicate X-ray detected sources. The Spearman correlation coefficient R and the corresponding p-value are highlighted in each panel. 
    We derive no correlation for the first diagram, and a mild correlation $>3\sigma$ for the second, thus only in the latter case we derive the best linear fit (blue line, with equation included the legend) and $1\sigma$ dispersion (blue shaded region). However, even in the first plot, the highest clumpiness values are found only for the less attenuated starbursts, while low clumpiness objects are systematically more obscured (A$_\text{V,tot}>15$). 
    Given that A$_\text{V,tot}$ and EW(H$\alpha$) have been used as merger stage indicators in \citet{calabro19}, our findings provide an indication for a possible clumpiness evolution, decreasing from early-intermediate phases to late stage mergers. This is corroborated by the subset of X-ray AGNs (likely close to the blow-out phase) showing preferentially a lower clumpiness.}\label{correlations_clumpiness}
\end{figure}

In Fig. \ref{correlations_clumpiness} we compare the clumpiness of this SB subset to A$_\text{V,tot}$ and the EW of H$\alpha$ and Pa$\beta$ lines, measured in \citep{calabro19}. 
In the first upper panel, we find no significant correlation between $c$ and A$_\text{V,tot}$, with a Spearman correlation coefficient R$=-0.34$ (p-value$=0.13$) and angular coefficient of the best-fit line consistent with $0$. However, if we assume no correlation, the four starbursts with the highest obscurations A$_\text{V,tot}$ above $15$ mag all have a low clumpiness below $0.04$. This confirms that an important fraction of galaxies ($50\%$ in our Magellan dataset) contributing to the first two low clumpiness bins in Fig. \ref{histogram_clumpiness1} may be actually late stage mergers observed after the coalescence. 
On the other hand, early and intermediate merger systems with $0<$ A$_\text{V,tot}<15$ mag can show the full variety of clumpiness values, and these are the only phases where we observe a substantial clumpiness enhancement above the average population level and above $0.1$.

In analogy to the former result, when comparing the clumpiness to the equivalent width of Pa$\beta$ and H$\alpha$, we also measure an angular coefficient slightly below $2\sigma$. However, in the latter case the Spearman coefficient R is equal to $0.53$ (p-value$=0.014$), indicating that the correlation is significant according to this statistical test. 
The existence of the latter (even though mild) correlation, and the position of the Magellan SBs in the first panel of Fig. \ref{correlations_clumpiness}, can shed light on the possible triggering mechanisms and fate of the observed clumps. Following the results of previous simulation works, clumps can form with the increasing compressive turbulence modes and subsequent fragmentation induced by the merger during early-intermediate stages \citep{renaud14}. %, following also the increase of the SFR above the SB level (as suggested by the correlation between SFR and clumpiness in our simulations, refer to the figure).
The absence of late stage mergers with high clumpiness suggests that, after the main triggering events, clumps could be rapidly destroyed by strong stellar radiation or AGN feedback, or they could be incorporated in the central galactic bulge. 

Intriguingly, we notice that $4$ out of $5$ X-ray detected AGNs in this small subset have a very low clumpiness below $0.02$. % and, if we do not consider these objects, we would obtain a higher statistical significance of the A$_\text{V,tot}$-$c$ correlation. 
As suggested in \citet{calabro19}, these systems may be in an advanced phase of the AGN activity and central black hole growth, and possibly approaching the blow-out phase, thus an impact on clumps survival cannot be excluded. If we do not consider these objects, we would obtain a higher statistical significance of the A$_\text{V,tot}$-$c$ correlation.

Another possibility is that clumps become too faint in i-band compared to the host galaxy, following the aging of the stellar populations or the higher dust obscurations expected in advanced phases, and they are not detected anymore in this band. However, we remind that testing the impact of feedback processes and studying the final fate of the clumps is beyond the scope of this work, and it will be investigated in future papers. 

Even though the time evolutionary sequence is a tantalizing interpretation, the absence of very strong correlations may indicate that other effects also play a role on clumps formation, such as the impact geometry, the dynamics and mass ratio of the two components, and the viewing angle toward the system, all of which would be in any case very difficult to quantify from current observational data. % both on the visibility of morphological merger features (i.e., affecting its classification in the Gini-M20 diagram) and on the formation or visibility of clumps. %, in particular to move it to the bottom-right part below the separation line, thus affecting its merger selection. 
In particular, the strong dependences on the orbital configuration (Section \ref{simulations}) suggest that we may have large object by object variations in the clumpiness while having still a strong SFR enhancement (Fig. \ref{simulations_results2}-\textit{top}).
%In particular, we checked with simulations that a retrograde-retrograde galaxy collision (Section \ref{simulations}) can have a strong influence as well, and may indeed be one of the main reason for reducing the significance of the above correlations. 

\subsection{Multi-wavelength morphology from CANDELS}\label{CANDELS}

We have identified clumps in single broad-band (F814W) images, which are available for the majority of galaxies in the 2 degree$^2$ COSMOS field. A possible limitation of our approach relies on the fact that we can probe the emission of clumpy structures only in a limited wavelength range, which, considering the filter transmission curve and our redshift interval, covers approximately the $3700$ \AA\ to $6400$ \AA\ rest-frame range. The morphology of clumpy features can potentially change if we go to the UV rest-frame or at longer wavelengths (near-IR).

\begin{figure}[t!]
    \centering
    \includegraphics[angle=0,width=\linewidth,trim={5.5cm 7.cm 6.5cm 1.5cm},clip]{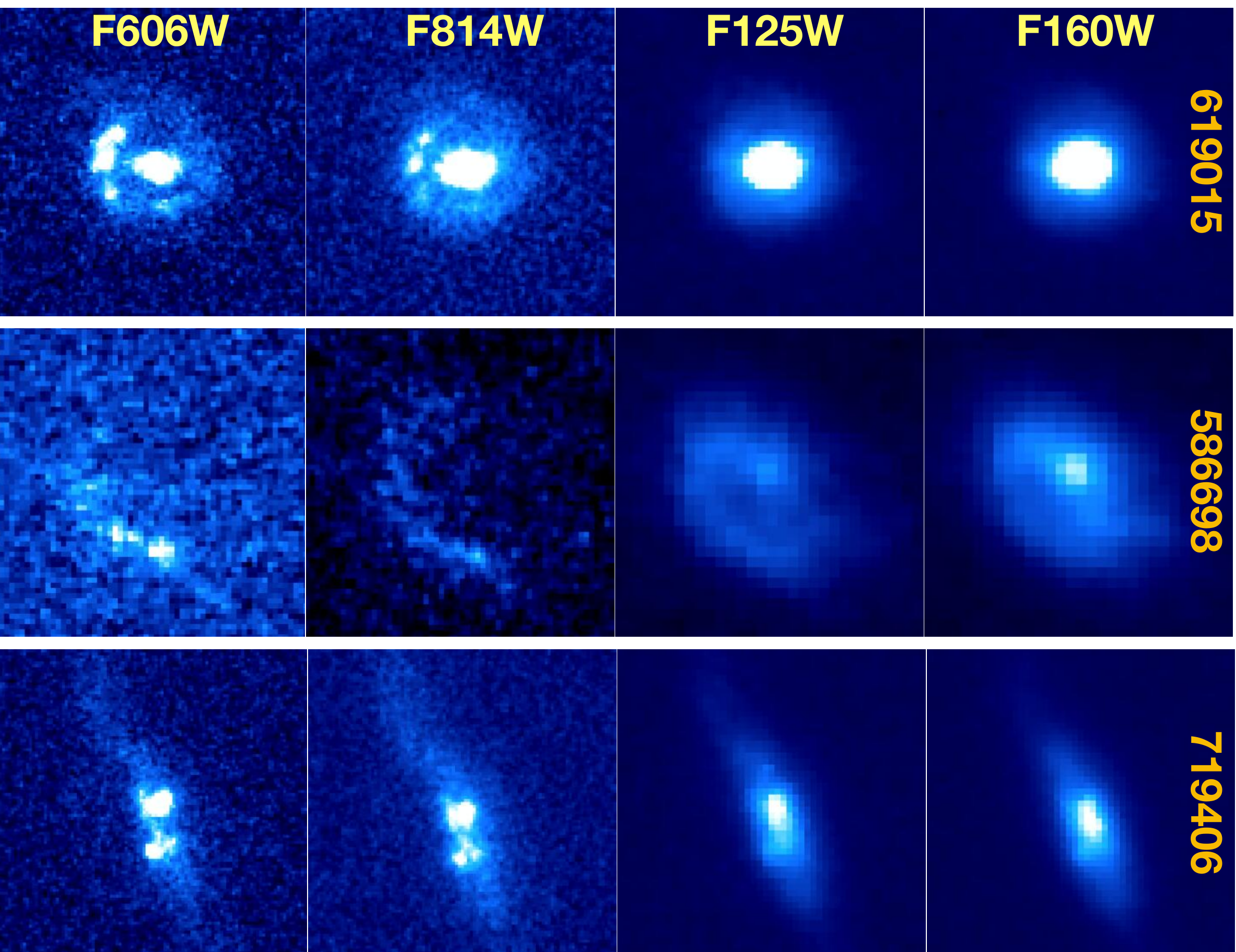}
    \caption{\small HST cutout images in the F606W, F814W, F125W and F160W broad band filters for three starbursts in the COSMOS-CANDELS field. As explained in the text, the second galaxy has a photometric redshift of $0.92$, while the third has a SFR a factor of $3$ above the main sequence at the same redshift $z=0.74$, but clearly shows a merger morphology (in spite of its edge-on profile). The multi-wavelength comparison shows that clumps become brighter at shorter wavelengths, thus they likely represent merger-induced young stellar associations rather than old pre-existing structures. The IDs shown in the right part of the figure come from \citet{laigle16}.}\label{CANDELS_figure}
    \vspace{-0.4cm}
\end{figure}

We can check the multi-wavelength behavior for the three starburst galaxies in our sample that are included in the COSMOS-CANDELS field, observed at high-resolution by HST \citep{koekemoer11} in four broad-band filters (F160W, F814W, F125W and F160W) at resolutions of $0.08''$, $0.09''$, $0.12''$ and $0.18''$, respectively (Fig. \ref{CANDELS_figure}). In order to have a larger statistics, we have considered in this analysis one galaxy in the subset analyzed so far (ID 619015), and other two galaxies (ID 586698 and 719406) belonging to the parent sample but excluded from the final selection adopted in this paper: while the first has a slighly higher redshift ($0.92$) than our selection criteria, the second has a distance from the MS of $0.51$ dex (above a factor of $3$). However, it has clearly a merger-like morphology given the presence of a long tidal feature in the upper part. 
Having similar clumpy morphologies (from i-band) to the population of galaxies considered in this work, they can yield important informations on clump properties in principle also for the remaining sample. % not included in CANDELS. 

Analyzing each starburst in more detail, we can clearly see in the first galaxy two clumps in i-band and four clumps in the UV, while all of them become fainter and undetected at longer wavelengths. They also occupy part of a ring structure surrounding the central nucleus. Simulations predict that this configuration could represent the late phase of a collision between disk galaxies \citep{bekki98}, or created by tidal accretion of material from a gas rich donor galaxy \citep{bournaud03}. 

Also in the second galaxy the clumps become brighter in the UV rest-frame and they are displaced along half a ring. Longer wavelength images reveal that this elongated structure is connected to a single bigger system with just one main nucleus, highly attenuated or very old, invisible in F606W and F814W filters. %\textcolor{blue}{Should derive age, stellar mass and attenuation for clumps and nucleus.}

Finally, the last system shows multiple clumps in $i$ and U band images, below the main central nucleus visible in all the bands. In particular, the brightest clump falls at the border of the main edge-on stellar disk. % of the newly formed system (or ongoing merger). 

Overall, the clumps shown in the above three galaxies become more prominent from redder to bluer bands (where sometimes they can outshine the nucleus itself), which would yield an increased clumpiness in the UV compared to i-band. On the opposite side, they become undetected in near-IR bands (even though the resolution is slightly lower).
This result indicates that they are likely young structures induced by the merger rather than pre-existing aggregations of older stars. 
Furthermore, they seem to have a low mass fraction and possibly a low SFR fraction, despite their UV prominence. %On the other hand, , suggesting that . This suggests that. This result suggest that our clumps are really young structures induced by the merger and not pre-existing aggregations of old stars. 

A systematic investigation of the physical properties of the clumps formed by merger events at these redshifts, including their dust attenuation, stellar ages and masses, gravitational stability and kinematics, could be possible in the future with the availability of high spatial resolution multi-wavelength bands for a larger subset of objects. This will additionally allow to compare their size and stellar mass distributions to those observed in main sequence systems and at higher redshifts.  

\subsection{Comparison with other studies}

As mentioned in the Introduction, several studies have investigated the origin of clumpy galaxies at redshifts overlapping with our range. \citet{puech10} claimed that mergers may be the dominant triggering mechanisms of clumpy galaxies at $z<1$. However, his limited sample of $11$ objects is representative of a very specific redshift ($\simeq0.6$), galaxy stellar mass (log$_{10}$ M$_\ast \sim 10.2$ [M$_\odot$]) and type (e.g., absence of a central bulge), which cannot be representative of all the star-forming population and of the same dynamic range spanned in our work. In addition, he found that the majority of UV clumps tend to vanish when looking at longer wavelengths, so they could be biased towards lower attenuations or higher ongoing SFRs. 
%Also Ribeiro et al. (2018) points out that mergers are likely the origin for the high fraction of galaxies at $2 < z < 6$ with two bright clumps. 
Also \citet{ribeiro17}, while focusing on much higher redshift than our work ($2 < z < 6$), interprets double clumps as possible merging systems. However, if it is true, the clumps likely represent the nuclei of the two colliding galaxies, thus they should not be considered anymore as clumpy galaxies by our definition. %, and it is not clear how this can affect their results. 

On the other hand, \citet{murata14} noticed that the redshift evolution of the fraction of clumpy galaxies is inconsistent with that expected from the merger rate, thus concluding that mergers do not contribute to the clumpy population at all epochs. 
However, they detected clumps directly on the images without an intermediate smoothing step, which is important to facilitate the identification of high spatial frequency components and separate them from equally bright regions with smoother profiles. In addition, they identified clumpy galaxies preferentially on the main sequence. Indeed, their SFRs are derived from SED fitting (from UV to mid-IR), not allowing to select systems with obscured star-formation, which is a dominant component in infrared-luminous mergers \citep{goldader02,calabro18}. %Finally, they do not perform a merger classification, but they , which is uncertain. 
A similar conclusion based on the same argument is reached by \citet{guo15} for clumpy galaxies at $0.5<z<3$. They suggest instead that violent disk instabilities are the main triggering mechanism at high stellar masses (M$_\ast > 10^{10.6}$ M$_\odot$), while minor mergers may contribute the most for intermediate mass systems with $10^{9.8}<$ M$_\ast < 10^{10.6}$ M$_\odot$. However, they consider UV clumps, which may disappear in the optical rest-frame, as shown by \citet{puech10}. 

Our result should be considered as complementary to all these studies. We are showing the importance of mergers as responsible for triggering clump formation in intermediate redshift galaxies, enhancing the clumpiness at higher levels compared to other mechanisms at this cosmic epoch. 
However, we do not claim that all clumps are induced by mergers. Indeed, a fraction of main sequence galaxies (which are the majority $96$-$98\%$ of the star-forming population) with higher clumpiness are not identified as mergers and may be consistent with other formation channels, such as minor mergers or disk instabilities. Therefore, we are not in contradiction with the two previous works. 
% Se hai tempo discuti anche Bournaud 12 e Hinojosa-Goni's paper.

Finally, \citet{lotz04} showed that local ULIRGs (which are all mergers) have an enhanced clumpiness compared to the main sequence star-forming population. Our paper thus suggests that this result can be extended up to redshift $\sim 1$ to starburst galaxies and to morphologically selected mergers.

\subsection{Interpreting high redshift clumpy galaxies}\label{highz}

We have demonstrated in previous sections that mergers can trigger the formation of stellar clumps in galaxies at $0.5<z<0.9$. Given this explanation at intermediate redshifts, we can wonder whether a similar connection also holds at earlier cosmic times. 

At high redshifts ($z>>1$), young massive clumps observed in the UV and optical rest-frame, having M$_\ast$ $\sim10^{8}$ M$_\odot$ and ages of $100$-$500$ Myr, are generally thought to be driven by violent gravitational instabilities in gas-rich highly turbulent primordial disks, typically lying in the star-forming MS \citep[e.g.,][]{noguchi99,immeli04a,immeli04b,elmegreen05A,elmegreen07a,elmegreen08B,bournaud07,forster06,forster09}. The fuel needed for star-formation and clumps triggering may be provided by relatively smooth accretion of cold gas from the cosmic web and from the CGM \citep{keres05,dekel09,aumer10}. %(Keres et al. 2005, Dekel et al. 2009, Aumer et al. 2010).
Furthermore, many studies have revealed that clumpy galaxies have kinematics consistent with rotating disks \citep{bournaud08,daddi08,shapiro08,epinat09}. %(Bournaud et al. 2008, Epinat et al. 2009, Daddi et al. 2008, Shapiro et al. 2008). 

Even though this scenario is physically motivated, we cannot rule out mergers as possible triggering mechanisms for clumps formation, in analogy to what has been shown at $z<1$. 
As the fraction of mergers increases monotonically at earlier times \citep{conselice03}, their contribution could be more important in the past, and, although they can be less efficient at rising the SFR to starburst levels \citep{fensch17}, our results indicate that a large clumpiness parameter can be obtained even without a strong enhancement of the star-formation activity.
In addition, it is possible to preserve some degree of global rotation even during a merger event, and also a disk could rapidly reform in the latest merger stages \citep{rothberg06,springel02,fensch17}. %(Rothberg et al. 2006, need more refs). 
Therefore, many high-$z$ systems with global rotation in the stellar or gaseous components can still be mergers. As complementary probes to recognize these systems, we could instead look for the presence of compact, highly obscured cores in the host galaxy, which can trace late or post-coalescence merger phases \citep{calabro19,puglisi19}. %, or directly for dust emission from the same clumpy structures through high-resolution sub-mm images.

Further clues to the origin of high-redshift clumpy galaxies come from recent ALMA observations, which show a dicotomy of clumps properties depending on the SFR level of the host galaxy.
For example, \citet{hodge18} and \citet{tadaki18} observed sub-kpc clumpy structures with ALMA in the dust continuum and CO emission for a small subset of luminous submillimeter galaxies (SMGs) in the redshift range $1.5$-$5$.
Among the sample of $11$ SMGs presented by \citet{hodge18}, the low Sersic index profile measured in one galaxy suggests it might be a late stage merger, while interacting signatures in the optical are revealed for some of their remaining systems. They showed that these structures are displaced in the inner $5$-$10$ kpc regions, analogous to the spatial distribution of our brightest clumps (in both observations and simulations), which form close to the nuclei and in the beginning of tidal tails. 
Additionally, the ALMA clumps produce $2$ to $10\%$ of the total galaxy emission, in agreement with the range of clumpiness that we found in the optical. This suggests that, being highly star-forming and dusty, their structures may represent still early phases of clump formation. 

On the contrary, normal star-forming isolated disks are smoother at the same sub-mm wavelengths. \citet{cibinel17} found that UV clumps in a main sequence galaxy at redshift $=1.5$ are not visible anymore with ALMA in the CO(5-4) transition, which could be due to a lower gas content (or equivalently, higher SFE), or to a lower SFR of the clumps. \citet{rujopakarn16} studied $11$ normal star-forming galaxies at redshifts $1.3$-$3$ with $\sim 0.4''$ resolution ALMA images. They also found no evidence of clumpy structures, which instead appear at UV wavelengths as unobscured regions, owing small SFR fractions from $0.1$ to $5\%$ of the whole systems. %Similarly to the other work, he also found no evidence of star-formation rate maps  

This dicotomy resembles the difference of clumpiness between IR-luminous starbursts and normal MS galaxies at $z<0.9$, and we may wonder whether it has the same physical explanation. Under the merger origin hypothesis for SMGs \citep[e.g.,][]{tacconi06,tacconi08,engel10,alaghbandzadeh12}, it seems reasonable to think that many of the ALMA clumps observed in high-$z$ infrared luminous galaxies may be actually produced by merger interactions. 
However, we warn the reader that the merger origin of SMGs has not been fully assessed yet, with alternative studies claiming they are just gas rich disks representing the most massive, luminous extension of the galaxy main sequence \citep[e.g.,][]{dave10,dunlop11,michalowski12,targett13}.
Future observations with ALMA could further constrain the different hypotheses and better characterize the clumps detected in our COSMOS sample for both the MS and SB population.

\section{Summary and conclusions}\label{conclusions}

Inspired by the merger nature of infrared luminous starbursts at $0.5<z<0.9$ shown in previous works, we have studied in the same redshift range the effects of mergers on clumps formation by comparing the high-resolution HST optical rest-frame morphologies of 79 starbursts to a control sample of 109 normal star-forming main sequence galaxies. 
We performed an additional visual merger identification among the main sequence population and applied the classical Gini-M$_{20}$ classification to select galaxies with merger-like shapes.
Our main results are summarized as follows:
\begin{itemize}
\item Starburst and main sequence galaxies have different clumpiness histogram distributions: the former dominate in the high clumpiness regime, while the latter are mostly found at lower clumpiness values. Given the merger nature of intermediate-$z$ starbursts, this suggests that mergers are likely responsible for clumps formation and their increased luminosity with respect to normal star-forming isolated disks.
\item The majority ($76\%$) of morphological mergers (according to Gini-M$_{20}$ based criterion by \citet{lotz08}) are starburst galaxies. However, $57\%$ of starbursts are not classified as mergers from their i-band morphology. This suggests that the Gini-M$_{20}$ merger selection is highly incomplete, likely due to multiple effects, including the galaxy inclination, the impact parameter, the dust attenuation, and the rapid fading of merger signatures after the coalescence.
\item A larger difference of the clumpiness histogram distributions is obtained when including visually selected mergers in the SB subset and when comparing morphological mergers to not mergers. 
In particular, Gini-M$_{20}$ mergers, regardless of their level of SFR, have a median clumpiness a factor of three higher compared to the rest of the population, and are almost entirely responsible for the high clumpiness tail observed among our sample.
We also found that the fraction of morphological mergers and their median clumpiness increase monotonically with the distance from the main sequence. 
\item From hydrodynamical simulations of merger galaxies with initial conditions typical for our redshift range, %From mock HST-ACS F814W observations derived at different time steps, 
we found that mergers can significantly enhance the clumpiness of the system compared to isolated main sequence galaxies, by a similar amount to that observed in real images. Different spin-orbit coupling of merging galaxies can fully explain the scatter of the observed clumpiness values from $0$ to $20\%$.
\item For a sample of $19$ SBs with Magellan-FIRE spectra, there is a mild correlation between the clumpiness and the equivalent width of Balmer and Paschen lines, suggesting a possible clumpiness evolution during the merger, decreasing from early-intermediate to later stages after the coalescence.
X-ray detected AGNs are preferentially found in low-clumpiness systems, suggesting a possible clump suppression induced by AGN feedback. However, other effects (including galaxy inclination, rotation, attenuation and impact parameter) are likely responsible for the low correlation strength ($\simeq$ $2\sigma$). 
\item Using four band high-resolution images for three clumpy galaxies in the COSMOS-CANDELS field, we have showed that merger induced clumps are generally young and UV-bright, likely formed during the merger rather than being older pre-existing stellar structures. However, a larger sample is needed to study the statistical properties of the clumps (e.g., sizes, stellar masses and ages) and investigate their evolution.
%However in the future, a detailed physical study  of these clumps and of a larger statistical dataset is needed to investigate their evolution.
\end{itemize}

Merger triggered gas compression and fragmentation can provide the physical explanation for the formation of stellar bright clumps. We expect that this mechanism is more frequent at high redshift, given the increasing fraction of mergers at earlier epochs. This work rises questions on the real nature of clumps observed in high redshift galaxies, suggesting that mergers could be an alternative, powerful channel for enhancing the clumpiness.  
%, and because the clumpiness during a merger event can be high even with modest enhancements of the SFR. 
%The question arises whether many clumps observed at high redshift (both in MS) and is SBs could be formed through mergers. 
If this is true, the clumpiness could be used as a complementary merger diagnostic (though still incomplete) to identify mergers from the morphology when the typical low-surface brightness interacting features (e.g. tidal tails) become too faint. % (with respect to the clumps) to be detected. 

Deeper images in the optical and near-IR rest-frame with Euclid and JWST will allow in the near future clumps detection and their physical characterization (through a multi-$\lambda$ approach) for larger statistical samples of clumpy galaxies at higher redshifts and similar spatial resolutions to those considered here. In addition, they will facilitate the study of the environment and the morphological properties (including merger signatures) of their host galaxies.  

\begin{acknowledgements}
We thank the anonymous referee for comments improving the quality of this manuscript. I.D. is supported by the European Union's Horizon 2020 research and innovation program under the Marie Sk\l{}odowska-Curie grant agreement No 788679.
\end{acknowledgements}

% for the bibliography, at the end
%\bibliographystyle{aa} % style aa.bst
%\bibliography{bibliography_clumpiness.bib} % your references Yourfile.bib

%\end{document}

%\newpage

\appendix

%\clearpage
\section{Comparison between the clumpiness and other quantities}\label{additional_plots}

\begin{figure}[h!]
    \centering
    \includegraphics[angle=0,width=0.9\linewidth,trim={1.1cm 0.cm 3.cm 2cm},clip]{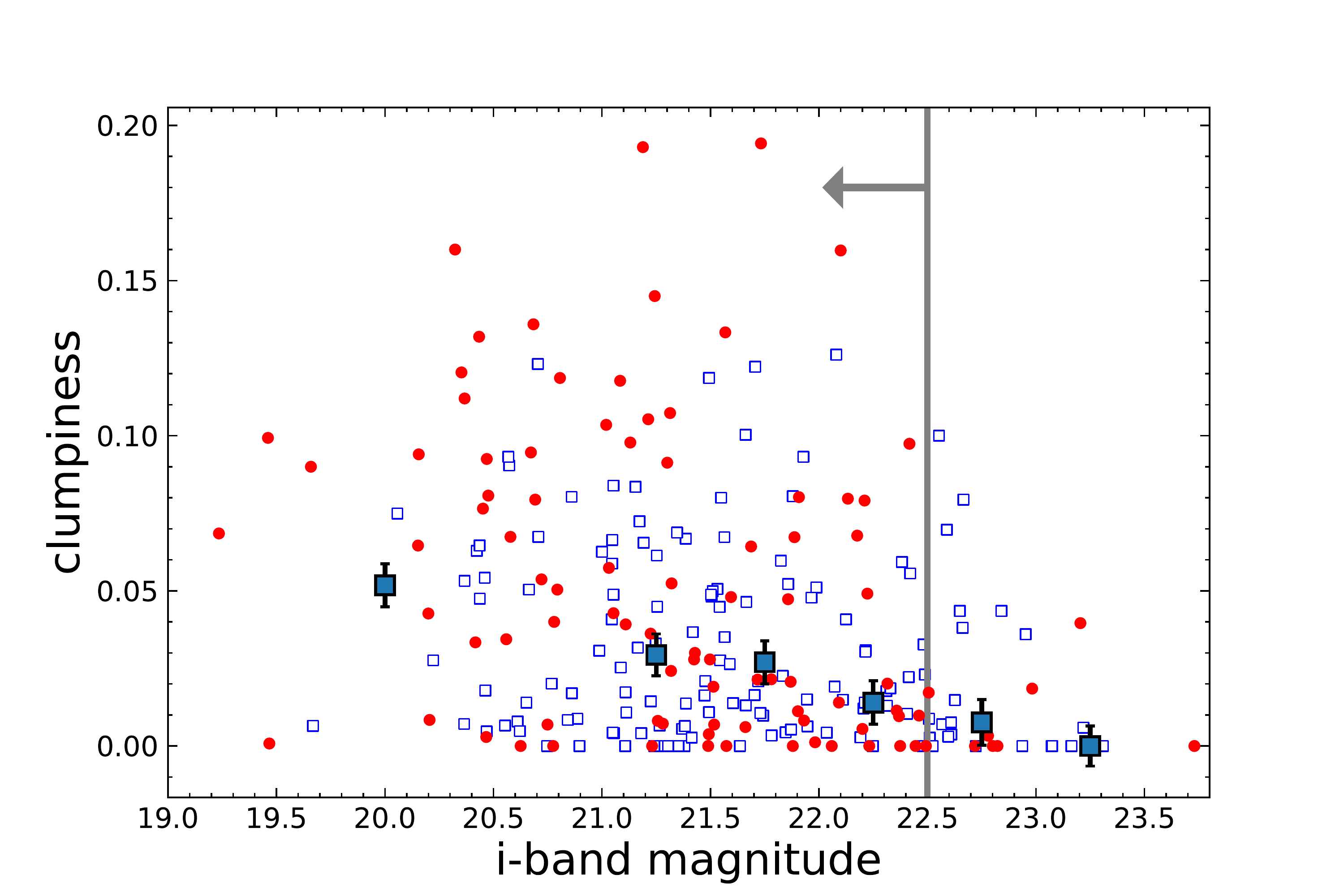}
    \includegraphics[angle=0,width=0.9\linewidth,trim={1.1cm 0.cm 3.cm 2cm},clip]{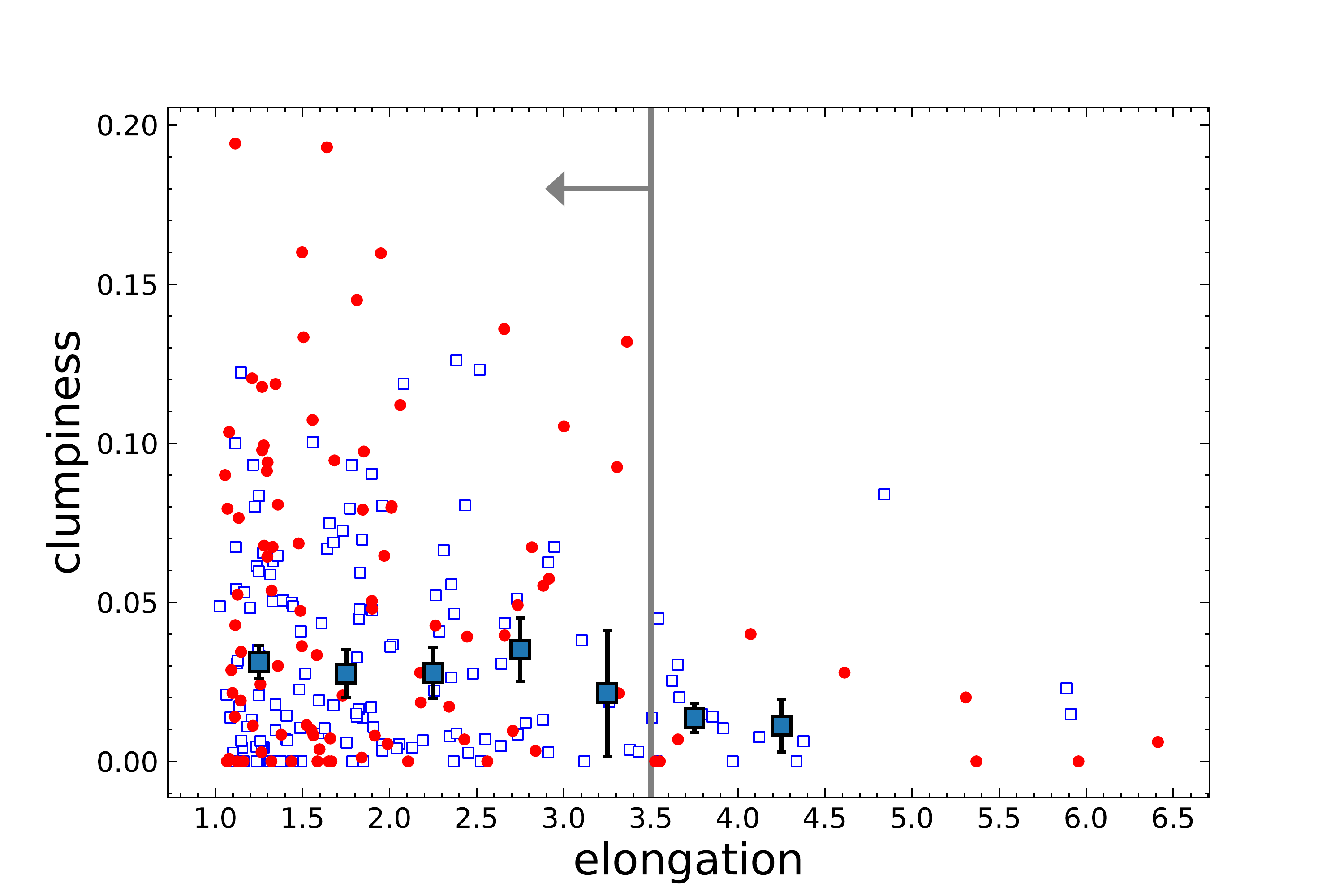}
    \caption{\small \textit{Top:} Clumpiness vs i-band total magnitude (i$_\text{mag}$) %(from \citet{laigle16}) 
    for our initial sample of $96$ starburst and $145$ main sequence galaxies (with red filled circles and blue empty squares, respectively). Median clumpiness values and errors (gray squares with black error bars) are derived for $6$ bins of i$_\text{mag}$. Our selection cut (i$_\text{mag}<22.5$) is highlighted with a vertical gray line. \textit{Bottom:} Clumpiness vs elongation for the same sample as above, with median clumpiness and errors estimated in $7$ bins of elongation. The vertical gray line indicates our cut for the final selection (elongation $<3.5$), even though we remember that one additional nearly edge-on MS galaxy was also removed by visual inspection.}\label{sample_cuts}
\end{figure}

\begin{figure*}[h!]
    \centering
    \includegraphics[angle=0,width=0.48\linewidth,trim={1.1cm 0.cm 3.cm 2cm},clip]{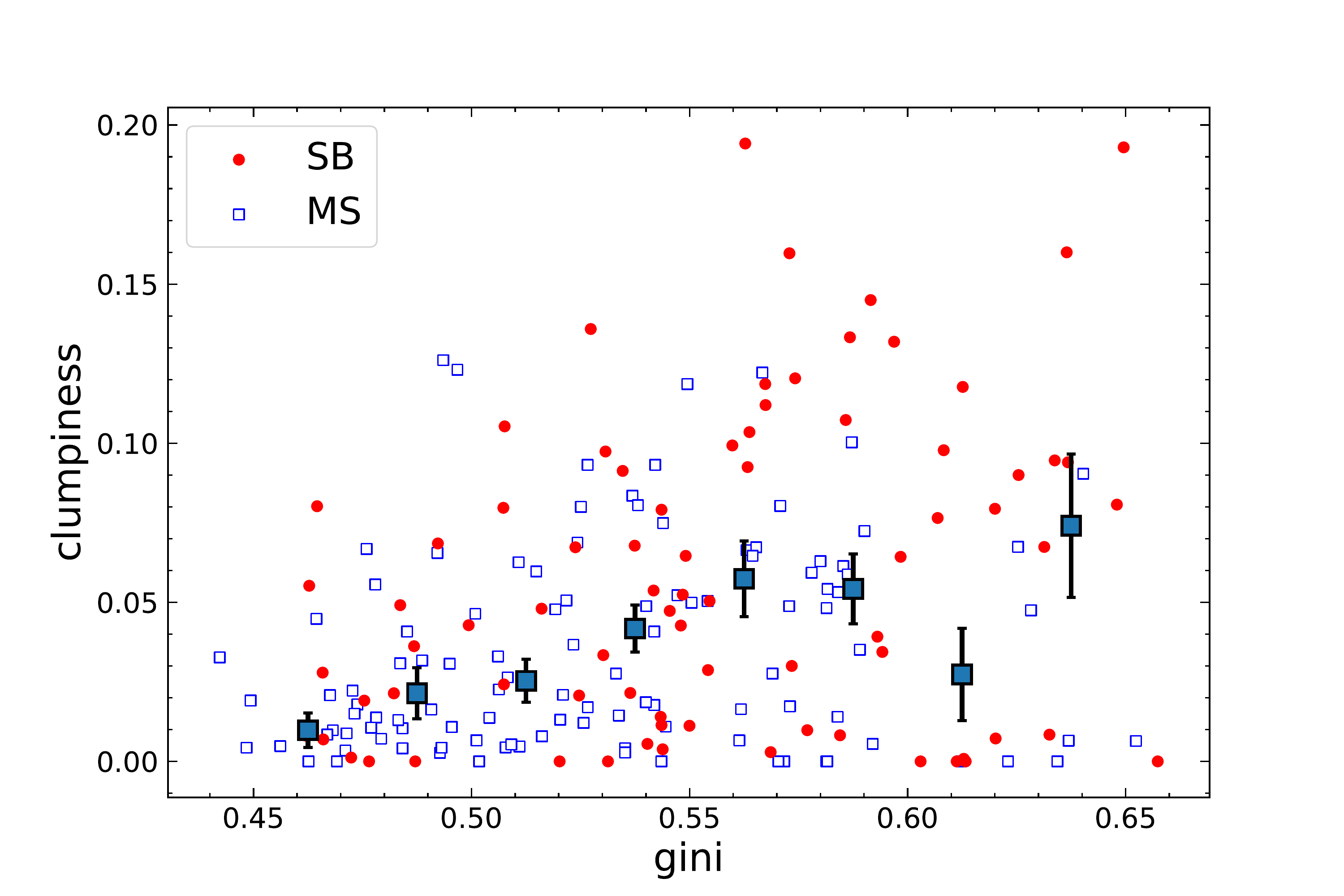}
    \includegraphics[angle=0,width=0.48\linewidth,trim={1.1cm 0.cm 3.cm 2cm},clip]{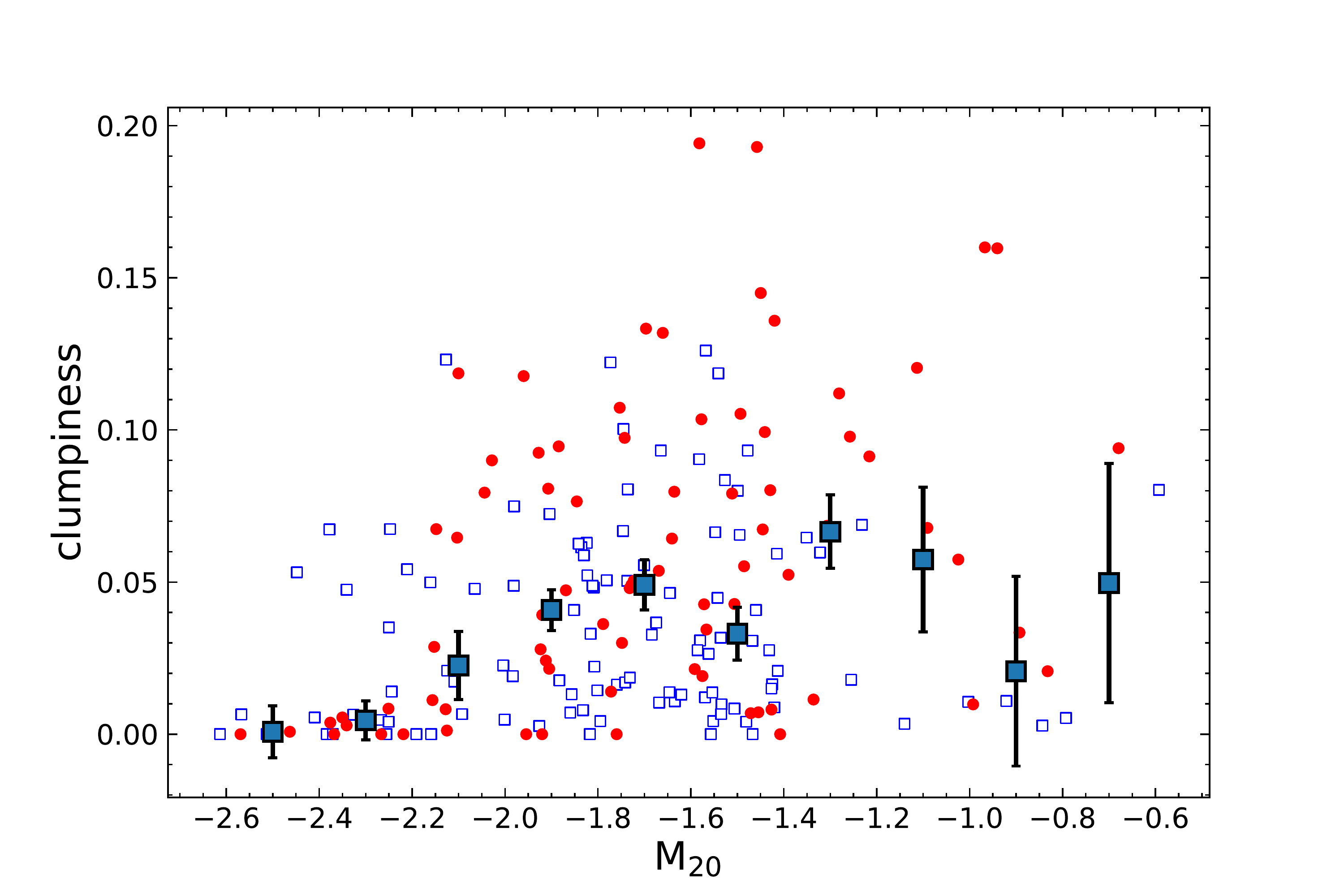}
    \caption{\small The three diagrams compare the clumpiness of our final galaxy sample to the Gini and M$_{20}$ coefficients (first and second panels, respectively), and to the $mergerness$ parameter, defined in Eq. \ref{mergerness}. Starbursts are shown with red filled circles, while main sequence galaxies are empty squares and color coded in blue. The median clumpiness (and errors) computed in bins of Gini, M$_{20}$ and $mergerness$ are represented with gray squares and black error bars.}\label{clumpiness_correlations}
\end{figure*}

\begin{figure}[h!]
    \centering
    \includegraphics[angle=0,width=0.9\linewidth,trim={0.4cm 0.8cm 6.9cm 3.4cm},clip]{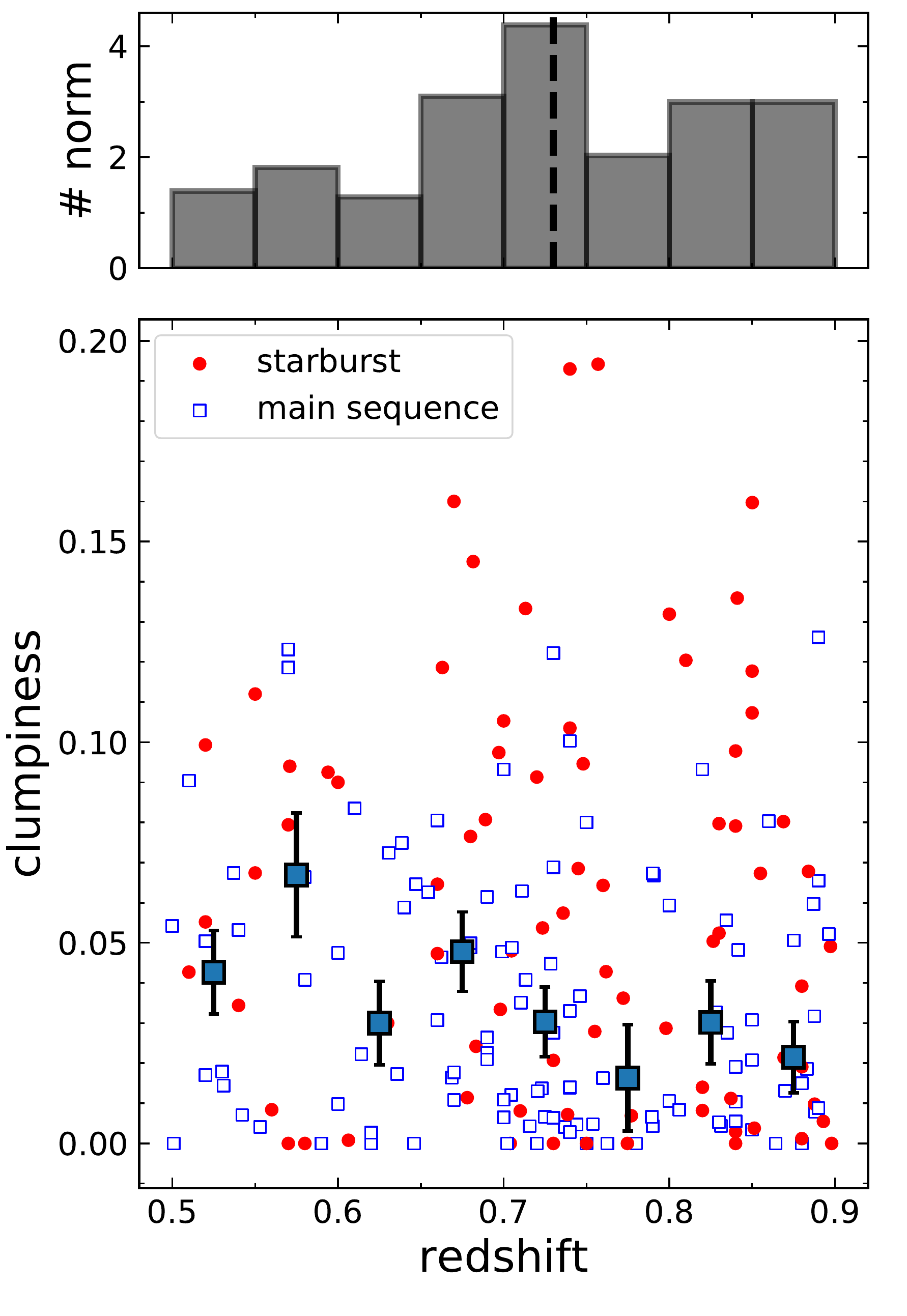}
    \caption{\small Redshift vs clumpiness for our final selected galaxies, with gray squares and black error bars representing the median clumpiness and error calculated inside $8$ redshift bins. On the top of the panel is shown the histogram distribution of the redshifts of our galaxies, with the median redshift ($z_{med}=0.73$) highlighted with a black dashed line.}\label{redshift_scatter}
\end{figure}

\begin{figure}[h!]
    \centering
    \includegraphics[angle=0,width=0.9\linewidth,trim={1cm 0.cm 2cm 2cm},clip]{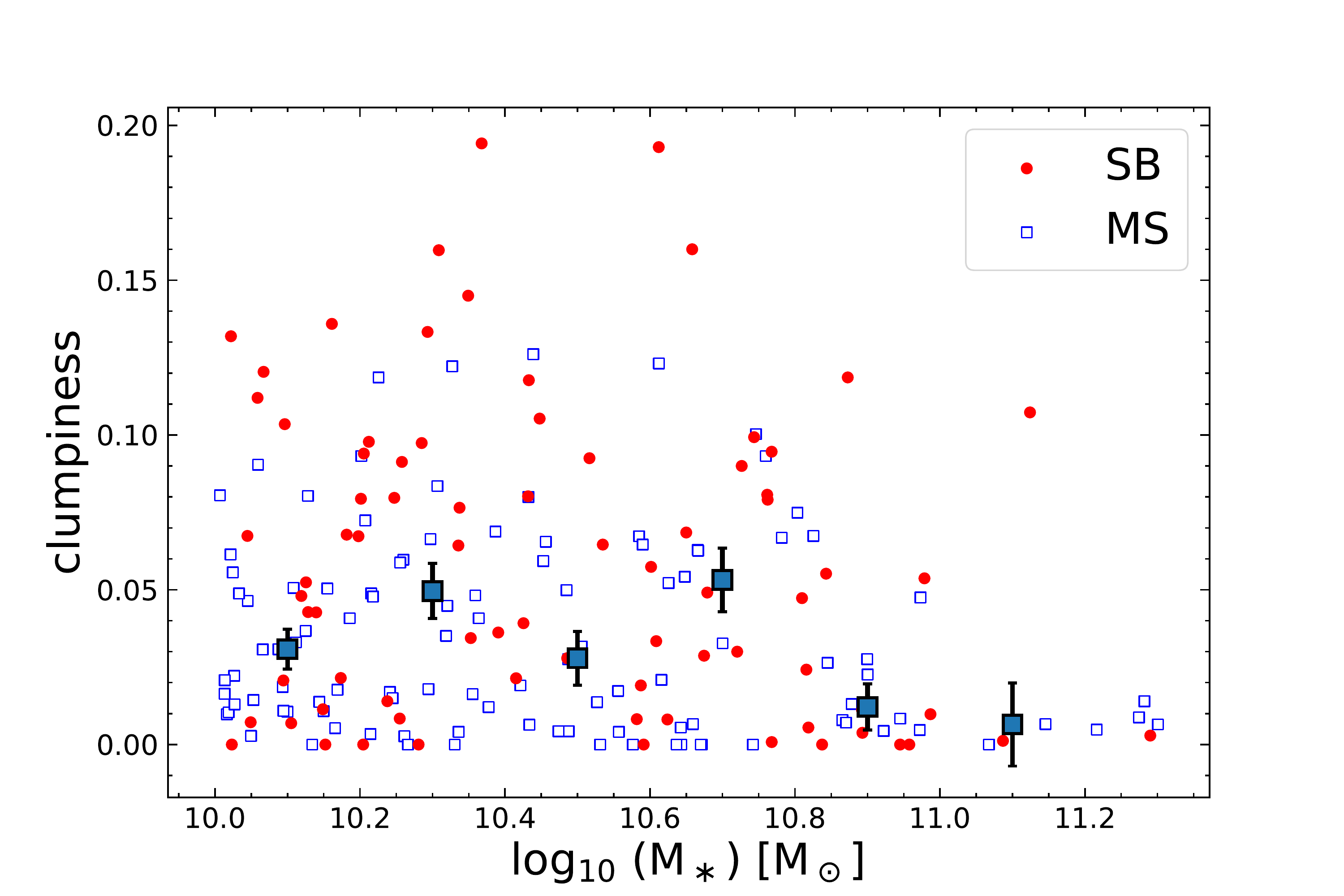}
    \caption{\small \textit{Right:} Stellar mass vs clumpiness for our final sample of SB and MS galaxies (with the same color coding and markers of Fig. \ref{clumpiness_correlations}), with median clumpiness (and error) in $6$ M$_\ast$ bins superimposed with gray squares.}\label{mass_scatter}
\end{figure}

In this Appendix we show how the clumpiness compares with other parameters considered in the main part of the paper. The Fig. \ref{sample_cuts} displays the clumpiness as a function of the i-band magnitude and the elongation of our targets. This explains the final sample selection made for our analysis (described in Sections \ref{cuts} and 3), highlighted by the gray vertical lines.  %In particular, XXX
In Fig. \ref{clumpiness_correlations} is presented the comparison of the clumpiness with different morphological indicators, that are, the Gini coefficient, the M$_{20}$ parameter and the $mergerness$. The first two diagrams show the lack of correlation between the two quantities in the $x$ and $y$-axis. %, even though a slightly increasing trend of $c$ is found at higher Gini. In the last diagram can be noticed a much steeper increase of the clumpiness with the $mergerness$, which is used to define morphological mergers in Section~2.3.
Finally, Fig. \ref{redshift_scatter}-top shows the redshift distribution of our sample, ranging $0.5<z<0.9$ and with a median $z$ of $0.73$. Fig. \ref{redshift_scatter}-bottom and Fig. \ref{mass_scatter} demonstrate, respectively, that the clumpiness does not depend on the stellar mass, and does not significantly evolve with redshift (within our uncertainties). 

\clearpage

%\begin{sidewaystable*}
\begin{table*}[t!]
%\rule{+2cm}{2.cm}
\centering
\vspace{-0.2cm}
%\Large{Morphological properties and clumpiness for the sample analyzed in this paper}
%\begin{adjustbox}{width={\textwidth},totalheight={8cm},angle=0,keepaspectratio} 
\rule{0cm}{-2.9cm}
\centering
%\rule{-5.15cm}{-30.101cm} 
%\centering
%\caption{Optical+NIR emission line fluxes, A$_{V,tot}$ and sizes (of the Magellan sample)}
%\label{table2}
\resizebox{\textwidth}{!}{%
\centering
\rule{+0.8cm}{0.cm}
\renewcommand{\arraystretch}{0.9}
\begin{tabular}{|l|l|l|l|l|l|l|l|l|}
\hline
%\rule{0pt}{1pt}
\vspace{-0.2cm}
& & & & & & & & \\
\textbf{ID} & \textbf{RA} & \textbf{DEC}  & \textbf{redshift} &  \textbf{c$'$}  &  \textbf{clumpiness} & \textbf{mergerness} & M$_\ast$ [M$_\odot$] & SFR$_\text{IR}$ [M$_\odot$/yr]  \\  
\hline
  217536   &      149.524108   &      1.606315   &      0.87   $\pm$ 0.03   &       0.02    &       0.02    &       0.07    &       10.4    &       74.25   $\pm$ 6.633   \\ 
  220514   &      149.58302    &      1.61253    &      0.89   $\pm$ 0.03   &       0.01    &       0.01    &       0.07    &       10.9    &       49.96   $\pm$ 4.284   \\ 
  222723   &      150.173213   &      1.616315   &      0.61   $\pm$ 0.01   &       0.0     &       0.0     &       0.06    &       10.8    &       151.2   $\pm$ 6.357   \\ 
  222977   &      150.414727   &      1.618189   &      0.75   $\pm$ 0.0    &       0.03    &       0.04    &       0.04    &       10.1    &       15.29   $\pm$ 2.6     \\ 
  223715   &      149.765371   &      1.617017   &      0.52   $\pm$ 0.0    &       0.09    &       0.1     &       -0.03   &       10.7    &       52.4    $\pm$ 0.805   \\ 
  228149   &      149.816844   &      1.627891   &      0.63   $\pm$ 0.12   &       0.06    &       0.07    &       0.01    &       10.2    &       14.57   $\pm$ 6.078   \\ 
  239406   &      150.338269   &      1.644141   &      0.73   $\pm$ 0.02   &       0.04    &       0.04    &       0.08    &       10.3    &       12.82   $\pm$ 1.733   \\ 
  244932   &      149.460788   &      1.651282   &      0.72   $\pm$ 0.02   &       0.01    &       0.01    &       0.06    &       10.7    &       19.61   $\pm$ 2.242   \\ 
  245158   &      150.188543   &      1.654977   &      0.52   $\pm$ 0.0    &       0.05    &       0.06    &       0.07    &       10.8    &       83.05   $\pm$ 1.227   \\ 
  247210   &      150.487543   &      1.656873   &      0.83   $\pm$ 0.0    &       0.05    &       0.05    &       -0.02   &       10.1    &       41.72   $\pm$ 1.106   \\ 
  249989   &      150.685402   &      1.661076   &      0.74   $\pm$ 0.04   &       0.06    &       0.06    &       -0.09   &       10.6    &       102.4   $\pm$ 14.72   \\ 
  259328   &      149.916036   &      1.676354   &      0.74   $\pm$ 0.0    &       0.03    &       0.03    &       0.08    &       10.1    &       26.09   $\pm$ 0.57    \\ 
  271645   &      149.607229   &      1.694341   &      0.74   $\pm$ 0.08   &       0.0     &       0.0     &       0.14    &       11.0    &       15.24   $\pm$ 3.196   \\ 
  275629   &      150.182349   &      1.700807   &      0.74   $\pm$ 0.0    &       0.14    &       0.19    &       -0.11   &       10.6    &       101.3   $\pm$ 0.924   \\ 
  280708   &      149.969049   &      1.710048   &      0.86   $\pm$ 0.0    &       0.09    &       0.1     &       0.02    &       10.1    &       36.63   $\pm$ 4.083   \\ 
  286337   &      149.808093   &      1.717684   &      0.75   $\pm$ 0.0    &       0.0     &       0.0     &       0.08    &       10.6    &       17.23   $\pm$ 1.73    \\ 
  287281   &      149.662106   &      1.719228   &      0.71   $\pm$ 0.02   &       0.06    &       0.06    &       0.01    &       10.7    &       44.15   $\pm$ 5.46    \\ 
  289993   &      149.951404   &      1.725684   &      0.68   $\pm$ 0.0    &       0.01    &       0.01    &       -0.03   &       10.1    &       42.29   $\pm$ 0.49    \\ 
  293450   &      149.496283   &      1.729198   &      0.86   $\pm$ 0.06   &       0.06    &       0.07    &       0.01    &       10.2    &       45.48   $\pm$ 7.794   \\ 
  294096   &      149.51561    &      1.731195   &      0.67   $\pm$ 0.06   &       0.01    &       0.02    &       -0.03   &       10.0    &       4.835   $\pm$ 1.635   \\ 
  295213   &      150.421274   &      1.731761   &      0.64   $\pm$ 0.04   &       0.07    &       0.07    &       0.06    &       10.8    &       47.17   $\pm$ 6.101   \\ 
  297678   &      149.582631   &      1.738114   &      0.87   $\pm$ 0.03   &       0.07    &       0.08    &       0.06    &       10.4    &       78.84   $\pm$ 5.764   \\ 
  300643   &      149.779375   &      1.741341   &      0.53   $\pm$ 0.0    &       0.02    &       0.02    &       0.03    &       10.3    &       5.695   $\pm$ 1.548   \\ 
  308894   &      150.259527   &      1.756639   &      0.88   $\pm$ 0.0    &       0.04    &       0.05    &       0.06    &       10.1    &       25.86   $\pm$ 3.1     \\ 
  314674   &      149.922125   &      1.764706   &      0.89   $\pm$ 0.01   &       0.06    &       0.07    &       0.05    &       10.5    &       22.89   $\pm$ 3.078   \\ 
  319784   &      150.073982   &      1.772639   &      0.73   $\pm$ 0.0    &       0.02    &       0.02    &       -0.08   &       10.1    &       47.72   $\pm$ 0.947   \\ 
  326384   &      149.517858   &      1.783572   &      0.71   $\pm$ 0.04   &       0.1     &       0.13    &       -0.02   &       10.3    &       125.0   $\pm$ 14.8    \\ 
  327954   &      149.493504   &      1.786618   &      0.61   $\pm$ 0.11   &       0.02    &       0.02    &       0.11    &       10.0    &       9.337   $\pm$ 4.839   \\ 
  331390   &      150.640087   &      1.78974    &      0.67   $\pm$ 0.0    &       0.01    &       0.01    &       0.06    &       10.1    &       10.15   $\pm$ 2.1     \\ 
  331485   &      149.494555   &      1.791539   &      0.71   $\pm$ 0.02   &       0.02    &       0.04    &       0.06    &       10.3    &       15.02   $\pm$ 2.365   \\ 
  332156   &      150.23588    &      1.791352   &      0.83   $\pm$ 0.16   &       0.0     &       0.0     &       0.14    &       10.9    &       29.63   $\pm$ 11.85   \\ 
  340778   &      150.458433   &      1.803992   &      0.75   $\pm$ 0.0    &       0.0     &       0.0     &       0.15    &       11.2    &       39.87   $\pm$ 1.179   \\ 
  350404   &      150.20613    &      1.822038   &      0.74   $\pm$ 0.0    &       0.0     &       0.0     &       0.05    &       10.3    &       14.75   $\pm$ 4.196   \\ 
  358661   &      149.710063   &      1.832255   &      0.5    $\pm$ 0.13   &       0.0     &       0.0     &       0.12    &       10.3    &       6.222   $\pm$ 3.722   \\ 
  366376   &      150.138331   &      1.844028   &      0.57   $\pm$ 0.0    &       0.08    &       0.09    &       -0.21   &       10.2    &       38.12   $\pm$ 0.414   \\ 
  368717   &      149.718723   &      1.849718   &      0.67   $\pm$ 0.0    &       0.14    &       0.16    &       -0.17   &       10.7    &       67.19   $\pm$ 0.758   \\ 
  371578   &      150.674055   &      1.85337    &      0.51   $\pm$ 0.0    &       0.04    &       0.04    &       0.0     &       10.1    &       26.9    $\pm$ 0.222   \\ 
  371675   &      149.901208   &      1.85448    &      0.74   $\pm$ 0.0    &       0.01    &       0.01    &       0.08    &       10.1    &       8.38    $\pm$ 2.119   \\ 
  371886   &      149.900477   &      1.854753   &      0.64   $\pm$ 0.0    &       0.01    &       0.02    &       0.05    &       10.6    &       7.781   $\pm$ 1.071   \\ 
  372591   &      150.503808   &      1.854601   &      0.75   $\pm$ 0.0    &       0.07    &       0.08    &       0.01    &       10.4    &       36.58   $\pm$ 1.03    \\ 
  %386128   &      150.52365    &      1.876848   &      0.89   $\pm$ 0.03   &       0.01    &       0.01    &       0.09    &       10.5    &       24.85   $\pm$ 2.378   \\ 
  387242   &      150.222431   &      1.878024   &      0.84   $\pm$ 0.0    &       0.08    &       0.1     &       -0.1    &       10.2    &       51.69   $\pm$ 1.204   \\ 
  387454   &      150.652196   &      1.877237   &      0.59   $\pm$ 0.0    &       0.08    &       0.09    &       0.04    &       10.5    &       64.74   $\pm$ 1.105   \\ 
  387747   &      149.916015   &      1.879961   &      0.84   $\pm$ 0.0    &       0.07    &       0.08    &       -0.0    &       10.8    &       102.4   $\pm$ 0.974   \\ 
  400118   &      150.051501   &      1.898621   &      0.57   $\pm$ 0.0    &       0.1     &       0.12    &       -0.0    &       10.2    &       11.69   $\pm$ 1.625   \\ 
  402258   &      150.648684   &      1.902372   &      0.74   $\pm$ 0.0    &       0.09    &       0.1     &       -0.01   &       10.1    &       42.46   $\pm$ 0.332   \\ 
  409602   &      149.578964   &      1.913372   &      0.83   $\pm$ 0.0    &       0.07    &       0.08    &       0.05    &       10.2    &       52.3    $\pm$ 0.685   \\ 
  409814   &      150.198597   &      1.914859   &      0.87   $\pm$ 0.0    &       0.02    &       0.02    &       0.06    &       10.2    &       46.7    $\pm$ 0.862   \\ 
  412250   &      150.741711   &      1.91764    &      0.68   $\pm$ 0.03   &       0.1     &       0.14    &       -0.06   &       10.3    &       109.4   $\pm$ 10.24   \\ 
  416314   &      150.663036   &      1.924054   &      0.83   $\pm$ 0.0    &       0.0     &       0.01    &       -0.07   &       10.2    &       12.06   $\pm$ 0.433   \\ 
  418804   &      149.935142   &      1.927629   &      0.68   $\pm$ 0.0    &       0.07    &       0.08    &       -0.02   &       10.3    &       47.2    $\pm$ 0.721   \\ 
  431551   &      149.568139   &      1.94866    &      0.54   $\pm$ 0.02   &       0.01    &       0.01    &       0.11    &       10.9    &       13.7    $\pm$ 1.814   \\ 
  431596   &      149.948236   &      1.950357   &      0.85   $\pm$ 0.0    &       0.03    &       0.03    &       0.07    &       10.1    &       15.5    $\pm$ 0.389   \\ 
  436769   &      150.617708   &      1.958166   &      0.65   $\pm$ 0.08   &       0.0     &       0.0     &        0.01   &       10.5    &       14.6    $\pm$ 5.141   \\ 
  439419   &      150.301785   &      1.962747   &      0.7    $\pm$ 0.02   &       0.01    &       0.01    &       0.02    &       10.4    &       15.06   $\pm$ 1.283   \\ 
  444878   &      150.687694   &      1.970867   &      0.76   $\pm$ 0.12   &       0.16    &       0.19    &       -0.01   &       10.4    &       85.03   $\pm$ 27.02   \\ 
  447374   &      150.543608   &      1.974443   &      0.79   $\pm$ 0.03   &       0.06    &       0.07    &       0.1     &       10.8    &       26.29   $\pm$ 3.279   \\ 
  451272   &      150.250512   &      1.980928   &      0.72   $\pm$ 0.0    &       0.08    &       0.09    &       -0.03   &       10.3    &       45.3    $\pm$ 0.57    \\ 
  451426   &      149.651824   &      1.981124   &      0.79   $\pm$ 0.0    &       0.0     &       0.0     &       0.1     &       10.5    &       14.85   $\pm$ 3.362   \\ 
  466112   &      149.999276   &      2.005995   &      0.76   $\pm$ 0.0    &       0.05    &       0.06    &       -0.04   &       10.3    &       144.0   $\pm$ 1.052   \\ 
  472775   &      150.481476   &      2.013621   &      0.68   $\pm$ 0.02   &       0.02    &       0.02    &       0.09    &       10.8    &       159.7   $\pm$ 11.82   \\ 
  473147   &      150.399682   &      2.014254   &      0.72   $\pm$ 0.0    &       0.05    &       0.05    &       0.02    &       11.0    &       107.1   $\pm$ 1.269   \\ 
  %473560   &      150.429777   &      2.01896    &      0.79   $\pm$ 0.1    &       0.08    &       0.1     &       0.05    &       10.5    &       22.55   $\pm$ 6.319   \\ 
  474838   &      149.543724   &      2.01889    &      0.77   $\pm$ 0.02   &       0.03    &       0.04    &       0.09    &       10.4    &       58.87   $\pm$ 3.831   \\ 
  475972   &      149.526013   &      2.020502   &      0.62   $\pm$ 0.0    &       0.0     &       0.0     &       0.05    &       10.7    &       8.241   $\pm$ 2.693   \\ 
  478933   &      150.666347   &      2.025354   &      0.8    $\pm$ 0.0    &       0.13    &       0.13    &       -0.03   &       10.0    &       53.94   $\pm$ 0.785   \\ 
  480125   &      149.749497   &      2.027741   &      0.7    $\pm$ 0.0    &       0.0     &       0.01    &       0.05    &       11.3    &       31.46   $\pm$ 1.037   \\ 
  %480339   &      150.205592   &      2.028766   &      0.72   $\pm$ 0.05   &       0.0     &       0.0     &       0.08    &       10.0    &       6.73    $\pm$ 1.602   \\ 
  486542   &      150.761399   &      2.040174   &      0.88   $\pm$ 0.0    &       0.06    &       0.07    &       -0.05   &       10.2    &       103.9   $\pm$ 2.043   \\
\hline
\end{tabular} 

}
\hfill
%\end{table}
%\caption{\normalsize Table columns: (1) Identifica}\label{table2}
%\end{adjustbox}
\end{table*}

\newpage
\pagebreak

\begin{table*}[t!]
%\rule{+2cm}{2.cm}
\centering
\vspace{-0.2cm}
%\Large{Morphological properties and clumpiness for the sample analyzed in this paper}
%\begin{adjustbox}{width={\textwidth},totalheight={8cm},angle=0,keepaspectratio} 
\rule{0cm}{-2.9cm}
\centering
%\rule{-5.15cm}{-30.101cm} 
%\centering
%\caption{Optical+NIR emission line fluxes, A$_{V,tot}$ and sizes (of the Magellan sample)}
%\label{table2}
\resizebox{\textwidth}{!}{%
\centering
\rule{+0.8cm}{0.cm}
\renewcommand{\arraystretch}{0.9}
\begin{tabular}{|l|l|l|l|l|l|l|l|l|}
\hline
%\rule{0pt}{1pt}
\vspace{-0.2cm}
& & & & & & & & \\
\textbf{ID} & \textbf{RA} & \textbf{DEC}  & \textbf{redshift} &  \textbf{c$'$}  &  \textbf{clumpiness} & \textbf{mergerness} & M$_\ast$ [M$_\odot$] & SFR$_\text{IR}$ [M$_\odot$/yr]  \\   
\hline 
  488940   &      149.831475   &      2.040881   &      0.53   $\pm$ 0.13   &       0.01    &       0.01    &       0.05    &       10.1    &       4.905   $\pm$ 2.899   \\ 
  493881   &      150.749674   &      2.047066   &      0.6    $\pm$ 0.0    &       0.08    &       0.09    &       -0.01   &       10.7    &       129.2   $\pm$ 1.006   \\ 
  500548   &      149.62879    &      2.05949    &      0.63   $\pm$ 0.0    &       0.02    &       0.03    &       0.0     &       10.7    &       67.56   $\pm$ 1.033   \\ 
  503971   &      150.65256    &      2.063559   &      0.6    $\pm$ 0.0    &       0.04    &       0.05    &       0.03    &       11.0    &       9.35    $\pm$ 1.21    \\ 
  505311   &      150.089164   &      2.065345   &      0.73   $\pm$ 0.0    &       0.03    &       0.03    &       -0.02   &       10.9    &       34.49   $\pm$ 3.37    \\ 
  506131   &      149.528615   &      2.0665     &      0.57   $\pm$ 0.0    &       0.06    &       0.08    &       -0.0    &       10.2    &       39.06   $\pm$ 0.442   \\ 
  508753   &      149.928339   &      2.07214    &      0.68   $\pm$ 0.0    &       0.04    &       0.05    &       0.03    &       10.2    &       23.69   $\pm$ 0.944   \\ 
  508907   &      149.856873   &      2.072084   &      0.69   $\pm$ 0.0    &       0.05    &       0.06    &       0.0     &       10.0    &       15.84   $\pm$ 1.112   \\ 
  %513839   &      150.321074   &      2.08064    &      0.89   $\pm$ 0.0    &       0.01    &       0.01    &       0.11    &       10.2    &       23.74   $\pm$ 5.231   \\ 
  515278   &      150.523885   &      2.082297   &      0.89   $\pm$ 0.0    &       0.11    &       0.13    &       0.06    &       10.4    &       28.73   $\pm$ 3.716   \\ 
  516551   &      150.205761   &      2.083436   &      0.78   $\pm$ 0.01   &       0.01    &       0.01    &       0.07    &       10.1    &       44.4    $\pm$ 2.191   \\ 
  518855   &      150.12551    &      2.08703    &      0.89   $\pm$ 0.0    &       0.0     &       0.01    &       0.12    &       10.8    &       142.3   $\pm$ 1.436   \\ 
  519651   &      150.430196   &      2.086883   &      0.66   $\pm$ 0.0    &       0.05    &       0.06    &       0.07    &       10.5    &       137.1   $\pm$ 1.312   \\ 
  529521   &      150.103938   &      2.104858   &      0.83   $\pm$ 0.0    &       0.03    &       0.03    &       0.12    &       10.7    &       15.19   $\pm$ 3.556   \\ 
  532321   &      149.893588   &      2.107745   &      0.74   $\pm$ 0.07   &       0.01    &       0.01    &       -0.09   &       10.0    &       52.17   $\pm$ 10.76   \\ 
  536590   &      149.918602   &      2.113077   &      0.72   $\pm$ 0.06   &       0.01    &       0.01    &       0.04    &       10.5    &       10.65   $\pm$ 3.274   \\ 
  539760   &      149.880566   &      2.120673   &      0.68   $\pm$ 0.0    &       0.04    &       0.05    &       0.08    &       10.5    &       15.08   $\pm$ 1.726   \\ 
  544522   &      150.052099   &      2.126677   &      0.66   $\pm$ 0.0    &       0.1     &       0.12    &       0.06    &       10.9    &       90.06   $\pm$ 0.794   \\ 
  545104   &      150.451931   &      2.127888   &      0.84   $\pm$ 0.0    &       0.0     &       0.0     &       0.12    &       10.3    &       85.6    $\pm$ 1.039   \\ 
  545185   &      149.528016   &      2.12725    &      0.54   $\pm$ 0.0    &       0.03    &       0.03    &       -0.04   &       10.4    &       127.3   $\pm$ 1.155   \\ 
  546483   &      150.674581   &      2.131899   &      0.9    $\pm$ 0.0    &       0.0     &       0.0     &       0.07    &       10.0    &       39.28   $\pm$ 1.527   \\ 
  562400   &      150.512919   &      2.152049   &      0.56   $\pm$ 0.0    &       0.01    &       0.01    &       0.01    &       10.3    &       40.5    $\pm$ 0.553   \\ 
  571040   &      149.509664   &      2.167659   &      0.84   $\pm$ 0.0    &       0.02    &       0.02    &       0.16    &       10.4    &       18.17   $\pm$ 1.262   \\ 
  574334   &      150.346743   &      2.170359   &      0.85   $\pm$ 0.0    &       0.1     &       0.12    &       -0.01   &       10.4    &       132.5   $\pm$ 1.197   \\ 
  577143   &      149.887635   &      2.175504   &      0.78   $\pm$ 0.0    &       0.0     &       0.0     &       0.11    &       10.7    &       15.76   $\pm$ 1.443   \\ 
  580153   &      150.259862   &      2.181773   &      0.67   $\pm$ 0.0    &       0.01    &       0.02    &       0.05    &       10.2    &       5.035   $\pm$ 1.073   \\ 
  581920   &      150.23722    &      2.183978   &      0.7    $\pm$ 0.0    &       0.08    &       0.09    &       0.01    &       10.2    &       17.91   $\pm$ 1.338   \\ 
  584271   &      149.729701   &      2.186421   &      0.59   $\pm$ 0.0    &       0.0     &       0.0     &       0.1     &       10.6    &       16.49   $\pm$ 0.88    \\ 
  586243   &      150.626379   &      2.187925   &      0.6    $\pm$ 0.0    &       0.01    &       0.01    &       0.08    &       10.0    &       5.475   $\pm$ 1.482   \\ 
  586666   &      149.68814    &      2.191839   &      0.89   $\pm$ 0.04   &       0.01    &       0.01    &       -0.11   &       11.0    &       146.4   $\pm$ 13.37   \\ 
  586799   &      150.024476   &      2.190772   &      0.76   $\pm$ 0.0    &       0.01    &       0.02    &       0.08    &       10.4    &       9.945   $\pm$ 1.957   \\ 
  587164   &      149.605886   &      2.190462   &      0.54   $\pm$ 0.0    &       0.06    &       0.07    &       0.02    &       10.8    &       13.38   $\pm$ 0.944   \\ 
  587556   &      150.211575   &      2.191062   &      0.87   $\pm$ 0.0    &       0.01    &       0.01    &       0.07    &       10.9    &       25.36   $\pm$ 0.477   \\ 
  588922   &      149.682152   &      2.193682   &      0.73   $\pm$ 0.0    &       0.07    &       0.07    &       -0.02   &       10.4    &       23.22   $\pm$ 0.597   \\ 
  601470   &      150.719972   &      2.211767   &      0.89   $\pm$ 0.03   &       0.03    &       0.03    &       0.06    &       10.5    &       34.8    $\pm$ 5.655   \\ 
  606235   &      150.508123   &      2.219968   &      0.82   $\pm$ 0.0    &       0.01    &       0.01    &       0.03    &       10.2    &       48.24   $\pm$ 0.716   \\ 
  607625   &      150.033442   &      2.221992   &      0.89   $\pm$ 0.03   &       0.05    &       0.06    &       0.0     &       10.3    &       14.44   $\pm$ 1.223   \\ 
  608991   &      150.478377   &      2.22183    &      0.83   $\pm$ 0.0    &       0.03    &       0.03    &       -0.0    &       10.5    &       45.74   $\pm$ 2.175   \\ 
  609835   &      150.678868   &      2.225219   &      0.77   $\pm$ 0.01   &       0.0     &       0.0     &       0.04    &       10.2    &       59.46   $\pm$ 2.104   \\ 
  619015   &      150.10376    &      2.237664   &      0.8    $\pm$ 0.0    &       0.02    &       0.03    &       0.08    &       10.7    &       100.7   $\pm$ 1.634   \\ 
  620032   &      150.626231   &      2.240874   &      0.66   $\pm$ 0.0    &       0.04    &       0.05    &       0.06    &       10.0    &       9.124   $\pm$ 0.871   \\ 
  622611   &      149.664939   &      2.241806   &      0.89   $\pm$ 0.0    &       0.01    &       0.01    &       0.06    &       11.3    &       12.98   $\pm$ 3.813   \\ 
  624991   &      149.9687     &      2.247664   &      0.85   $\pm$ 0.0    &       0.02    &       0.02    &       0.06    &       10.0    &       17.43   $\pm$ 1.755   \\ 
  625435   &      149.538265   &      2.248029   &      0.84   $\pm$ 0.0    &       0.01    &       0.01    &       0.08    &       10.0    &       26.65   $\pm$ 1.369   \\ 
  629919   &      150.477167   &      2.252186   &      0.84   $\pm$ 0.0    &       0.0     &       0.0     &       0.09    &       11.3    &       137.9   $\pm$ 9.384   \\ 
  %636240   &      149.575218   &      2.264885   &      0.71   $\pm$ 0.12   &       0.0     &       0.0     &       0.12    &       10.1    &       12.61   $\pm$ 4.708   \\ 
  638793   &      149.529624   &      2.266751   &      0.58   $\pm$ 0.0    &       0.06    &       0.07    &       -0.02   &       10.3    &       23.98   $\pm$ 1.356   \\ 
  650054   &      150.400428   &      2.286241   &      0.83   $\pm$ 0.16   &       0.05    &       0.06    &       0.09    &       10.0    &       20.26   $\pm$ 6.12    \\ 
  654259   &      149.872198   &      2.289675   &      0.7    $\pm$ 0.0    &       0.0     &       0.0     &       0.08    &       10.8    &       103.4   $\pm$ 1.199   \\ 
  %654362   &      149.863031   &      2.292513   &      0.87   $\pm$ 0.05   &       0.0     &       0.0     &       0.07    &       10.2    &       51.53   $\pm$ 9.556   \\ 
  666872   &      150.537907   &      2.310505   &      0.84   $\pm$ 0.0    &       0.0     &       0.01    &       0.07    &       10.6    &       21.74   $\pm$ 3.212   \\ 
  668065   &      149.937063   &      2.312279   &      0.52   $\pm$ 0.0    &       0.05    &       0.05    &       0.02    &       10.2    &       7.974   $\pm$ 1.205   \\ 
  668738   &      150.210203   &      2.31168    &      0.75   $\pm$ 0.0    &       0.07    &       0.09    &       -0.04   &       10.8    &       158.9   $\pm$ 1.188   \\ 
  668769   &      149.67653    &      2.313653   &      0.69   $\pm$ 0.0    &       0.02    &       0.02    &       0.1     &       10.9    &       41.91   $\pm$ 0.512   \\ 
  %671274   &      149.946693   &      2.317185   &      0.81   $\pm$ 0.01   &       0.02    &       0.02    &       0.1     &       10.0    &       54.33   $\pm$ 2.275   \\ 
  %680418   &      149.819751   &      2.331018   &      0.73   $\pm$ 0.0    &       0.06    &       0.08    &       0.08    &       10.4    &       33.44   $\pm$ 0.823   \\ 
  695086   &      150.05225    &      2.351018   &      0.72   $\pm$ 0.13   &       0.01    &       0.01    &       0.07    &       10.0    &       2.633   $\pm$ 1.312   \\ 
  %701586   &      150.170535   &      2.362247   &      0.73   $\pm$ 0.08   &       0.03    &       0.04    &       0.11    &       10.4    &       30.66   $\pm$ 7.542   \\ 
  706850   &      149.931017   &      2.370359   &      0.85   $\pm$ 0.03   &       0.15    &       0.16    &       -0.11   &       10.3    &       38.69   $\pm$ 1.226   \\ 
  708427   &      149.578884   &      2.370704   &      0.75   $\pm$ 0.0    &       0.06    &       0.07    &       0.02    &       10.6    &       70.09   $\pm$ 0.971   \\ 
  711307   &      150.53029    &      2.373861   &      0.66   $\pm$ 0.0    &       0.03    &       0.05    &       0.05    &       10.8    &       102.0   $\pm$ 1.356   \\ 
  720848   &      150.261788   &      2.391079   &      0.8    $\pm$ 0.0    &       0.01    &       0.01    &       -0.01   &       10.1    &       12.21   $\pm$ 1.839   \\ 
  721398   &      150.670506   &      2.391459   &      0.61   $\pm$ 0.0    &       0.08    &       0.08    &       0.01    &       10.3    &       14.31   $\pm$ 1.796   \\ 
  724796   &      150.597688   &      2.396014   &      0.58   $\pm$ 0.0    &       0.04    &       0.04    &       0.05    &       10.4    &       19.33   $\pm$ 0.723   \\ 
  739926   &      150.470841   &      2.419552   &      0.55   $\pm$ 0.0    &       0.0     &       0.0     &       0.11    &       10.6    &       4.908   $\pm$ 1.447   \\
  742281   &      150.641614   &      2.423434   &      0.73   $\pm$ 0.0    &       0.0     &       0.0     &       0.12    &       11.0    &       112.7   $\pm$ 0.43    \\ 
  743322   &      149.657291   &      2.425131   &      0.57   $\pm$ 0.0    &       0.1     &       0.12    &       0.13    &       10.6    &       9.331   $\pm$ 0.836   \\ 
  749126   &      150.427092   &      2.431647   &      0.88   $\pm$ 0.0    &       0.03    &       0.04    &       0.01    &       10.4    &       71.11   $\pm$ 0.756   \\ 
  753450   &      150.327658   &      2.44088    &      0.52   $\pm$ 0.0    &       0.02    &       0.02    &       0.05    &       10.2    &       5.353   $\pm$ 0.727   \\ 
\hline
\end{tabular} 
}
\hfill
%\end{table}
%\caption{\normalsize Table columns: (1) Identifica}\label{table2}
%\end{adjustbox}
\end{table*}

\newpage
\pagebreak

\begin{table*}[t!]
%\rule{+2cm}{2.cm}
\centering
\vspace{-0.2cm}
%\Large{Morphological properties and clumpiness for the sample analyzed in this paper}
%\begin{adjustbox}{width={\textwidth},totalheight={8cm},angle=0,keepaspectratio} 
\rule{0cm}{-2.9cm}
\centering
%\rule{-5.15cm}{-30.101cm} 
%\centering
%\caption{Optical+NIR emission line fluxes, A$_{V,tot}$ and sizes (of the Magellan sample)}
%\label{table2}
\resizebox{\textwidth}{!}{%
\centering
\rule{+0.8cm}{0.cm}
\renewcommand{\arraystretch}{0.9}
\begin{tabular}{|l|l|l|l|l|l|l|l|l|}
\hline
%\rule{0pt}{1pt}
\vspace{-0.2cm}
& & & & & & & & \\
\textbf{ID} & \textbf{RA} & \textbf{DEC}  & \textbf{redshift} &  \textbf{c$'$}  &  \textbf{clumpiness} & \textbf{mergerness} & M$_\ast$ [M$_\odot$] & SFR$_\text{IR}$ [M$_\odot$/yr]  \\ 
\hline
  764862   &      150.189794   &      2.45758    &      0.51   $\pm$ 0.0    &       0.08    &       0.09    &       -0.09   &       10.1    &       6.139   $\pm$ 0.292   \\ 
  773897   &      149.833747   &      2.481315   &      0.84   $\pm$ 0.0    &       0.12    &       0.14    &       0.0     &       10.2    &       40.46   $\pm$ 1.969   \\ 
  777034   &      150.150252   &      2.475166   &      0.69   $\pm$ 0.0    &       0.06    &       0.08    &       -0.05   &       10.8    &       211.7   $\pm$ 0.764   \\ 
  778756   &      150.548939   &      2.480459   &      0.8    $\pm$ 0.0    &       0.05    &       0.06    &       -0.05   &       10.5    &       41.11   $\pm$ 1.792   \\ 
  780365   &      150.398266   &      2.480753   &      0.85   $\pm$ 0.0    &       0.0     &       0.0     &       0.02    &       10.2    &       8.178   $\pm$ 3.15    \\ 
  783499   &      150.506268   &      2.486894   &      0.62   $\pm$ 0.0    &       0.0     &       0.0     &       0.11    &       10.3    &       10.15   $\pm$ 0.465   \\ 
  783743   &      149.93968    &      2.488034   &      0.72   $\pm$ 0.15   &       0.0     &       0.0     &       0.09    &       10.5    &       10.73   $\pm$ 3.42    \\ 
  %784789   &      149.786036   &      2.489983   &      0.84   $\pm$ 0.02   &       0.0     &       0.0     &       0.09    &       10.0    &       8.033   $\pm$ 1.431   \\ 
  789914   &      149.979212   &      2.498304   &      0.88   $\pm$ 0.0    &       0.0     &       0.0     &       0.08    &       10.1    &       9.062   $\pm$ 1.257   \\ 
  790685   &      150.370657   &      2.498115   &      0.82   $\pm$ 0.0    &       0.09    &       0.09    &       0.02    &       10.8    &       44.95   $\pm$ 0.6     \\ 
  794024   &      150.055643   &      2.504436   &      0.7    $\pm$ 0.0    &       0.07    &       0.1     &       0.04    &       10.3    &       50.4    $\pm$ 0.998   \\ 
  811857   &      149.913557   &      2.530162   &      0.69   $\pm$ 0.0    &       0.02    &       0.02    &       0.11    &       10.6    &       7.52    $\pm$ 1.842   \\ 
  819000   &      149.457523   &      2.539333   &      0.83   $\pm$ 0.02   &       0.05    &       0.05    &       0.02    &       10.3    &       58.92   $\pm$ 2.595   \\ 
  824508   &      149.862472   &      2.548484   &      0.74   $\pm$ 0.0    &       0.01    &       0.01    &       0.06    &       11.3    &       14.81   $\pm$ 0.813   \\ 
  830418   &      149.562546   &      2.557087   &      0.76   $\pm$ 0.03   &       0.0     &       0.0     &       0.03    &       10.7    &       39.12   $\pm$ 4.827   \\ 
  834449   &      150.1201     &      2.561479   &      0.5    $\pm$ 0.0    &       0.04    &       0.05    &       0.06    &       10.6    &       14.31   $\pm$ 0.228   \\ 
  837890   &      150.054801   &      2.569459   &      0.75   $\pm$ 0.0    &       0.03    &       0.03    &       0.13    &       10.5    &       156.8   $\pm$ 1.615   \\ 
  838188   &      149.77795    &      2.567252   &      0.72   $\pm$ 0.0    &       0.0     &       0.0     &       0.07    &       10.3    &       10.45   $\pm$ 0.524   \\ 
  840232   &      149.716601   &      2.572475   &      0.9    $\pm$ 0.01   &       0.04    &       0.05    &       0.09    &       10.7    &       117.4   $\pm$ 4.215   \\ 
  842149   &      150.342316   &      2.575943   &      0.75   $\pm$ 0.0    &       0.0     &       0.0     &       -0.03   &       10.2    &       54.86   $\pm$ 0.529   \\ 
  842173   &      149.733742   &      2.574995   &      0.71   $\pm$ 0.0    &       0.01    &       0.01    &       0.07    &       10.6    &       93.81   $\pm$ 1.129   \\ 
  844990   &      149.926835   &      2.580581   &      0.7    $\pm$ 0.0    &       0.04    &       0.05    &       0.1     &       10.2    &       14.41   $\pm$ 1.532   \\ 
  %845788   &      149.903112   &      2.582074   &      0.69   $\pm$ 0.0    &       0.0     &       0.0     &       0.14    &       10.5    &       8.617   $\pm$ 0.728   \\ 
  848785   &      150.42295    &      2.583241   &      0.81   $\pm$ 0.0    &       0.12    &       0.12    &       -0.09   &       10.1    &       83.81   $\pm$ 1.125   \\ 
  849397   &      149.976994   &      2.586831   &      0.73   $\pm$ 0.0    &       0.09    &       0.12    &       0.01    &       10.3    &       13.25   $\pm$ 1.373   \\ 
  %852698   &      150.655744   &      2.592759   &      0.57   $\pm$ 0.07   &       0.05    &       0.07    &       0.04    &       10.0    &       8.331   $\pm$ 2.692   \\ 
  853769   &      150.33873    &      2.593234   &      0.85   $\pm$ 0.0    &       0.0     &       0.0     &       0.12    &       10.9    &       154.6   $\pm$ 2.136   \\ 
  860071   &      150.27635    &      2.603899   &      0.7    $\pm$ 0.0    &       0.01    &       0.01    &       -0.08   &       10.1    &       7.279   $\pm$ 2.386   \\ 
  864706   &      150.531072   &      2.610881   &      0.79   $\pm$ 0.0    &       0.05    &       0.07    &       0.1     &       10.6    &       23.57   $\pm$ 1.955   \\ 
  865896   &      150.054261   &      2.612264   &      0.74   $\pm$ 0.0    &       0.0     &       0.0     &       -0.09   &       10.0    &       8.219   $\pm$ 0.209   \\ 
  866054   &      149.811584   &      2.610245   &      0.54   $\pm$ 0.0    &       0.04    &       0.05    &       0.09    &       10.7    &       14.02   $\pm$ 0.637   \\ 
  %871544   &      150.609538   &      2.622533   &      0.82   $\pm$ 0.16   &       0.03    &       0.04    &       0.09    &       10.4    &       20.38   $\pm$ 11.81   \\ 
  876155   &      150.564918   &      2.625331   &      0.65   $\pm$ 0.02   &       0.06    &       0.06    &       0.08    &       10.7    &       24.88   $\pm$ 1.983   \\ 
  878476   &      150.752013   &      2.62904    &      0.84   $\pm$ 0.11   &       0.03    &       0.05    &       0.0     &       10.4    &       18.84   $\pm$ 6.223   \\ 
  878551   &      150.046263   &      2.630447   &      0.76   $\pm$ 0.03   &       0.04    &       0.04    &       0.04    &       10.1    &       38.72   $\pm$ 4.129   \\ 
  880787   &      149.550391   &      2.632955   &      0.81   $\pm$ 0.06   &       0.01    &       0.01    &       0.07    &       10.9    &       23.8    $\pm$ 3.235   \\ 
  880925   &      149.937367   &      2.634136   &      0.71   $\pm$ 0.03   &       0.03    &       0.04    &       0.05    &       10.2    &       11.79   $\pm$ 1.779   \\ 
  886498   &      149.513205   &      2.643508   &      0.64   $\pm$ 0.0    &       0.05    &       0.06    &       +0.0    &       10.3    &       11.81   $\pm$ 0.759   \\ 
  887351   &      150.04403    &      2.643885   &      0.7    $\pm$ 0.0    &       0.03    &       0.03    &       -0.07   &       10.6    &       86.33   $\pm$ 0.814   \\ 
  893857   &      150.159952   &      2.654341   &      0.85   $\pm$ 0.0    &       0.09    &       0.11    &       -0.01   &       11.1    &       217.3   $\pm$ 2.864   \\ 
  894779   &      150.427097   &      2.656439   &      0.55   $\pm$ 0.0    &       0.05    &       0.07    &       -0.0    &       10.0    &       61.12   $\pm$ 0.869   \\ 
  901597   &      150.709924   &      2.667676   &      0.7    $\pm$ 0.08   &       0.04    &       0.05    &       0.06    &       10.1    &       44.59   $\pm$ 9.812   \\ 
  902885   &      150.136763   &      2.669235   &      0.7    $\pm$ 0.0    &       0.04    &       0.05    &       0.04    &       10.0    &       12.35   $\pm$ 0.37    \\ 
  909617   &      149.970409   &      2.677765   &      0.69   $\pm$ 0.0    &       0.02    &       0.03    &       0.04    &       10.8    &       22.88   $\pm$ 1.869   \\ 
  911723   &      149.681335   &      2.681084   &      0.58   $\pm$ 0.0    &       0.0     &       0.0     &       0.0     &       10.9    &       82.03   $\pm$ 0.502   \\ 
  912969   &      150.589787   &      2.684536   &      0.74   $\pm$ 0.0    &       0.1     &       0.1     &       -0.01   &       10.7    &       44.6    $\pm$ 2.707   \\ 
  915913   &      150.557165   &      2.688555   &      0.88   $\pm$ 0.0    &       0.02    &       0.02    &       0.03    &       10.1    &       20.01   $\pm$ 0.293   \\ 
  917600   &      150.029291   &      2.690141   &      0.79   $\pm$ 0.09   &       0.01    &       0.01    &       0.04    &       11.1    &       23.59   $\pm$ 6.409   \\ 
  921254   &      149.910328   &      2.695579   &      0.86   $\pm$ 0.14   &       0.0     &       0.0     &       0.08    &       11.1    &       39.36   $\pm$ 14.69   \\ 
  922401   &      150.035101   &      2.698678   &      0.65   $\pm$ 0.0    &       0.06    &       0.06    &       -0.04   &       10.6    &       18.22   $\pm$ 0.675   \\ 
  925981   &      150.433978   &      2.705011   &      0.88   $\pm$ 0.0    &       0.01    &       0.01    &       0.06    &       10.2    &       15.2    $\pm$ 1.928   \\ 
  927031   &      149.716671   &      2.705636   &      0.88   $\pm$ 0.0    &       0.0     &       0.0     &       0.15    &       11.1    &       155.3   $\pm$ 1.796   \\ 
  931390   &      150.586062   &      2.712214   &      0.66   $\pm$ 0.0    &       0.07    &       0.08    &       0.03    &       10.0    &       9.482   $\pm$ 2.196   \\ 
  937909   &      149.79431    &      2.722162   &      0.9    $\pm$ 0.02   &       0.05    &       0.05    &       0.04    &       10.6    &       28.88   $\pm$ 2.201   \\ 
  940851   &      150.539461   &      2.728218   &      0.66   $\pm$ 0.0    &       0.03    &       0.03    &       0.04    &       10.1    &       6.119   $\pm$ 0.435   \\ 
  946233   &      150.227312   &      2.737645   &      0.7    $\pm$ 0.09   &       0.0     &       0.0     &       0.07    &       10.6    &       29.97   $\pm$ 8.641   \\ 
  948557   &      149.627616   &      2.739277   &      0.55   $\pm$ 0.0    &       0.11    &       0.11    &       -0.06   &       10.1    &       25.99   $\pm$ 0.689   \\ 
  952356   &      149.886951   &      2.747311   &      0.88   $\pm$ 0.0    &       0.02    &       0.02    &       0.07    &       10.6    &       109.3   $\pm$ 1.016   \\ 
  960767   &      149.848315   &      2.760829   &      0.7    $\pm$ 0.0    &       0.1     &       0.11    &       0.03    &       10.4    &       78.83   $\pm$ 1.575   \\ 
  %964122   &      150.438196   &      2.766868   &      0.71   $\pm$ 0.0    &       0.0     &       0.0     &       0.08    &       10.8    &       18.74   $\pm$ 0.757   \\ 
  964786   &      150.126928   &      2.7654     &      0.57   $\pm$ 0.0    &       0.0     &       0.0     &       0.05    &       10.6    &       56.32   $\pm$ 0.274   \\ 
  970636   &      150.408721   &      2.777253   &      0.84   $\pm$ 0.09   &       0.01    &       0.01    &       0.08    &       10.9    &       114.8   $\pm$ 26.57   \\ 
  978011   &      150.155692   &      2.787663   &      0.73   $\pm$ 0.0    &       0.0     &       0.01    &       0.0     &       10.4    &       18.89   $\pm$ 0.559   \\ 
  981123   &      150.399371   &      2.794152   &      0.82   $\pm$ 0.0    &       0.01    &       0.01    &       0.04    &       10.6    &       170.4   $\pm$ 3.618   \\     
\hline
\end{tabular} 

}
\hfill
%\end{table}
%\caption{\normalsize Table columns: (1) Identifica}\label{table2}
%\end{adjustbox}
\caption{\normalsize Table columns: (1,2,3) Identification number, RA and DEC (in degree) from Laigle et al. (2016); (4) photometric redshift (or spectroscopic if available) from Jin et al. (2018); (5,6) clumpiness parameters (clumpiness is estimated as in equation~4, while in $c'$ the flux of the clumps is divided by the total flux of the galaxy (including the nuclei) inside the segmentation map; (7) $mergerness$ parameter, estimated as in equation~3; (8) stellar mass from Laigle et al. (2016), with typical uncertainty of $0.1$ dex; (9) SFR$_\text{IR}$ from Jin et al. (2018).}\label{table3}
%\end{adjustbox}
\end{table*}
%\end{table}

\end{document}